\documentclass[letterpaper,11pt]{article}
\pdfoutput=1

\usepackage{jheppub} 

\usepackage[utf8]{inputenc}
\usepackage{color}
\usepackage{tikz}
\usetikzlibrary{trees, arrows, decorations.pathmorphing, decorations.markings, shapes.geometric}
\usepackage{hepunits}
\usepackage[section]{placeins}
\usepackage{floatrow}
\usepackage{subfig}
\usepackage{units}

\makeatletter
\g@addto@macro\bfseries{\boldmath}
\makeatother

\newcommand{\Ref}[1]{Ref.~\cite{#1}}
\newcommand{\Refs}[1]{Refs.~\cite{#1}}
\newcommand{\Fig}[1]{Fig.~\ref{#1}}
\newcommand{\Figs}[2]{Figs.~\ref{#1} and \ref{#2}}
\newcommand{\Figss}[3]{Figs.~\ref{#1}, \ref{#2}, and \ref{#3}}

\newcommand{\Sec}[1]{Sec.~\ref{#1}}

\newcommand{\App}[1]{App.~\ref{#1}}

\newcommand{\Eq}[1]{Eq.~\eqref{#1}}
\newcommand{\Eqs}[2]{Eqs.~\eqref{#1} and \eqref{#2}}

\tikzset{
	particle/.style={draw=blue,thick, postaction={decorate},
		decoration={markings,mark=at position .5 with {\arrow{>}}}},
	antiparticle/.style={draw=blue,thick, postaction={decorate},
		decoration={markings,mark=at position .5 with {\arrow{<}}}},
	photon/.style={decorate, decoration={snake}, draw=black},
	electron/.style={draw=blue, postaction={decorate},
		decoration={markings,mark=at position .55 with {\arrow[draw=black]{>}}}},
	gluon/.style={decorate, draw=red,
		decoration={coil,amplitude=4pt, segment length=5pt}},
	scalar/.style={dashed, draw=black}
}

\preprint{\vbox{\hbox{MIT-CTP/5021}}}

\title{Aspects of Track-Assisted Mass}

\author{Benjamin T. Elder and}
\author{Jesse Thaler}

\affiliation{Center for Theoretical Physics, Massachusetts Institute of Technology, Cambridge, MA 02139, USA}
\emailAdd{benjamin.elder11@gmail.com}
\emailAdd{jthaler@mit.edu}

\abstract{
Track-assisted mass is a proxy for jet mass that only uses direction information from charged particles, allowing it to be measured at the Large Hadron Collider with very fine angular resolution. 
In this paper, we introduce a generalization of track-assisted mass and analyze its performance in both parton shower generators and resummed calculations. 
For the original track-assisted mass, the track-only mass is rescaled by the charged energy fraction of the jet.
In our generalization, the rescaling factor includes both per-jet and ensemble-averaged information, facilitating a closer correspondence to ordinary jet mass.
Using the track function formalism in electron-positron collisions, we calculate the spectrum of generalized track-assisted mass to next-to-leading-logarithmic order with leading-order matching. 
These resummed calculations provide theoretical insight into the close correspondence between track-assisted mass and ordinary jet mass.
With the growing importance of jet grooming algorithms, we also calculate track-assisted mass on soft-drop groomed jets.
}

\begin{document} 
\maketitle

\section{Introduction}


The Large Hadron Collider (LHC) is currently operating at a collision energy of 13 TeV, allowing it to produce electroweak-scale resonances---like $W$/$Z$ bosons, Higgs bosons, and top quarks---with very high Lorentz boosts. 
The typical angular separation between the products of a two-body decay $A \rightarrow BC$ is $\Delta R_{BC} \approx 2m_A/p_{T,A}$, so boosted resonances are often reconstructed as a single hadronic jet.
At the most extreme kinematics, the decay products can become so collimated that their separation is even below the typical hadronic (electromagnetic) calorimeter resolution of $0.1\times 0.1$ ($0.02\times 0.02$) in the rapidity-azimuth plane.
For example, the products of a decaying $W$ boson would become indistinguishable to a hadronic calorimeter cell at $p_T \approx 1.5~\TeV$.

On the other hand, the charged particle tracking detectors at the LHC experiments offer 10--100 times better angular resolution than the electromagnetic and hadronic calorimeters~\cite{Asquith:2018igt,Larkoski:2017jix}.
This has motivated the design of jet substructure observables which require direction information from only charged particles
\cite{Katz:2010mr,Son:2012mb,Calkins:2013ega,Schaetzel:2013vka,Larkoski:2015yqa,Spannowsky:2015eba,Bressler:2015uma}.\footnote{CMS performed a study of vector boson tagging at high $p_T$ and found that the hadronic calorimeter resolution was too coarse to effectively identify vector bosons at $p_T > 1.5$ TeV~\cite{CMS-PAS-JME-14-002}.  This study concluded that momenta reconstructed from the electromagnetic calorimeter would be the key component of high-$p_T$ vector boson tagging, but did not examine a purely track-based measurement.}
With the goal of improving the mass resolution of boosted objects and improving the stability of the calibrations, the ATLAS collaboration defined the track-assisted mass as~\cite{ATLAS:2016vmy}
\begin{equation}
\label{eq:tam-definition}
M_{\rm TA}  = M_{\rm track} \left(\frac{p_{T,\rm calo}}{p_{T,\rm track}}\right),
\end{equation}
where the track-only mass $M_{\rm track}$ is computed from charged particle tracking information, while the charged-to-neutral fraction $p_{T,\rm track}/p_{T,\rm calo}$ requires input from both tracking and calorimetry.
Throughout this paper, we use ``track'' to refer to just charged particles and ``calo'' to refer to all particles, even through algorithms like particle flow~\cite{Aaboud:2017aca,Sirunyan:2017ulk} determine ``calo'' quantities through a combination of tracking and calorimetry.
Because of approximate isospin conservation, $M_{\rm TA}$ is a good proxy for ordinary jet mass $M_{\rm calo}$. 
In addition, quantities like $M_{\rm track}$ defined in terms of just charged particles are more resilient to the impact of secondary pileup collisions~\cite{Asquith:2018igt,Larkoski:2017jix}.


In this paper, we introduce the generalized track-assisted mass (GTAM) and study its properties for ordinary quark and gluon jets.
Taking \Eq{eq:tam-definition} as a starting point, we define a two-parameter family of GTAM observables,
\begin{equation}
\label{eq:gtam-intro}
M_{\rm TA}^{(\kappa,\lambda)} = M_{\rm track} \left(\frac{p_{T,\rm calo}}{p_{T,\rm track}}\right)^\kappa \bigg\langle \frac{p_{T,\rm calo}}{p_{T,\rm track}} \bigg\rangle^\lambda\,, \qquad M_{\rm TA}^{(1,0)} \equiv M_{\rm TA}\,,
\end{equation}
where $\langle \cdot \rangle$ denotes an average over an ensemble of jets. 
The parameters $\kappa$ and $\lambda$ determine whether the charged-to-neutral fraction is estimated jet-by-jet or ensemble-by-ensemble.
One expects the dimensionless ratios $M_{\rm calo}/p_{T,\rm calo}$ and $M_{\rm track}/p_{T,\rm track}$ to be comparable, which suggests that the best proxy for $M_{\rm calo}$ should satisfy $\kappa + \lambda = 1$.
Through a combination of parton-shower studies and resummed calculations, we confirm this expectation.
Furthermore, we find that parameter values around $\kappa \simeq 0.5$ and $\lambda \simeq 0.5$ produce an observable which outperforms $M_{\rm TA}$ as a proxy for jet mass, at least for quark- and gluon-initiated jets.
We also study GTAM with soft-drop grooming, and find that it remains a good substitute for jet mass, but with a shift in the optimal parameter values which depends on the degree of grooming.

\begin{figure}[t]
	\centering
	\subfloat[]{
		\includegraphics[width=0.45\textwidth]{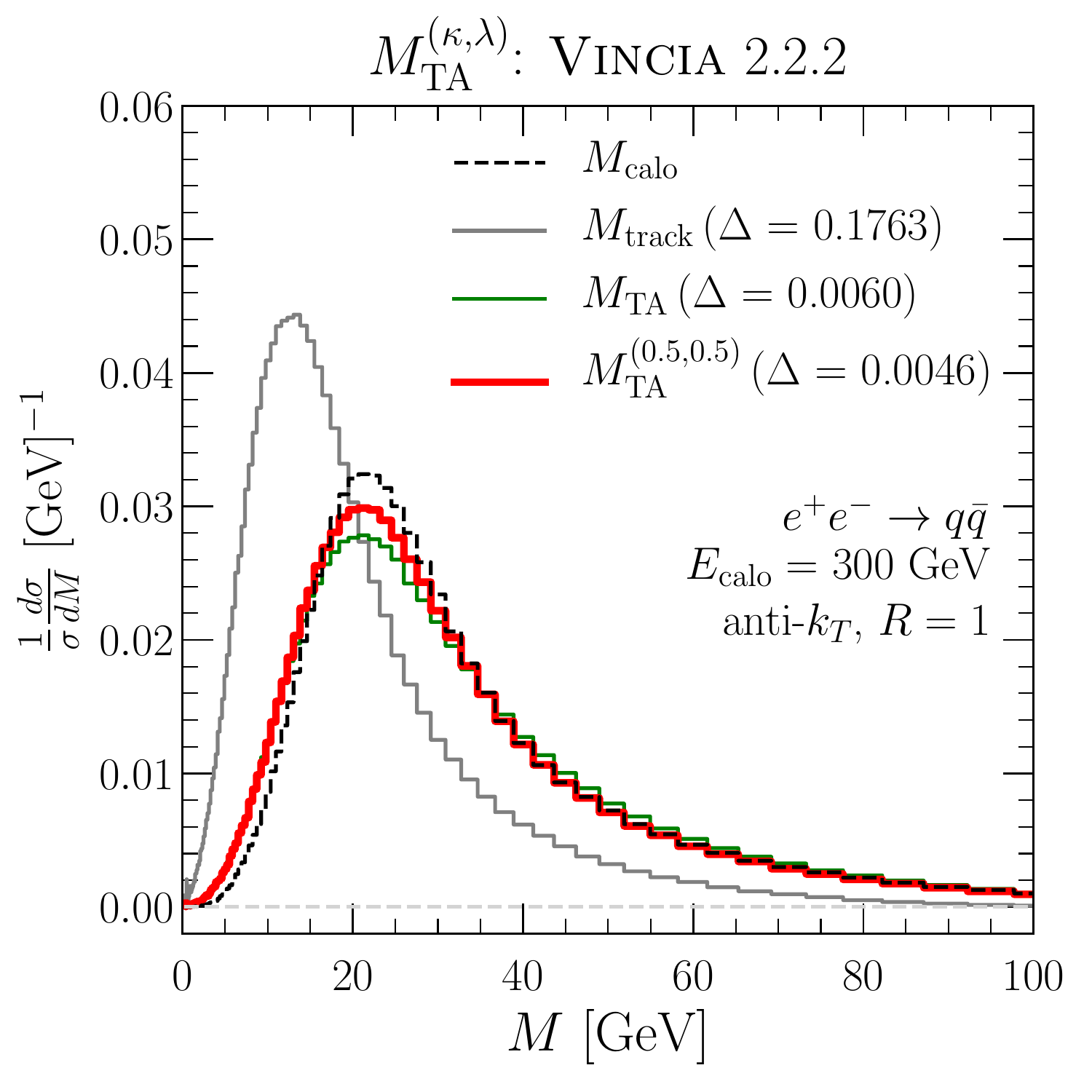}
		\label{fig:intro-comp-ps}
	}
	\subfloat[]{
		\includegraphics[width=0.45\textwidth]{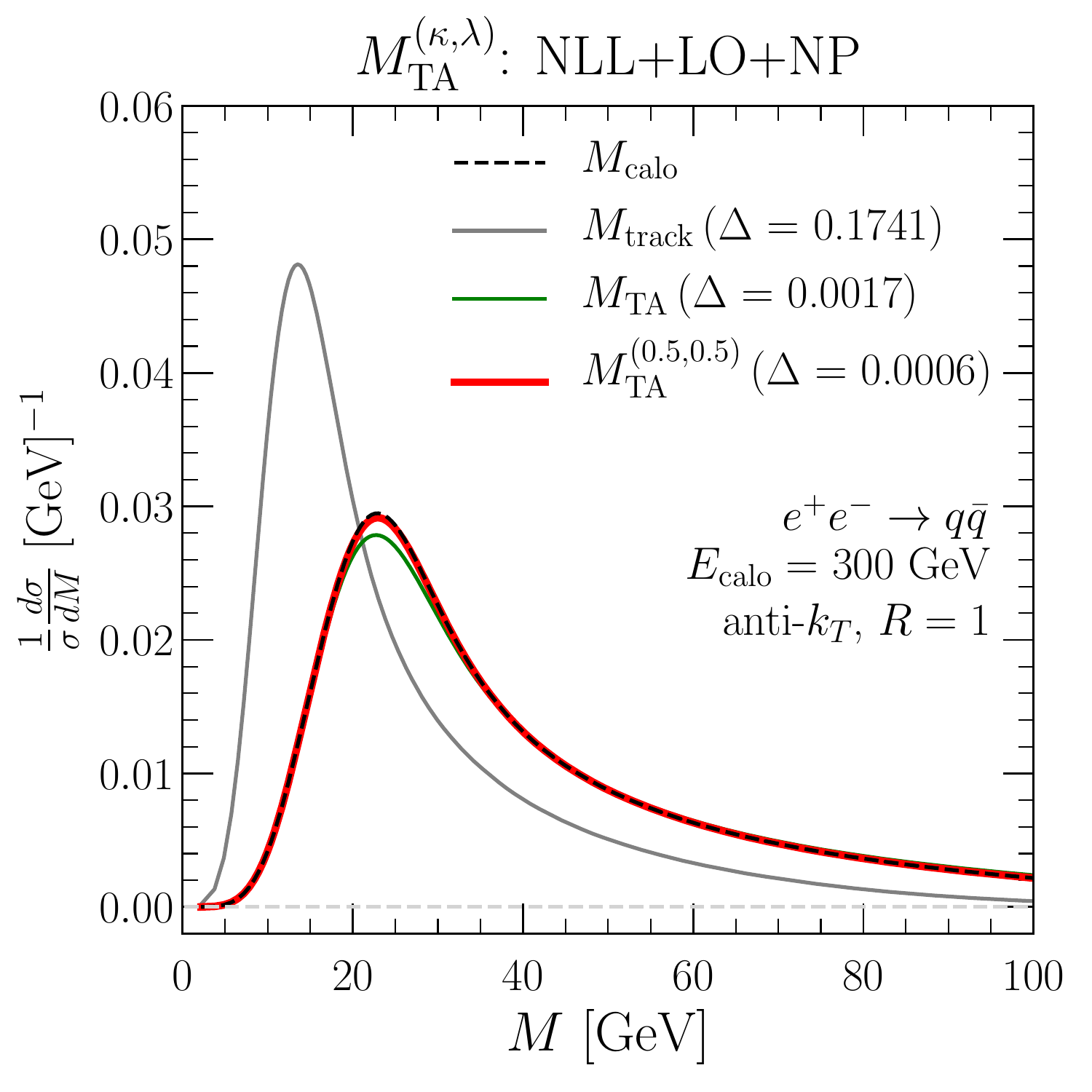}
		\label{fig:intro-comp-analytic}
	}
	\caption{Distributions of jet mass $M_{\rm calo}$ and GTAM $M_{\rm TA}^{(\kappa,\lambda)}$ in $e^+e^-$ collisions, extracted from (a) $\textsc{Vincia}$ 2.2.2 and (b) NLL+LO calculations convolved with a non-perturbative shape function.  The values of $\Delta$ in the legend correspond to the similarity measure introduced in \Eq{eq:dist}, with small values of $\Delta$ indicating a closer match to $M_{\rm calo}$.}
	\label{fig:intro-comp}
\end{figure}

To preview our results, distributions of jet mass $M_{\rm calo}$ and GTAM $M_{\rm TA}^{(\kappa,\lambda)}$ are plotted in \Fig{fig:intro-comp} for anti-$k_t$ jets~\cite{Cacciari:2008gp} in $e^+e^-$ annihilation.
Here, we show distributions (a) from the $\textsc{Vincia}$ 2.2.2~\cite{Giele:2007di,Fischer:2016vfv} parton shower plugin to $\textsc{Pythia}$ 8.230~\cite{Sjostrand:2006za,Sjostrand:2014zea} and (b) from our analytic calculations described below. 
As expected, $M_{\rm track}$ differs from $M_{\rm calo}$ by roughly a factor of $2/3$ (corresponding to equal fractions of $\pi^+$, $\pi^-$, and $\pi^0$).
Using standard track-assisted mass $M_{\rm TA}$ restores the desired peak location, but with some degree of smearing.
Our recommended GTAM default of $M_{\rm TA}^{(0.5,0.5)}$ gets even closer to matching the $M_{\rm calo}$ distribution.
By dimensional analysis, one could already guess that $\kappa + \lambda = 1$ would be preferred, and this intuition is borne out in our analytic calculations.
The precise relationship between $\kappa$ and $\lambda$ is sensitive to the details of the event sample and the accuracy of the calculation.
Eventually, experimental measurements will be needed to determine whether our recommendation of $\kappa \simeq \lambda\simeq 0.5$ indeed has the best performance as a jet-mass proxy.
In addition, GTAM can be studied as an observable in its own right, and our GTAM analytic calculations can be compared directly to experimental measurements for a range of $\kappa$ and $\lambda$ values.


The importance of jet mass as a collider observable cannot be overstated.
When the decay products of boosted objects become collimated, jet production cross sections cannot distinguish between boosted signal jets and QCD background jets. 
This challenge has spurred the development of many substructure techniques for tagging highly boosted objects
\cite{Butterworth:2008iy,Kaplan:2008ie,Almeida:2008yp,Ellis:2009su,Ellis:2009me,Krohn:2009th,Chekanov:2010vc,Soper:2010xk,Plehn:2010st,Thaler:2010tr,Cui:2010km,Hook:2011cq,Jankowiak:2011qa,Thaler:2011gf,Jouttenus:2013hs,Larkoski:2013eya,Dasgupta:2013ihk,Anders:2013oga,Larkoski:2014wba,Kasieczka:2015jma,Thaler:2015xaa,Moult:2016cvt,Komiske:2017aww}.
The most fundamental substructure observable is the jet mass, which has been computed at fixed order, and in resummed calculations to next-to-leading logarithmic (NLL) order \cite{Catani:1991bd,Li:2011hy,Dasgupta:2012hg,Chien:2012ur,Liu:2014oog}. 
The robustness of jet mass measurements can be improved using grooming techniques \cite{Butterworth:2008iy,Ellis:2009su,Ellis:2009me,Krohn:2009th,Dasgupta:2013ihk,Larkoski:2014wba}, which also serve to simplify the structure of theoretical calculations \cite{Frye:2016okc,Frye:2016aiz,Marzani:2017mva,Marzani:2017kqd}.
The mass of a boosted signal jet originates primarily from the decaying heavy resonance, while background jets produced by light quarks and gluons gain mass from collinear parton splitting governed by the DGLAP evolution equations \cite{Gribov:1972ri,Lipatov:1974qm,Dokshitzer:1977sg,Altarelli:1977zs}. 
A cut on the value of the jet mass can therefore be an important discriminant between signal and background jets~\cite{Khachatryan:2014vla,Aad:2015rpa}, which is why having excellent jet mass resolution is of paramount importance, perhaps achieved through track-assisted measurements.


We now give a detailed outline of the remainder of this paper.
In \Sec{sec:MCexplore}, we review the definition of track-assisted mass and then perform an exploratory parton shower study with \textsc{Vincia}, using ensembles of quark and gluon jets from $pp$ collisions at $E_{\rm CM} = 14$ TeV. 
We compare GTAM and ordinary jet mass for a range of $\kappa$ and $\lambda$ parameters.
The closest correspondence between the two occurs for $\kappa \approx 0.5$ and $\lambda \approx 0.5$, where we define the degree of similarity by a symmetric version of the $\chi^2$ statistic.
In \App{app:pure-quark-gluon}, we repeat this study for pure samples of quark/gluon jets (as defined by the \textsc{Vincia} hard process), and find that the optimal GTAM parameters are insensitive to the parton content of the jet.


In \Sec{sec:calculation}, we calculate the GTAM spectrum for quark- and gluon-initiated jets in $e^+e^-$ collisions.
The use of observables depending on only charged particles is theoretically complicated since these observables are not infrared and collinear (IRC) safe.
IRC safety guarantees a finite perturbative expansion order-by-order in $\alpha_s$~\cite{Sterman:1977wj}, whereas the perturbative spectra of unsafe observables exhibit unphysical divergences. 
Perturbative calculations of cross sections for a large class of collinear-unsafe observables can be performed with the track function formalism, or the broader generalized fragmentation function (GFF) formalism~\cite{Waalewijn:2012sv,Krohn:2012fg,Chang:2013rca,Chang:2013iba,Larkoski:2014pca,Elder:2017bkd}.
Just like ordinary fragmentation functions for inclusive single-hadron cross sections~\cite{Berman:1971xz,Mueller:1978xu,Mueller:1981sg,Collins:1981uk,Collins:1981uw,Brock:1993sz,Collins:2011zzd}, these methods absorb collinear singularities from the fixed-order calculation into non-perturbative GFFs, which can be extracted from global fits to experimental data.\footnote{See \Ref{Metz:2016swz} for a recent review of these extractions and the experimental datasets used to perform them.} 
For the purposes of this study, we used track functions extracted from $\textsc{Pythia}$ 8.230 as described in \Ref{Elder:2017bkd}. 

The details of the resummed calculation in \Sec{sec:calculation} closely follow those of track thrust in \Ref{Chang:2013iba}, with additional details provided in \App{app:details-resummed} and \App{app:details}.
The analytic calculations include resummation to NLL order, excluding the effects of non-global logarithms (NGLs)~\cite{Dasgupta:2001sh}. 
To perform leading fixed-order (LO) matching, we use the processes $e^+e^-\rightarrow \gamma^*/Z \rightarrow q\bar{q}g$ and $e^+e^-\rightarrow H \rightarrow ggg (gq\bar{q})$.
Non-perturbative (NP) effects are modeled by convolution with a shape function~\cite{Korchemsky:1999kt,Korchemsky:2000kp}, giving a final accuracy we call NLL+LO+NP.  
This calculation broadly supports our conclusions from the \textsc{Vincia} study and offers additional insight into the correspondence between jet mass and track-assisted mass. 
Appropriate convolutions with parton distribution functions (PDFs) and the replacement $E\rightarrow p_T$ would allow these results to be translated to the LHC.


We examine the effect of jet grooming in \Sec{sec:soft-drop}, using \textsc{Vincia} and an NLL+LO calculation to assess the correspondence between groomed GTAM and groomed jet mass.
In the noisy environment of the LHC, jet grooming techniques are essential for removing radiation from sources besides the parton initiating the jet~\cite{Asquith:2018igt,Larkoski:2017jix}. 
Pileup contamination in the upcoming high-luminosity runs will make grooming even more indispensable~\cite{Contardo:2020886,ATLAS-CONF-2017-065}. 
Soft-drop grooming~\cite{Larkoski:2014wba} can be easily incorporated into our analytic calculation, allowing us to compute the spectrum of groomed GTAM. 
In fact, soft-drop grooming removes radiation associated with NGLs to all perturbative orders~\cite{Frye:2016okc,Frye:2016aiz}, greatly simplifying analytic calculations. 
The absence of NGL contributions to the groomed jet mass distribution also implies that this resummed distribution is a complete NLL calculation. 
We find that the optimal values of the GTAM parameters $\kappa$ and $\lambda$ have a mild dependence on the values of the soft-drop parameters, so the optimal GTAM observable is not entirely independent of the grooming procedure. 
\App{app:alternate-sd-implementation} describes another possibility for track-assisted soft-drop grooming. 

Our main conclusions are summarized in \Sec{sec:conclusion}.


\section{Exploration of Generalized Track-Assisted Mass}
\label{sec:MCexplore}


\subsection{Observable and Statistic Definitions}
\label{sec:MCexplore-definitions}

Track mass is the mass of a jet computed only using charged particles,
\begin{equation}
\label{eq:jet-mass-definition}
M_{\rm track} = \sqrt{\left(\sum_{i \in {\rm tracks}} E_i\right)^2 - \left( \sum_{i  \in {\rm tracks}}\vec{p}_i\right)^2}\,, \qquad p_{T,\rm track} = \sqrt{\left(\sum_{i \in {\rm tracks}} \vec{p}_{T,i} \right)^2} \,,
\end{equation}
where we have also defined the track transverse momentum.
The ordinary jet mass ($M_{\rm calo}$) and jet transverse momentum ($p_{T,\rm calo}$) are defined analogously, with the sum running over all particles in a jet.
In practice, the energy $E_i$ might be replaced by the magnitude $\vert \vec{p}_i\vert$ if particle mass information is not available.%
\footnote{It is common in both ATLAS and CMS to assume that each track is a charged pion with mass $m_{\pi^\pm}$.  There is also a minimum $p_T$ threshold for a track to be detectable, though we will not account for this in our study.} 

By using angular information only from charged-particle tracks, measurements of track mass can achieve better particle-for-particle angular resolution than for ordinary jet mass.
This benefit comes with a clear drawback, though, since the distribution of track mass for heavy resonance jets will no longer peak sharply at the mass of the decaying resonance.
Because of the removal of neutral radiation, the track mass will be shifted to lower values compared to ordinary mass, but such an overall shift could easily be corrected through calibration.
More importantly, fluctuations in the fraction of energy carried by charged particles will widen the distribution. 
For the quark/gluon jet ensemble with $p_{T,\rm min} > 300$ GeV studied below, the fractional standard deviations of the jet mass distributions are 
\begin{equation}
\frac{\sigma_{\rm calo}}{\langle M_{\rm calo}\rangle } = 0.39\,, \qquad \frac{\sigma_{\rm track}}{\langle M_{\rm track}\rangle } = 0.46\,,
\end{equation}
so there is an intrinsic loss of resolution by using charged particles compared to all particles.

An observable that is more closely related to the ordinary jet mass can be constructed using a re-weighting factor involving $p_{T,{\rm calo}}$ and $p_{T,{\rm track}}$. 
This is the motivation for track-assisted mass in \Eq{eq:tam-definition} and our generalized version in \Eq{eq:gtam-intro}, repeated for convenience:
\begin{equation}
\label{eq:tam-generalized}
M_{\rm TA}^{(\kappa,\lambda)} = M_{\rm track} \left(\frac{p_{T,\rm calo}}{p_{T,\rm track}}\right)^\kappa \bigg\langle \frac{p_{T,\rm calo}}{p_{T,\rm track}} \bigg\rangle^\lambda\,, \qquad M_{\rm TA}^{(1,0)} \equiv M_{\rm TA}\,.
\end{equation}
If the jet-by-jet fluctuations in the charged-to-neutral mass fraction were identical to the jet-by-jet fluctuations in the charged-to-neutral energy fraction, then just using the per-jet rescaling factor $p_{T,\rm calo}/p_{T,\rm track}$ (i.e.~$\kappa = 1$ and $\lambda = 0$) would yield a good jet-mass proxy.
More generally, one can include charged-to-neutral fraction information from some ensemble of jets $\mathcal{E}$,
\begin{equation}
\label{eq:ratio_average}
\left\langle \frac{p_{T,\rm calo}}{p_{T,\rm track}} \right\rangle = \frac{1}{\vert \mathcal{E} \vert}\sum_{e\in \mathcal{E}} \frac{p^e_{T,\rm calo}}{p^e_{T,\rm track}}\,,
\end{equation}
which has the potential to smooth out jet-by-jet fluctuations to produce a narrower GTAM distribution closer to the width of the ordinary jet mass distribution.
In general, the optimal choices of $\kappa$ and $\lambda$ will depend on the jet samples of interest, since there can be different correlations between energy and mass for different types of jets.

To avoid biasing the GTAM distribution, the appropriate ensemble for computing the average in \Eq{eq:ratio_average} should be as similar as possible to the jets being measured, but still large enough to have acceptable statistical uncertainties. 
For example, all jets in a relatively narrow $p_T$ and rapidity range would be a reasonable choice.
Fortunately, the charged particle momentum fraction is rather scale insensitive, so averaging over a wide $p_T$ and rapidity range turns out to not have much of an effect.
This is true even accounting for differences in the quark/gluon composition of the ensemble; see further discussion in \App{app:pure-quark-gluon}.
We also examined the impact of adding a shift parameter $M_{\rm TA}^{(\kappa,\lambda)} \rightarrow M_{\rm TA}^{(\kappa,\lambda)} + B \, p_{T,\rm calo}$, but found that this additional parameter did not improve the correspondence to ordinary jet mass. 

To quantify the statistical difference between the distributions of $M_{\rm TA}^{(\kappa,\lambda)}$ and $M_{\rm calo}$, we use the statistic\footnote{Without the normalizing factor of $\tfrac{1}{2}$, this is known in the information theory literature as triangular discrimination~\cite{edseee.85070320000101}. In the high-energy physics literature, this is often called the separation power.}
\begin{equation}
\label{eq:dist}
\Delta(p,q) = \sum_{a \in \text{bins}} \frac{1}{2}\frac{(p_a-q_a)^2}{p_a+q_a}\,, \qquad \Delta \in [0,1]\,.
\end{equation}
The sum is over histogram bins, and $p_a$ and $q_a$ are the probability weights of bin $a$ in the probability distributions $p$ and $q$. 
A value of $\Delta = 0$ indicates that the distributions $p$ and $q$ are identical, and a value of 1 occurs when they have no overlap.
This statistic is appealing because it is simple, symmetric between $p$ and $q$, and does not rely on assumptions about the underlying distribution of the data, aside from statistical independence of the samples.
In this paper, we only compare the distributions $M_{\rm TA}^{(\kappa,\lambda)}$ and $M_{\rm calo}$, so for simplicity of notation, we label $\Delta$ by the $\kappa,\lambda$ parameters,
\begin{equation}
\label{eq:dist_kappa_lambda}
\Delta(\kappa,\lambda) \equiv \Delta \left(M_{\rm TA}^{(\kappa,\lambda)},M_{\rm calo}\right)\,.
\end{equation}

It is worth emphasizing that \Eq{eq:dist} is defined at the level of probability distributions, which is not the same as comparing observables on a jet-by-jet basis.
For single-differential jet mass cross sections, the similarity of the probability distributions is what matters, since that is what is being directly measured.
On the other hand, for multi-differential distributions, for calibration purposes, or in the presence of additional jet substructure cuts, the jet-by-jet comparison of GTAM to ordinary jet mass might be more meaningful.
Because our analytic calculations in \Sec{sec:calculation} can only handle single-differential distributions, we focus on the statistic in \Eq{eq:dist}, which favors $\kappa \simeq 0.5$ and $\lambda \simeq 0.5$. 
We have some evidence that the ATLAS default of $\kappa=1$ and $\lambda = 0$ may be preferable for multi-differential cross sections or when there is a narrow cut on jet mass, but this conclusion depends on the precise statistical metric used.


\subsection{Parton Shower Results}
\label{sec:MCexplore-results}

To gain some intuition for the performance of GTAM, we perform a parton shower study relevant for the LHC.
We generate the process $pp\rightarrow$ dijets (including the underlying event) at a center-of-mass energy of 14 TeV, using the $\textsc{Vincia}$ 2.2.2~\cite{Giele:2007di,Fischer:2016vfv} parton shower plugin to $\textsc{Pythia}$ 8.230~\cite{Sjostrand:2006za,Sjostrand:2014zea}.
We verified that similar results could be obtained using the $\textsc{Pythia}$ default parton shower as well.
Jets are identified using the anti-$k_t$ algorithm \cite{Cacciari:2008gp} with jet radius $R=0.4$, as implemented by \textsc{FastJet} 3.3.0 \cite{Cacciari:2005hq,Cacciari:2011ma}.
The jet mass and GTAM are measured on every jet in the event satisfying the pseudorapidity requirement $\vert \eta \vert < 2.0$.
The results are presented for jets with $p_{T,\rm calo} > 100$ GeV, $p_{T,\rm calo} > 300$ GeV, and $p_{T,\rm calo} > 1000$ GeV.
Results for separate quark and gluon distributions are provided in \App{app:pure-quark-gluon}, where the differences are shown to be small.
We start by looking at the impact of $\kappa$ and $\lambda$ separately (setting the other parameter equal to zero), before focusing on the preferred combination of $\kappa + \lambda = 1$.

\begin{figure}
	\centering
	\subfloat[]{
		\includegraphics[width=0.4\textwidth]{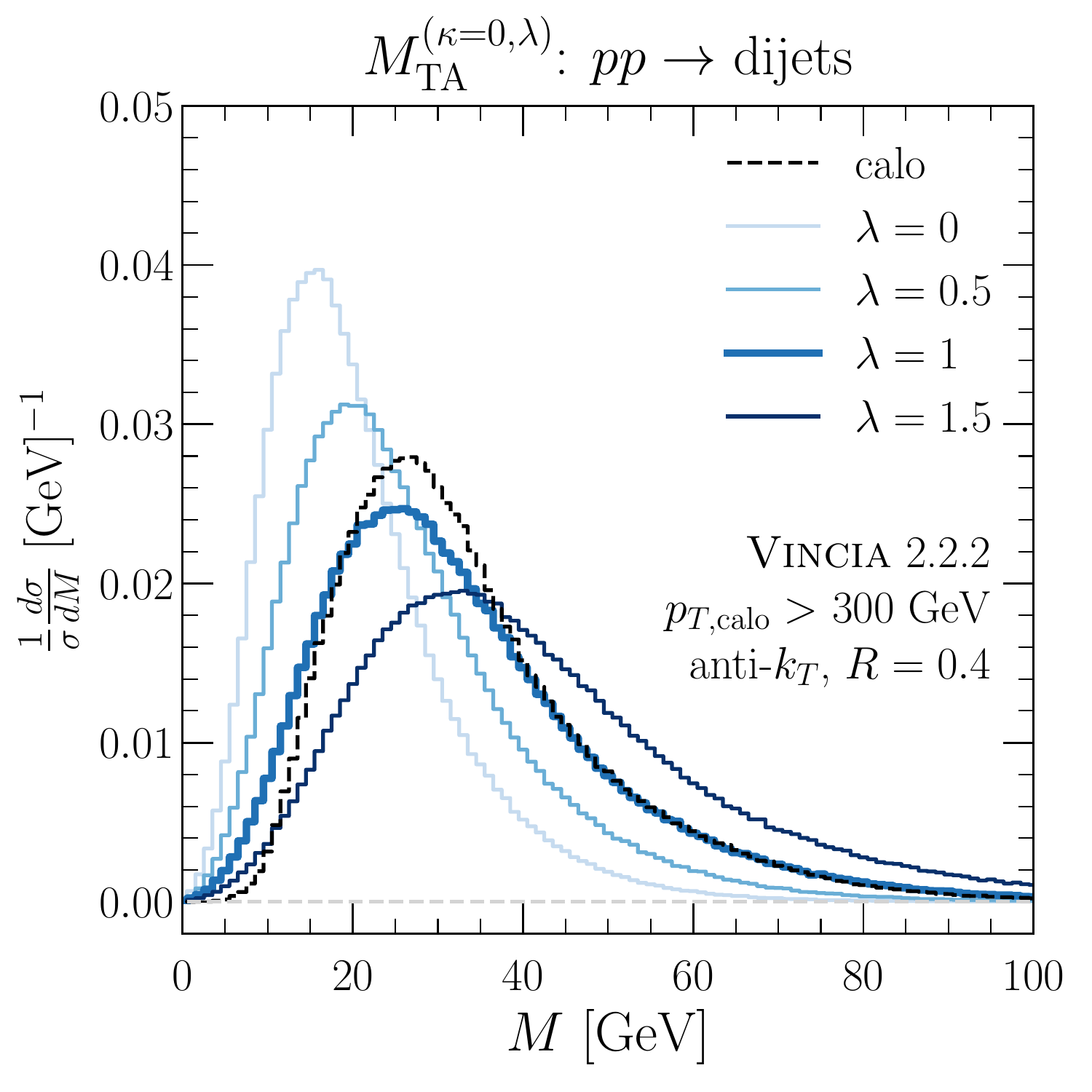}
		\label{fig:mc-panel-a}
	}
	\subfloat[]{
		\includegraphics[width=0.4\textwidth]{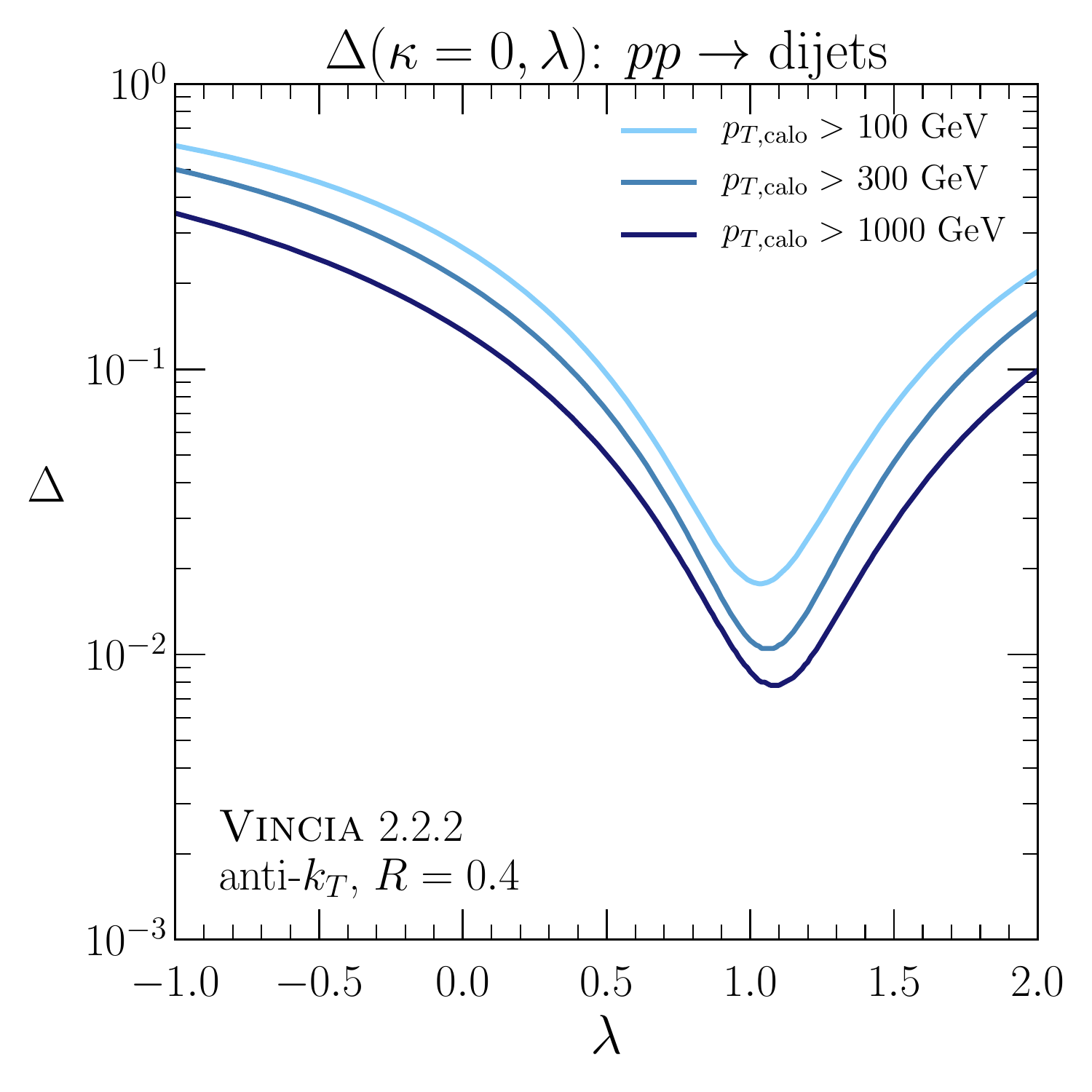}
		\label{fig:mc-panel-b}
	}
	\caption{(a) GTAM distributions for $p_{T,\rm calo} > 300$ GeV with $\kappa = 0$ and $\lambda = \{0,0.5,1,1.5 \}$, with ordinary jet mass plotted as a dashed black curve for comparison.  (b) $\Delta(0,\lambda)$ as a function of $\lambda$, in the $p_{T,\rm calo} > 100$ GeV, $p_{T,\rm calo} > 300$ GeV, and $p_{T,\rm calo} > 1000$ GeV ensembles.}
	\label{fig:mc-panel-1}
\end{figure}
	
\begin{figure}
	\centering
	\subfloat[]{
		\includegraphics[width=0.4\textwidth]{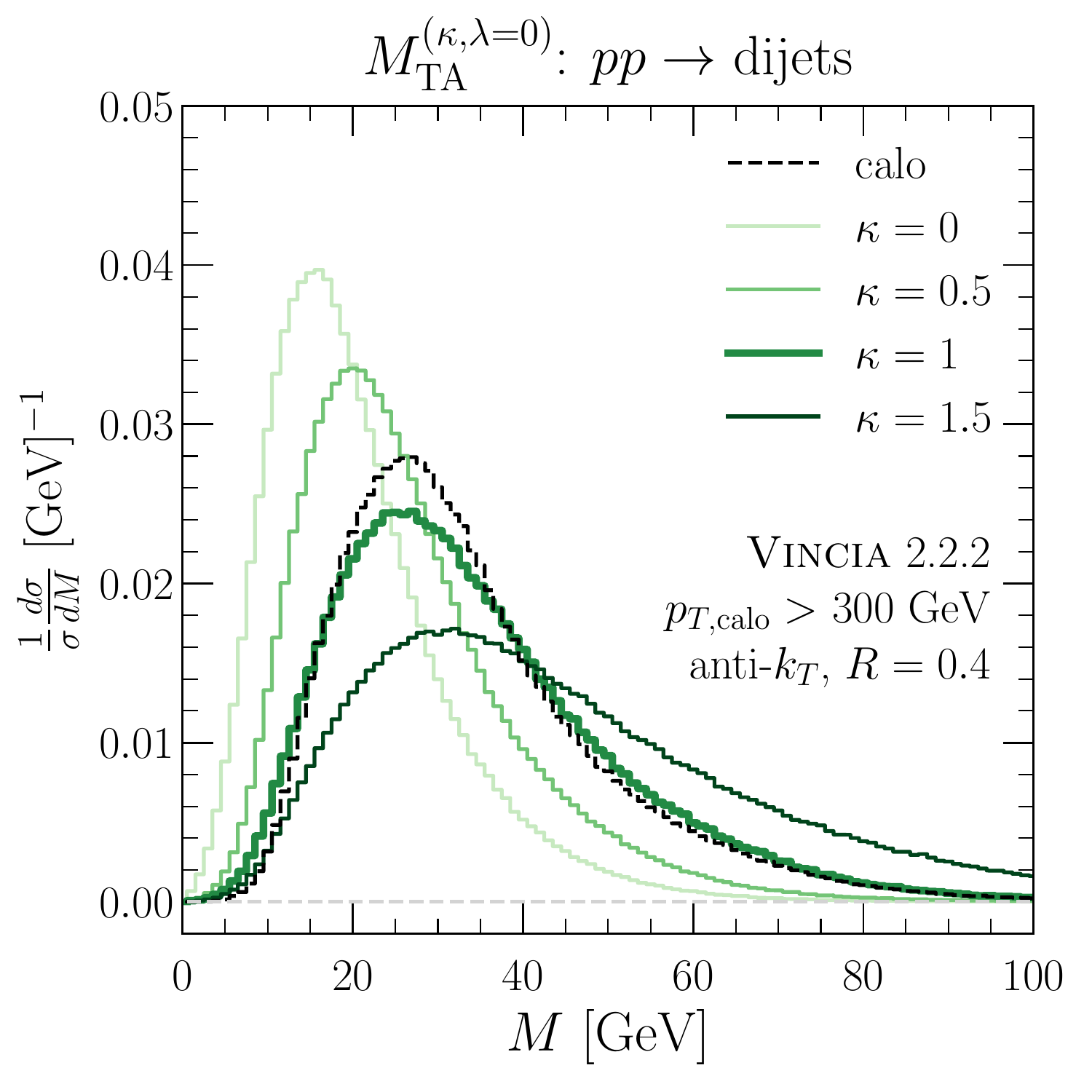}
		\label{fig:mc-panel-c}
	}
	\subfloat[]{
		\includegraphics[width=0.4\textwidth]{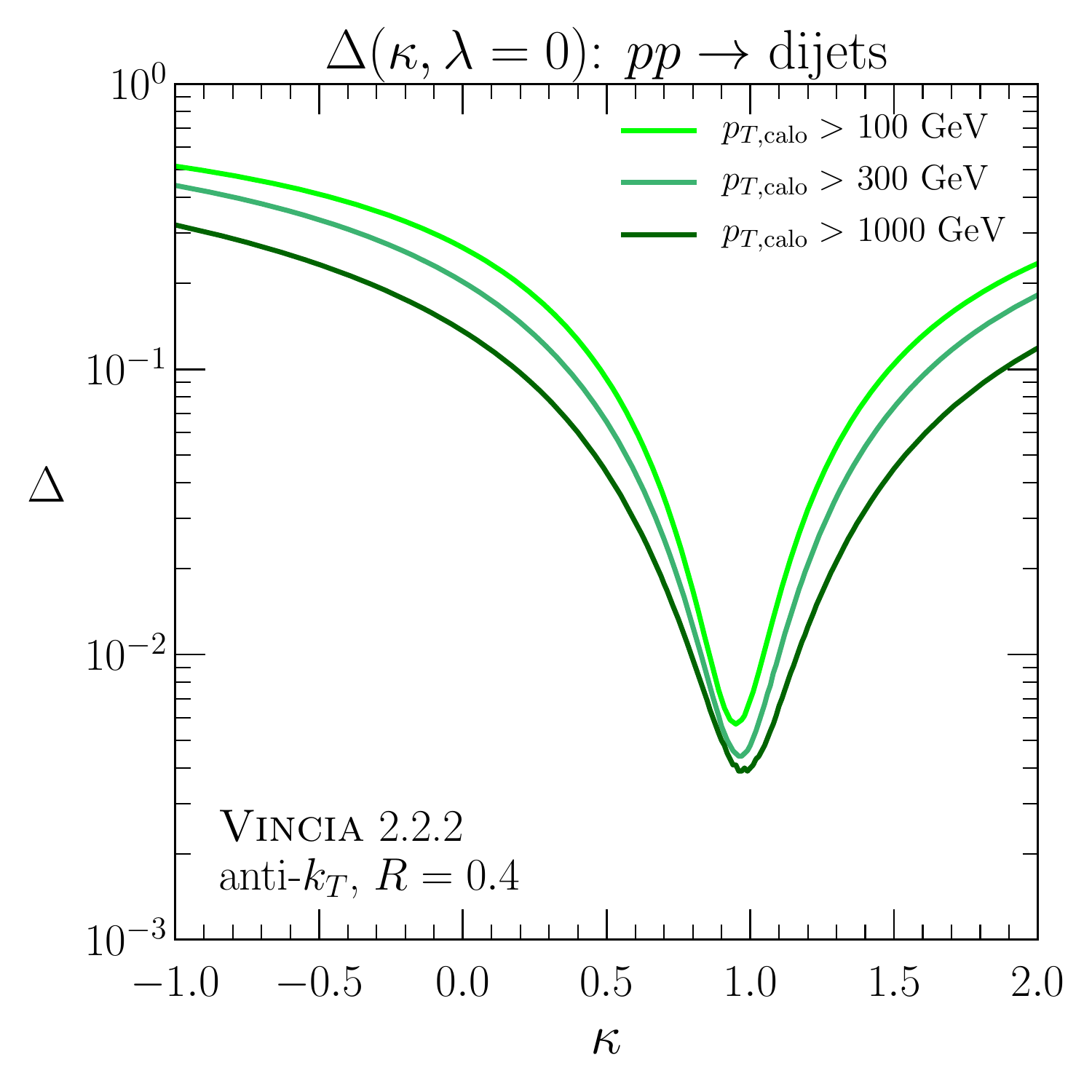}
		\label{fig:mc-panel-d}
	}
	\caption{Same as \Fig{fig:mc-panel-1} but (a) with GTAM parameters $\kappa = \{0,0.5,1,1.5 \}$ and $\lambda = 0$ and (b) $\Delta(\kappa,\lambda=0)$ as a function of $\kappa$. }
	\label{fig:mc-panel-2}
\end{figure}

\begin{figure}
	\centering
	\subfloat[]{
		\includegraphics[width=0.4\textwidth]{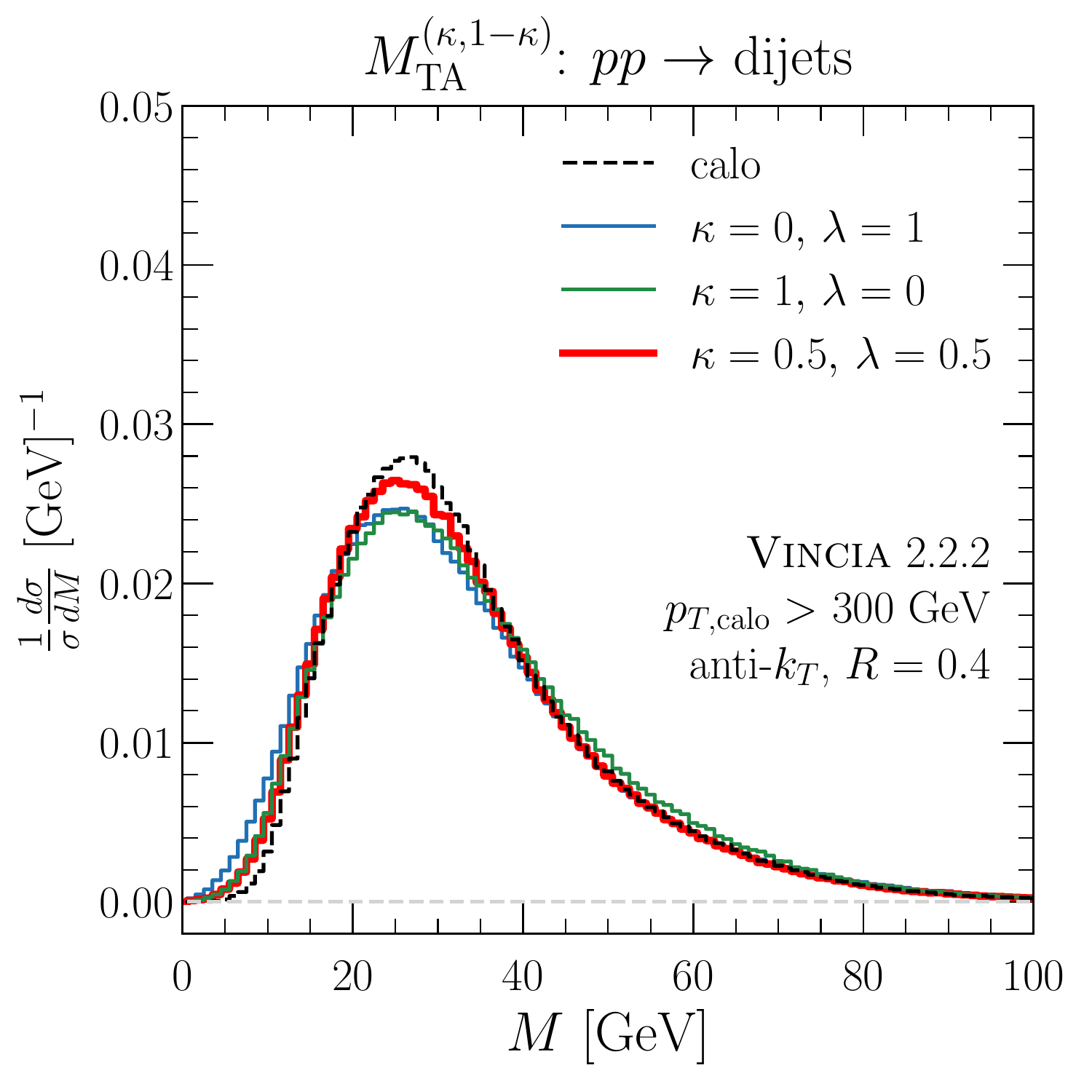}
		\label{fig:mc-panel-e}
	}
	\subfloat[]{
		\includegraphics[width=0.4\textwidth]{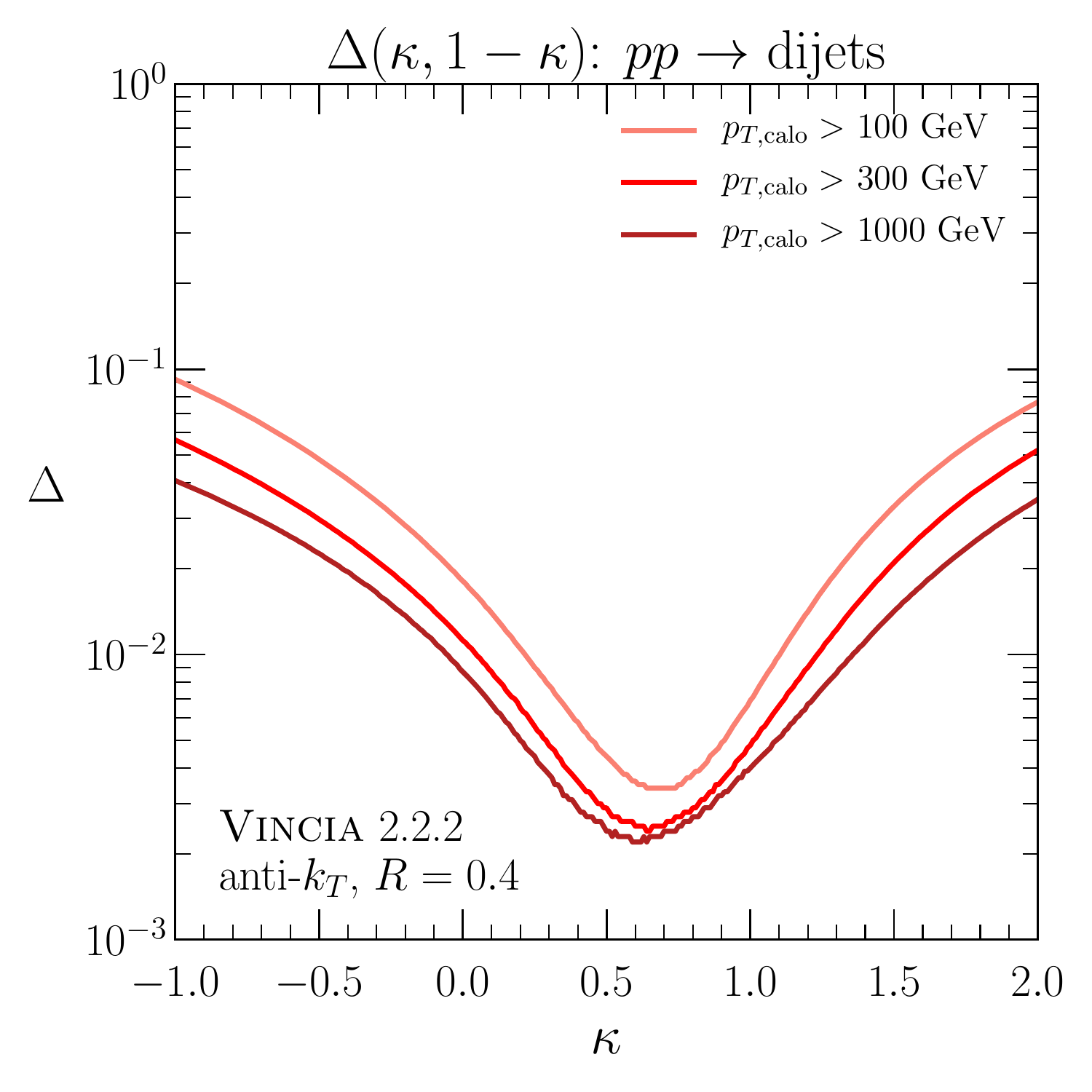}
		\label{fig:mc-panel-f}
	}
	\caption{Same as \Fig{fig:mc-panel-1} but (a) with GTAM parameters $(\kappa,\lambda) = \{(0,1),(0.5,0.5),(1,0) \}$ and (b) $\Delta(\kappa,1-\kappa)$ as a function of $\kappa$. }
	\label{fig:mc-panel-3}
\end{figure}

As a simple proxy for jet mass, one can rescale the track mass by a constant factor $M_{\rm track} \rightarrow C\, M_{\rm track}$. 
If QCD jets were made entirely of pions with exact isospin symmetry, we would expect $M_{\rm track}$ to be a rescaling of $M_{\rm calo}$ by a constant factor of $2/3$, which could be corrected using $C = 3/2$.
The jet ensembles considered here have an ensemble-averaged $p_T$ ratio of $\langle p_{T,\rm calo}/p_{T,\rm track}\rangle = 1.6$, which is close to the $3/2$ predicted by isospin symmetry.
Thus, a useful way to parametrize a constant rescaling is to fix $\kappa = 0$ in \Eq{eq:tam-generalized} and scan over $\lambda$ values, which is the same as rescaling $M_{\rm track}$ by an overall multiplicative constant.
As shown in \Fig{fig:mc-panel-a} for $p_{T,\rm calo} > 300$ GeV, this crude rescaling is reasonably successful in practice.
Scanning over $\lambda = \{0,0.5,1,1.5\}$, the best correspondence between GTAM and $M_{\rm calo}$ is for the thick blue curve with $\lambda = 1$.
In \Fig{fig:mc-panel-b}, we plot the $\Delta(0,\lambda)$ statistic for jets with $p_T$ cuts of 100 GeV, 300 GeV, and 1000 GeV, where again the best jet-mass proxy is achieved for $\lambda$ close to one.

In the ATLAS approach to track-assisted mass, the track mass is rescaled by the per-jet ratio $p_{T,\rm calo}/p_{T,\rm track}$.
This is equivalent to fixing $\lambda = 0$ in \Eq{eq:tam-generalized}.
The motivation for this strategy is that the charged-to-total $p_T$ fraction can vary jet by jet, which would leave an imprint on the $M_{\rm track}/M_{\rm calo}$ ratio. 
In \Fig{fig:mc-panel-c}, we show the distributions of GTAM for $\kappa = \{0,0.5,1,1.5\}$ in the $p_{T,\rm calo} > 300$ GeV sample.
The best fit is obtained from the thick green curve with $\kappa = 1$, which is the value used by ATLAS.
We plot $\Delta(\kappa,0)$ in \Fig{fig:mc-panel-d}, where $\kappa = 1$ is preferred in all three $p_T$ ranges considered.
Comparing \Figs{fig:mc-panel-b}{fig:mc-panel-d}, we see that the per-jet charged $p_T$ fraction is a significantly more effective rescaling factor than the ensemble-average $p_T$ fraction.

As a hybrid of the two above approaches, we can consider $\lambda = 1-\kappa$.
Reweighting the track mass by the per-jet $p_{T,\rm calo}/p_{T,\rm track}$ ratio does correct for the removal of neutral-particle energies, but it does not account for fluctuations in the \emph{angular} distribution of neutral particles.
Therefore, one expects that using ensemble-averaged information can help reduce this angular variability.
For the $p_{T,\rm calo} > 300$ GeV sample, distributions of $M_{\rm TA}^{(\kappa,1-\kappa)}$ are plotted in \Fig{fig:mc-panel-e}.
Intriguingly, the choice $(\kappa, \lambda) = (0.5,0.5)$ interpolates between the per-jet rescaling at low mass and the ensemble-averaged rescaling at high mass, giving an overall better correspondence to $M_{\rm calo}$.
The values of $\Delta(\kappa,1-\kappa)$ are shown in \Fig{fig:mc-panel-f}, where the overall best fit is close to $\kappa = 1- \lambda \approx 0.6$--$0.7$ in the three $p_T$ ranges considered.

\begin{figure}[t]
	\centering
	\subfloat[]{
		\includegraphics[width=0.4\textwidth]{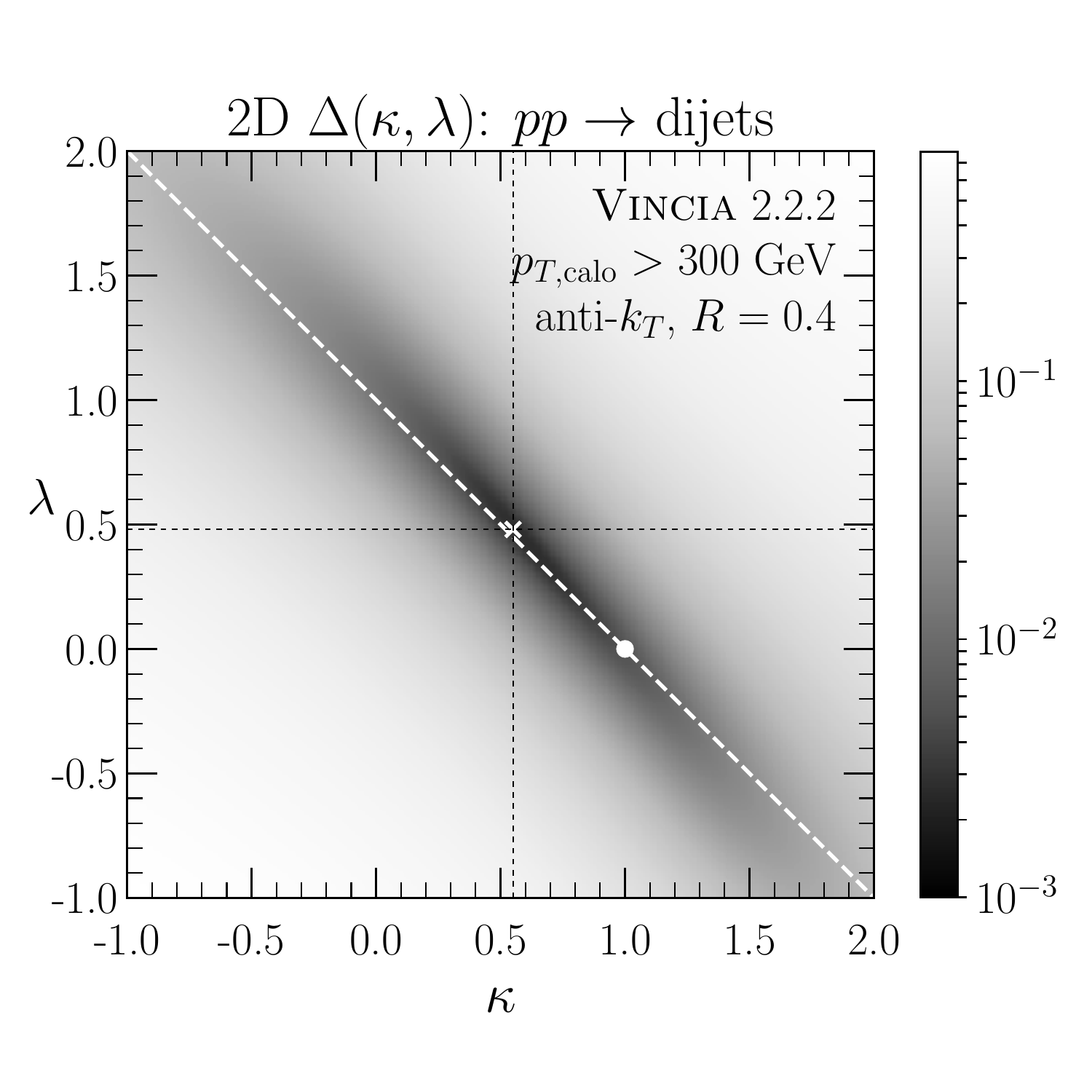}
		\label{fig:heatmap-pp}
	}
	\subfloat[]{
		\includegraphics[width=0.4\textwidth]{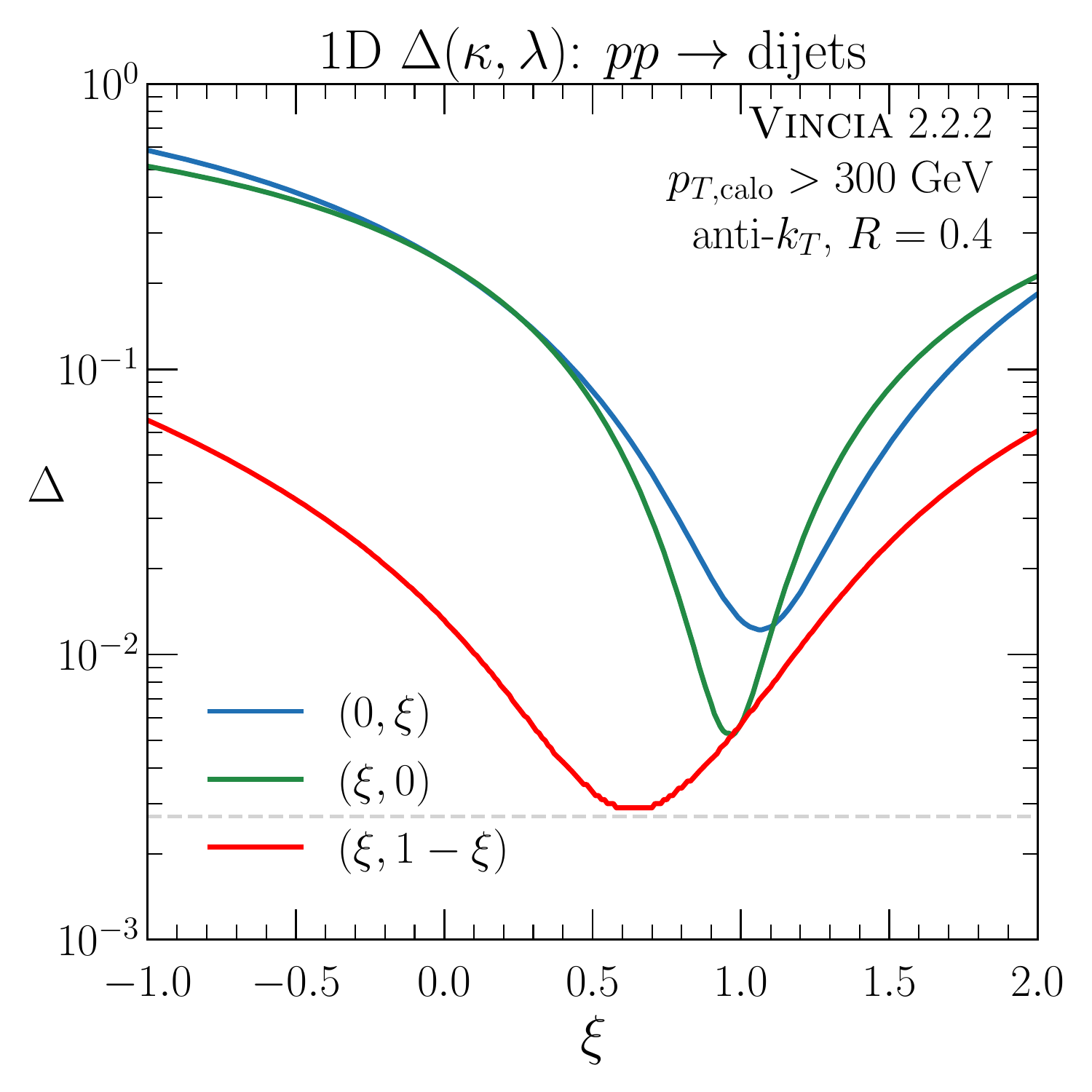}
		\label{fig:mc-allcurves}
	}
	\caption{(a) The $\Delta$ statistic in the two-dimensional $(\kappa,\lambda)$ space, for the $p_{T,\rm calo} > 300$ GeV sample. The minimum of $\Delta$ occurs at $\kappa = 0.55$, $\lambda = 0.48$.  (b) One-dimensional slices of $\Delta(\kappa,\lambda)$, corresponding to \Figss{fig:mc-panel-b}{fig:mc-panel-d}{fig:mc-panel-f}, parametrized by a common variable $\xi$.  The horizontal dashed line indicates the global minimum value of $\Delta(\kappa,\lambda)$.}
	\label{fig:heatmap}
\end{figure}

The full two-dimensional distribution of $\Delta$ as a function of $(\kappa,\lambda)$ is shown in \Fig{fig:heatmap-pp} for the $p_{T,\rm calo} > 300$ GeV sample. 
Similar results are obtained in the other $p_T$ ranges as well.
As expected by dimensional analysis, the relationship $\lambda_{\rm best} \simeq 1-\kappa$ holds to an excellent approximation, with the minimum of this $\Delta(\kappa,\lambda)$ distribution at $(\kappa=0.55, \lambda=0.48)$. 
In \Fig{fig:mc-allcurves}, we show three one-dimensional slices of the full two-dimensional parameter space, $(\kappa,0)$, $(0,\lambda)$, and $(\kappa, 1-\kappa)$, equivalent to the middle curves in \Figss{fig:mc-panel-b}{fig:mc-panel-d}{fig:mc-panel-f}. 
We also checked that turning off underlying event in \textsc{Vincia} does not change the optimal value of $\kappa$ and $\lambda$, though the absolute value of $\Delta$ does change by 5--10\%.
We conclude that a combination of per-jet and ensemble-averaged charged-fraction information provides a statistically closer proxy to calorimeter jet mass than the original track-assisted definition in \Eq{eq:tam-definition}, motivating further studies of GTAM at the LHC.

\begin{figure}[t]
	\centering
	\includegraphics[scale=0.5]{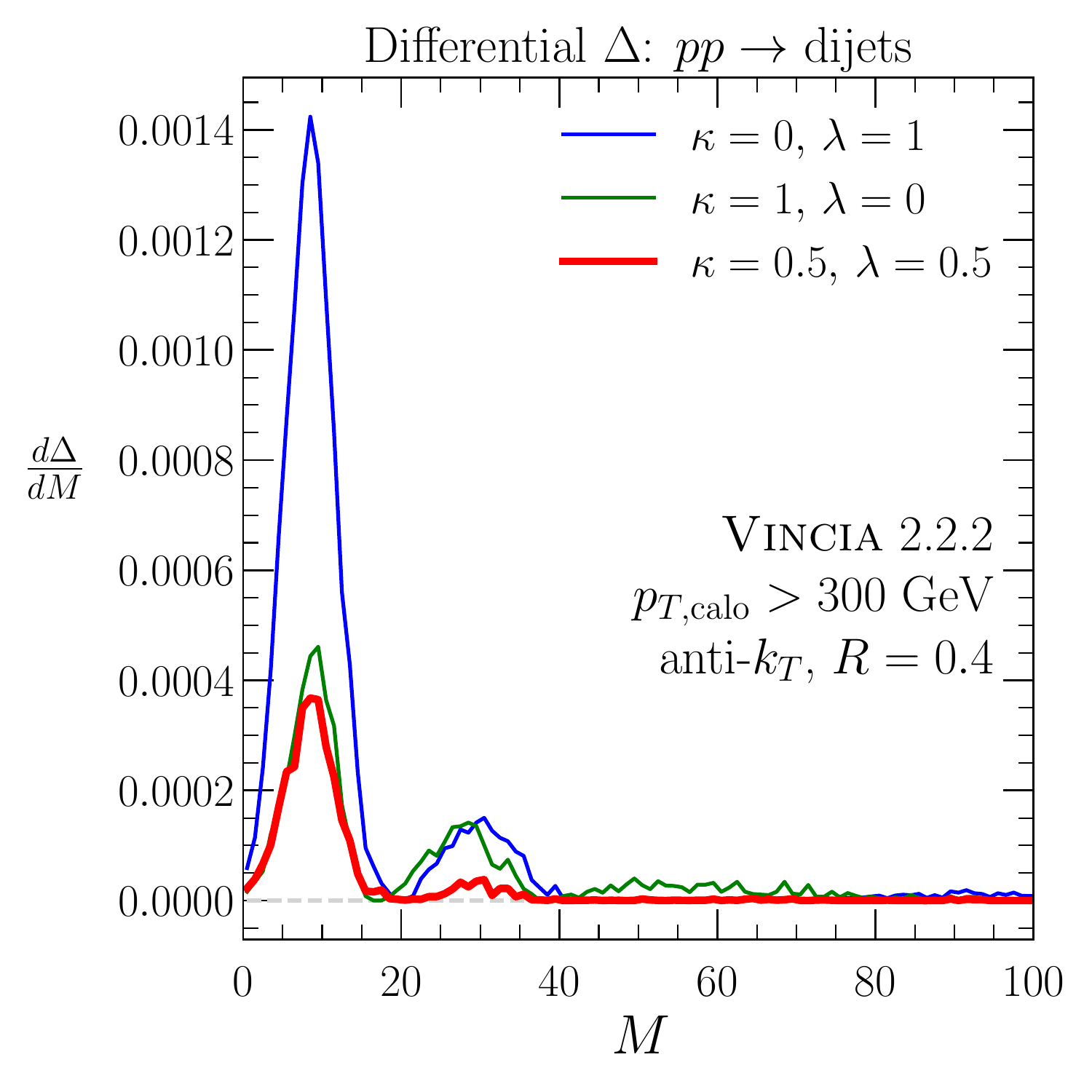}
	\caption{Differential contributions to $\Delta$ as a function of $M$ for the three GTAM variants in \Fig{fig:mc-panel-e}.  The choice $(\kappa, \lambda) = (0.5,0.5)$ minimizes $\text{d} \Delta / \text{d} M$ across the non-perturbative, resummation, and fixed-order regions. 
	}
	\label{fig:differential-delta}
\end{figure}

To gain an intuition for which regions of phase space contribute to the (dis)similarity between GTAM and calorimeter mass, we show differential distributions of $\Delta$ in \Fig{fig:differential-delta} (i.e.\ the summand in \Eq{eq:dist}).
The ordinary track-assisted mass with $(\kappa, \lambda) = (1,0)$ gets contributions to $\Delta$ from the non-perturbative region of small mass, the resummation region characterized by the Sudakov peak, and the tail region dominated by fixed-order emissions.
Taking $(\kappa, \lambda) = (0,1)$ essentially eliminates the $\Delta$ contribution from the fixed-order region, at the expense of introducing large contributions in the non-peturbative region.
With the compromise choice of $(\kappa, \lambda) = (0.5,0.5)$, one can decrease the $\Delta$ contribution from both the resummation and fixed-order regions, without degrading the behavior in the non-perturbative region.

It is worth emphasizing that these best-fit values of $\kappa \approx 0.5$ and $\lambda \approx 0.5$ are derived from quark/gluon ensembles.
Different optimal parameters may be found for boosted electroweak-scale resonances, though we expect the relation $\lambda_{\rm best} \simeq 1-\kappa$ to always hold to an excellent approximation.
In preliminary studies, we find that boosted $W$ jets have a $\Delta(\kappa,1-\kappa)$ minimum closer to $\kappa = 0.95$, more in keeping with the ATLAS default strategy, though this a relatively shallow minimum.
That said, an advantage in $W$ mass resolution might be offset by an increase in QCD background contamination, since larger values of $\kappa$ yield a larger high-side mass tail, as evident from \Fig{fig:mc-panel-e}.
As discussed in \Sec{sec:soft-drop}, the optimal GTAM observable is also affected by jet grooming, with the most aggressive grooming strategies favoring $\kappa \simeq 1.0$.%
\footnote{It is interesting that both boosted $W$ jets and groomed QCD jets prefer larger values of $\kappa$.  When soft gluon radiation is absent, the jet mass arises dominantly from two hard prongs of energy.  Thus, jet-by-jet fluctuations in the charged-to-neutral mass fraction are more similar to the jet-by-jet fluctuations in the charged-to-neutral energy fraction, favoring $\kappa = 1$.}
Moreover, the optimal choice of $\kappa$ and $\lambda$ will be affected by the use of substructure discriminants, so a more detailed study of track-assisted boosted object tagging is warranted in the context of signal/background discrimination.


\section{Track-Assisted Mass in $e^+e^-$ Annihilation}
\label{sec:calculation}


In this section, we perform a first-principles QCD calculation of the GTAM distribution.
To avoid the myriad complications from hadronic collisions, we focus on the process $e^+e^-\rightarrow$ hadrons with center-of-mass energy $E_{\rm CM}$, though many of the results here have a straightforward extension to the LHC.
We focus on large radius jets with $R=1$ such that $E_{\rm calo} \simeq E_{\rm CM}/2$.

To simplify the presentation, we start with the original track-assisted mass with $(\kappa, \lambda)= (1, 0)$.
After defining the necessary quantities in \Sec{sec:calculation-definitions}, we perform a resummed calculation of the $M_{\rm TA}$ distribution in \Sec{sec:calculation-resummed}, and use this to understand the close correspondence to ordinary jet mass in \Sec{sec:insights}.
We match to fixed-order calculations in \Sec{sec:calculation-matching-fo}.
Here, we neglect virtual terms, which only contribute to the overall normalization, and fix the normalization at the end of the calculation. 

We then extend these calculations to GTAM with general $\kappa$ and $\lambda$ in \Sec{sec:calculation-gtam}.
Non-perturbative corrections are included in \Sec{sec:calculation-np} using a shape function.
Finally, we present the best-fit values of $\kappa$ and $\lambda$ in \Sec{sec:calculation-best-fit}.
Details of the resummed calculation appear in \App{app:details-resummed}, and details of the fixed-order calculation appear in \App{app:details}.


\subsection{Defining Track-Based Observables}
\label{sec:calculation-definitions}


In $e^+e^-$ collisions, a modified version of track-assisted mass is appropriate, with energy replacing transverse momentum, 
\begin{equation}
M_{\rm TA} = M_{\rm track}\left(\frac{E_{\rm calo}}{E_{\rm track}}\right) \,.
\end{equation}
For convenience, we define the dimensionless rescaled (squared) track-assisted mass $\rho_{\mbox{\tiny TA}}$ and the equivalent $\rho_{\rm calo}$,
\begin{equation}\label{eq:obs-ee-def}
\rho_{\mbox{\tiny TA}} = \frac{M_{\text{track}}^2}{E_{\rm calo}^2R^2} \left(\frac{E_{\rm calo}}{E_{\rm track}}\right)^2\,,\qquad\qquad \rho_{\rm calo} = \frac{M_{\rm calo}^2}{E_{\rm calo}^2R^2}\,,
\end{equation}
which take values in the interval [0,1].\footnote{As discussed in \Sec{sec:calculation-matching-fo}, the upper kinematic boundary for a two parton jet is actually $\approx \frac{1}{4}$.}
For the calculations in this section and \Sec{sec:soft-drop}, we always use a jet radius of $R=1$ unless otherwise noted. 

The track fraction $x_i$ is the fraction of parton $i$'s momentum carried by charged particles after hadronization~\cite{Larkoski:2014pca,Chang:2013iba}. 
At the partonic level, each parton momentum $k_i$ is rescaled by its corresponding track fraction, $k_i^{\mu,\rm charged} = x_i \, k_i^\mu + \mathcal{O}(\Lambda_{\rm QCD})$. 
Writing the rescaled track mass then only requires making the replacement $k_i \rightarrow x_i \, k_i$, yielding
\begin{equation}
\rho_{\mbox{\tiny TA}} = \frac{\left(\sum_{i\in \text{jet}} x_i k_i\right)^2}{\left(\sum_{i\in \text{jet}} x_i E_i\right)^2R^2}\,.
\end{equation}

\begin{figure}
	\centering
	\subfloat[]{
		\includegraphics[width=0.45\textwidth]{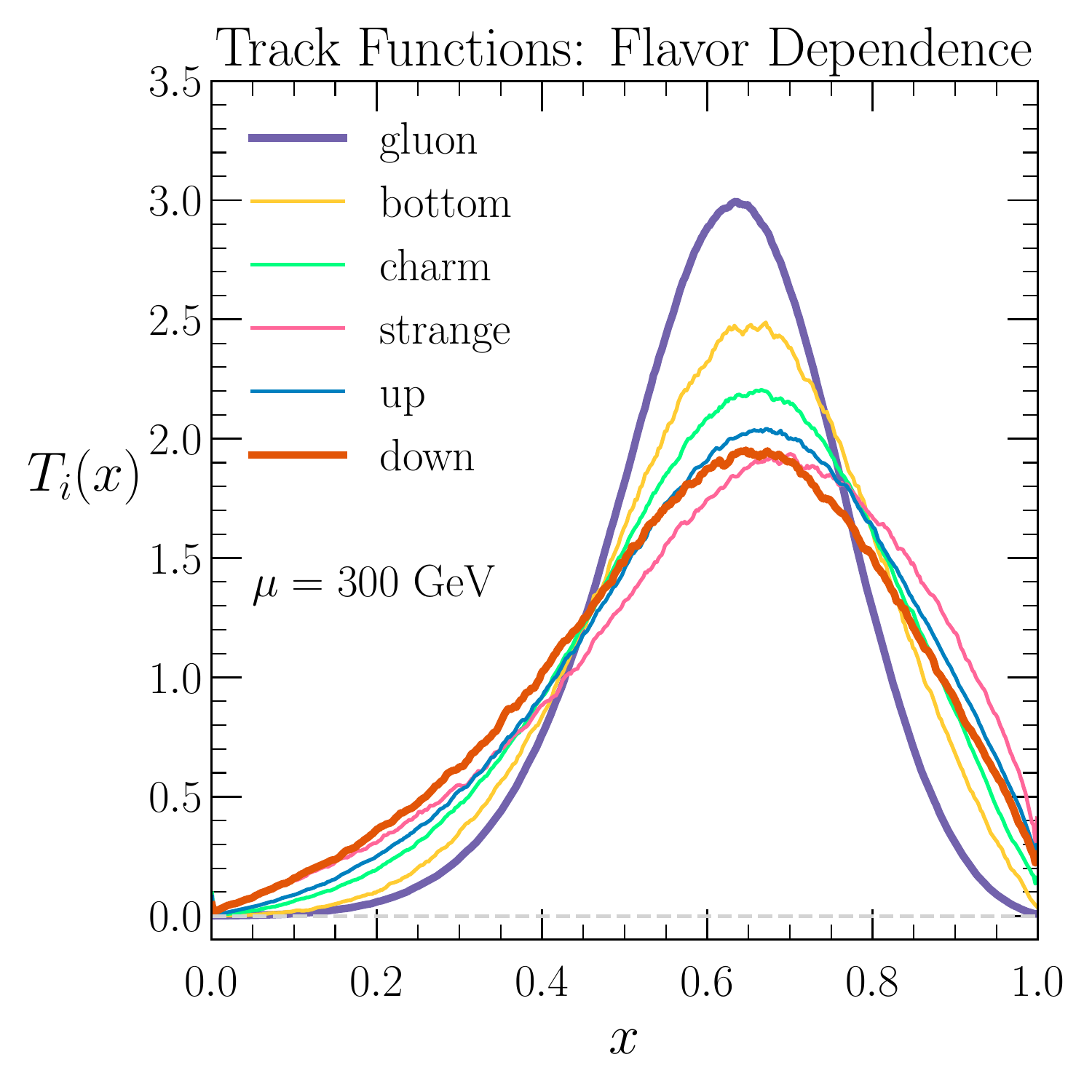}
		\label{fig:track-functions-a}
	}
	\subfloat[]{
		\includegraphics[width=0.45\textwidth]{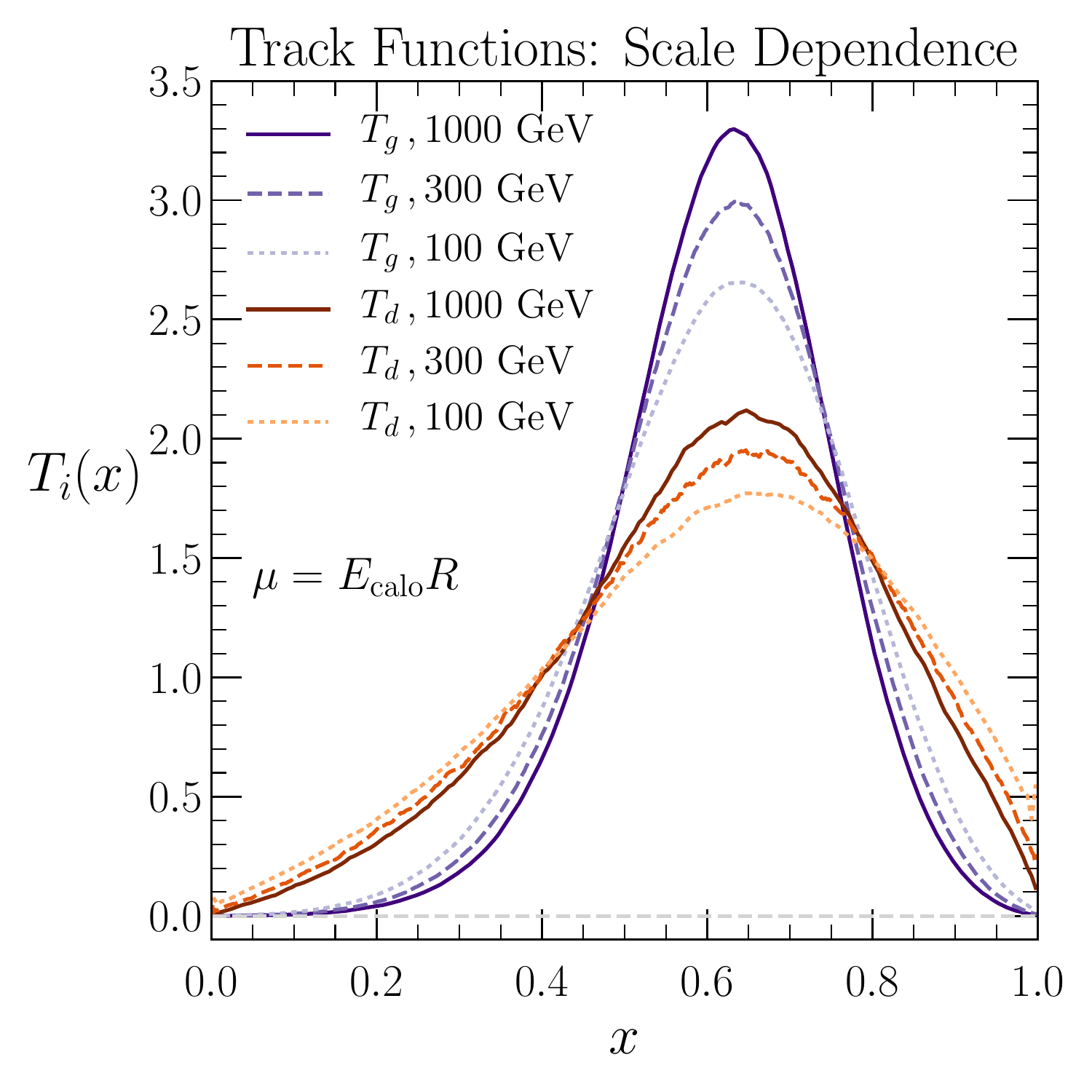}
		\label{fig:track-functions-b}
	}
	\caption{\label{fig:track-functions}  (a) Track functions extracted from \textsc{Pythia} 8.230 (see \Ref{Elder:2017bkd}) for gluons and active quark flavors at a scale of $\mu = 300$ GeV.  (b) Track functions from $\textsc{Pythia}$ for gluon jets (blue) and down-quark jets (red) at scales $\mu = \{100, 300, 1000\}$ GeV.}
\end{figure}

The track functions $T_i(x_i)$ are the distributions of the track fraction, where there is a process-independent track function $T_i$ for each parton flavor. 
As shown in \Ref{Chang:2013iba}, the track functions have well defined field-theoretic definitions.
At lowest order in $\alpha_s$, $T_i(x_i)$ is just the empirical distribution of $x_i$ extracted from a global fit to experimental data, or, as in this work, extracted from the \textsc{Pythia} parton shower.
Of course, all track functions have support only on $x \in [0,1]$.  
Some sample track functions are plotted for gluons and active quark flavors in \Fig{fig:track-functions-a}. 
The anti-quark track functions are the same by charge conjugation invariance. 
For the remainder of this work we focus on jets initiated by gluons and down quarks. 
Our calculation for quark-initiated jets is done in the limit of massless quarks, so the only dependence on the quark flavor comes from these distributions.

As illustrated in \Fig{fig:track-functions}, the track functions are smooth distributions with most of their probability weight parametrically far from zero.
We can parametrize the gluon and down-quark track functions as $T_g(x) \propto x^{\lambda_g}$ and $T_d(x) \propto x^{\lambda_d}$ as $x \rightarrow 0$, where a fit to \textsc{Pythia} yields $\lambda_g \approx 5.7$ and $\lambda_d \approx 2.4$.
This fact will be useful in \Sec{sec:insights} when convolving the track functions with integrands which have logarithmic divergences in the $x\rightarrow 0$ limit.

While the track functions are non-perturbative objects, they have a perturbative renormalization group evolution~\cite{Chang:2013iba,Elder:2017bkd}, which is described by a non-linear version of the DGLAP equations~\cite{Gribov:1972ri,Lipatov:1974qm,Dokshitzer:1977sg,Altarelli:1977zs}. 
When applied to a jet of energy $E_{\rm calo}$, the appropriate scale to evaluate the track function is typically $\mu = E_{\rm calo} R$. 
\Fig{fig:track-functions-b} illustrates the scale dependence of the gluon and down-quark track functions.
The ordinary jet mass $\rho_{\rm calo}$ can always be recovered from the track-assisted version simply by setting the track functions to be $T_i(x) = \delta(1-x)$, which then sets the energy fraction $x$ equal to one in all expressions.

At parton level, for a single splitting $i\rightarrow jk$, we can write the two-parton form of the observable $\hat{\rho}$ in terms of the track fractions $x_j$ and $x_k$, the momentum fraction $z$ carried by parton $k$, and the angle between the splitting products $\theta_{jk}$,
\begin{equation}
\label{eq:obs-ee-parton-def}
\hat{\rho}_{\mbox{\tiny TA}} = \left(\frac{2x_jx_kz(1-z)(1-\cos\theta_{jk})}{x_j^2(1-z)^2 + 2x_jx_kz(1-z) + x_k^2z^2}\right) \frac{1}{R^2}\,.
\end{equation}
To perform the resummed calculation for track-assisted mass, we need the soft-collinear limit of this expression. 
Expanding to lowest order in $z$ and $\theta_{ij}\equiv \theta$, we obtain 
\begin{equation}
\label{eq:observable-sc-limit}
\hat{\rho}_{\mbox{\tiny TA}} \simeq \frac{x_k}{x_j} \frac{z \theta^2}{R^2} + \dots \,. \simeq \frac{x_k}{x_j}\hat{\rho}_{\mbox{\tiny calo}}.
\end{equation}
Subleading terms in this expansion contribute starting at NNLL, beyond the accuracy of our resummed calculation.
We expect formally power-suppressed corrections proportional to $R \log R$ arising from this truncation to appear in the fixed-order matching (see \Sec{sec:calculation-matching-fo}).



\subsection{Resummed Calculation}
\label{sec:calculation-resummed}


To resum large logarithms of $\rho$, we must compute the cumulative distribution $\Sigma(\rho)$. 
Since $\rho$ is an additive observable in the limit $\rho \rightarrow 0$, the cumulative distribution can be written in the form
\begin{equation}
\Sigma(\rho) \equiv \int_0^\rho \text{d}\rho'\, \frac{\text{d}\sigma}{\text{d}\rho'} = C(\alpha_s) \tilde{\Sigma}(\alpha_s,\rho) + D(\alpha_s,\rho)\,.
\end{equation}
The function $C(\alpha_s)$ can be expanded in powers of $\alpha_s$ and is independent of the observable $\rho$. 
The remainder function $D(\alpha_s,\rho) \rightarrow 0$ as $\rho \rightarrow 0$. 
The large logarithms of $\rho$ appear in the function $\tilde{\Sigma}(\alpha_s,\rho)$, which can be expanded in powers of $\alpha_s$ according to~\cite{Almeida:2014uva}
\begin{align}
\label{eq:rs-log-counting}
\ln \tilde{\Sigma}(\alpha_s,\rho) &= \sum_{n=1}^\infty \sum_{m=1}^{n+1}\left(\frac{\alpha_s}{2\pi}\right)^n G_{nm} \ln^{m}\frac{1}{\rho}\\ \nonumber
&=\hspace{0.2cm} \left(\frac{\alpha_s}{2\pi}\right) \left(G_{12}\ln^2 \frac{1}{\rho} + G_{11}\ln \frac{1}{\rho} \right) \\ \nonumber
&\hspace{0.2cm}+ \left(\frac{\alpha_s}{2\pi}\right)^2 \left(G_{23}\ln^3 \frac{1}{\rho} + G_{22}\ln^2 \frac{1}{\rho} + G_{21}\ln \frac{1}{\rho} \right) \\ \nonumber
&\hspace{0.2cm}+ \left(\frac{\alpha_s}{2\pi}\right)^3 \left(G_{34}\ln^4 \frac{1}{\rho} + G_{33}\ln^3 \frac{1}{\rho} + G_{32}\ln^2 \frac{1}{\rho} + G_{31}\ln \frac{1}{\rho} \right)\\ \nonumber
&\hspace{0.2cm} + \ldots\,.
\end{align}
Our calculation at NLL order accuracy resums terms in $\ln \tilde{\Sigma}$ in the first two columns of \Eq{eq:rs-log-counting}, that is, terms of the form $\left(\tfrac{\alpha_s}{2\pi}\right)^n G_{nm}\ln^m\tfrac{1}{\rho}$ for $m=n+1$ and $m=n$. 
When combined with fixed-order corrections necessary in the large $\rho$ region, this gives a calculation that is valid for $\alpha_s\ln\tfrac{1}{\rho} \lesssim1$, which is much larger than the region $\alpha_s\ln^2\tfrac{1}{\rho} \ll 1$.

For a non-track-based observable, we can write this cumulative distribution as~\cite{Catani:1991bd,Catani:1992ua,Banfi:2004yd}
\begin{equation}
\label{eq:cal-cumulative}
\Sigma_{\rm calo}(\rho) = \frac{e^{-\gamma_E R_{\rm calo}'(\rho)}}{\Gamma(1+R_{\rm calo}'(\rho))} e^{-R_{\rm calo}(\rho)}\mathcal{N}(\rho), \ \ \ \ R_{\rm calo}'(\rho) = - \frac{\text{d}R_{\rm calo}(\rho)}{\text{d}\ln(\rho)}\,,
\end{equation}
valid to NLL order.
The factor $\mathcal{N}(\rho)$ is the result of NGLs~\cite{Dasgupta:2001sh}.
Although progress has been made towards computing this contribution,\footnote{For a recent review see, for example, \Ref{Larkoski:2017jix}. } it still presents a substantial complication to a full NLL calculation of the jet mass. 
Treating this factor is beyond the scope of this work, though it has been shown \cite{Larkoski:2014wba,Frye:2016aiz} that soft-drop grooming removes contributions from NGLs to all orders in $\alpha_s$, so this factor will not be present for the groomed distributions in \Sec{sec:sd-calculation}.
For the ungroomed case, we expect the relative impact of NGLs to be similar between ordinary jet mass and track-assisted mass.

The radiator $R_{\rm calo}(\rho)$ can be interpreted as the probability that the hard quark (or gluon) will radiate a gluon such that the mass of the resulting two-parton quark-initiated (gluon-initiated) jet is greater than some value $\rho$.
The factor $e^{-R_{\rm calo}}$ is the exponential of the single-emission probability, and the factors which depend on $R'$ describe the sensitivity of $\rho$ to multiple independent emissions. 
Explicitly, the radiator for a jet initiated by a parton of flavor $i$ is given by
\begin{equation}
\label{eq:radiator-calorimeter}
R_{\rm calo}(\rho) = \int_0^1 \text{d}z\, P_{i}(z) \int_0^{R}\frac{\text{d}\theta}{\theta}\frac{\alpha_s(E_{\rm calo}z\theta)}{\pi} \Theta(\hat{\rho}_{\rm calo}-\rho)\,,
\end{equation}
where $\hat{\rho}_{\rm calo}$ is given by \Eq{eq:observable-sc-limit} with $x_j = x_k = 1$.
The reduced splitting functions $P_i$ for $i=q,g$ are given in \App{app:details-resummed}. 

\begin{figure}[t]
	\centering
	\begin{tikzpicture}[scale=0.5]
		\draw[ultra thick,particle,orange] (0,0) -- (15,0);
		\draw[thick,gluon,purple] (1,0) -- (3,1);
		\draw[thick,gluon,purple] (4.5,0) -- (6.5,1);
		\draw[thick,gluon,purple] (8,0) -- (10,1);
		\draw[thick,gluon,purple] (11.5,0) -- (13.5,1);
		\draw[color=purple, fill,rotate around={45:(3,1)}] (3,1) ellipse (6pt and 12pt);
		\draw[color=purple, fill,rotate around={45:(6.5,1)}] (6.5,1) ellipse (6pt and 12pt);
		\draw[color=purple, fill,rotate around={45:(10,1)}] (10,1) ellipse (6pt and 12pt);
		\draw[color=purple, fill,rotate around={45:(13.5,1)}] (13.5,1) ellipse (6pt and 12pt);
		\draw[color=orange, fill] (15,0) ellipse (6pt and 12pt);
		\node[] at (16,-1) {$T_q(x_q)$};
		\node[] at (3,2) {$T_g(x_1)$};
		\node[] at (6.5,2) {$T_g(x_2)$};
		\node[] at (10,2) {$T_g(x_3)$};
		\node[] at (13.5,2) {$T_g(x_4)$};
	\end{tikzpicture}
	\caption{\label{fig:soft-collinear-emissions}An eikonal quark emitting soft, collinear gluons. For the calculation of a track-based observable, each gluon emission, as well as the final quark, must be weighted with the appropriate track function. }
\end{figure}
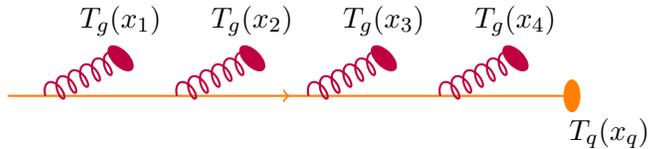

To calculate the cross section for a track-based observable, one has to convolve the all-particle results with a track function for each final-state parton~\cite{Chang:2013iba,Elder:2017bkd}.
This means that processes with $N$ independent gluon emissions from a parton of flavor $i$ must involve $N$ gluon track functions, plus a single track function for parton $i$, as in \Fig{fig:soft-collinear-emissions}. 
As discussed in \Ref{Chang:2013iba}, the perturbative part of the calculation will not depend on higher-order correlation functions, for instance of the form $T(x_1,x_2)$.
This is because of the universality of collinear singularities and the factorization of partons if their pairwise invariant mass is larger than $\Lambda_\text{QCD}$.%
\footnote{Note that there are no higher-order correlation functions associated with perturbative soft radiation at large angles.  The reason is that the track-based observables we consider are IR safe, and therefore do not exhibit uncanceled soft singularities.  For the calculation of track thrust in \Ref{Chang:2013iba}, soft radiation at large angles is captured by the soft function, which can be refactorized in the perturbative region into expressions involving only single-parton track functions.}
Non-perturbative power corrections from hadronization will in general involve higher-order correlations, though we will model these using a simple shape function in \Sec{sec:calculation-np}.

Analogous to the calculation for generalized angularities \cite{Larkoski:2014pca}, the gluon track functions will exponentiate with the radiator, while the initiating parton's track function will not.
Therefore for a track-based observable, we include the gluon track function in the definition of the radiator.
This leads to the modified cumulative distribution for track-assisted mass
\begin{equation}
\label{eq:track-cumulative}
\Sigma_{\mbox{\tiny TA}}(\rho) = \int_0^1 \text{d}x_j \, T_j(x_j,\mu)\, \frac{e^{-\gamma_E R_{\mbox{\tiny TA}}'(\rho,x_j)}}{\Gamma(1+R_{\mbox{\tiny TA}}'(\rho,x_j))} e^{-R_{\mbox{\tiny TA}}(\rho,x_j)}\,,
\end{equation}
where the radiator is
\begin{equation}
\label{eq:track-radiator}
R_{\mbox{\tiny TA}}(\rho,x_j) = \int_0^1 \text{d}x_k\, T_g(x_k,\mu) \int_0^1 \text{d}z\,P_i(z)\, \int_0^R \frac{\text{d}\theta}{\theta} \,\frac{\alpha_s(E_{\rm calo}z\theta)}{\pi}\, \Theta\left(\hat{\rho}_{\mbox{\tiny TA}}-\rho\right)\,. 
\end{equation}
Both track fractions $x_j$ and $x_k$ appear in the expression of the observable \Eq{eq:obs-ee-parton-def}, but since the gluon track function $T_g(x_k)$ exponentiates with the radiator, $x_k$ is integrated over in \Eq{eq:track-radiator}. 
This leaves the radiator as a function of $x_j$, and the integral over $x_j$ is only performed in \Eq{eq:track-cumulative}.

To achieve NLL accuracy in $R$ (neglecting NGLs), it is sufficient to include the running of $\alpha_s$ up to two-loop accuracy in the calculation of \Eqs{eq:radiator-calorimeter}{eq:track-radiator}. 
It is also sufficient to ignore the change in the track function from $g \to q \bar{q}$ splittings.
Strictly speaking, \Eq{eq:track-radiator} should be a matrix in flavor space \cite{Dasgupta:2013ihk,Marzani:2017mva} since $g \to q \bar{q}$ changes the flavor of the hard parton.
Because $g \to q \bar{q}$ does not have a soft divergence, though, the change in the track function does not yield a logarithmically-enhanced contribution, such that this flavor change can be ignored to NLL accuracy.
Because it has a collinear divergence, the $g \to q \bar{q}$ process must of course still be included in the splitting function $P_g(z)$.

To determine $R'$, we calculate the logarithmic derivative numerically, even though this includes terms that are formally beyond NLL accuracy.
As described above, the track functions have a scale dependence described at leading order $(\mathcal{O}(\alpha_s))$ by a nonlinear version of the DGLAP equations. 
As we shall see in the next section, the radiator in \Eq{eq:track-radiator} is independent of the track functions at LL, and therefore this running only contributes starting at NNLL. 
This is beyond the precision of our calculation, and so we freeze all track functions in $\Sigma_{\mbox{\tiny TA}}$ at the scale $\mu_{\rm hard}$. 
Pushing the precision of this calculation to NNLL order would require including the evolution of the track functions, evaluated at the scale $E_{\rm jet}z\theta$, as well as the flavor-changing process $g \to q \bar{q}$.

\begin{figure}[t]
	\centering
	\subfloat[]{
		\includegraphics[width=0.45\textwidth]{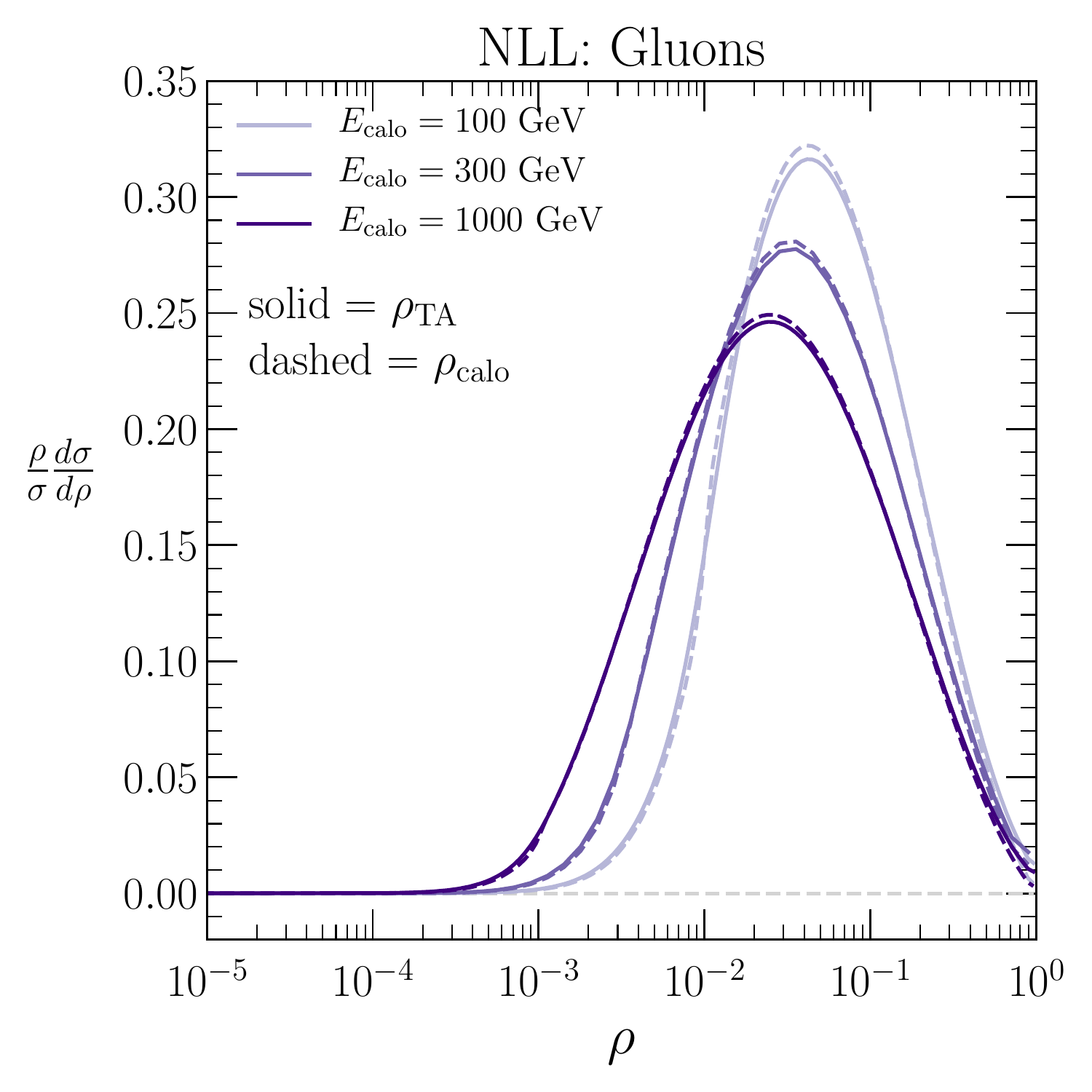}
		\label{fig:resummed-gluon-nogroom}
	}
	\subfloat[]{
		\includegraphics[width=0.45\textwidth]{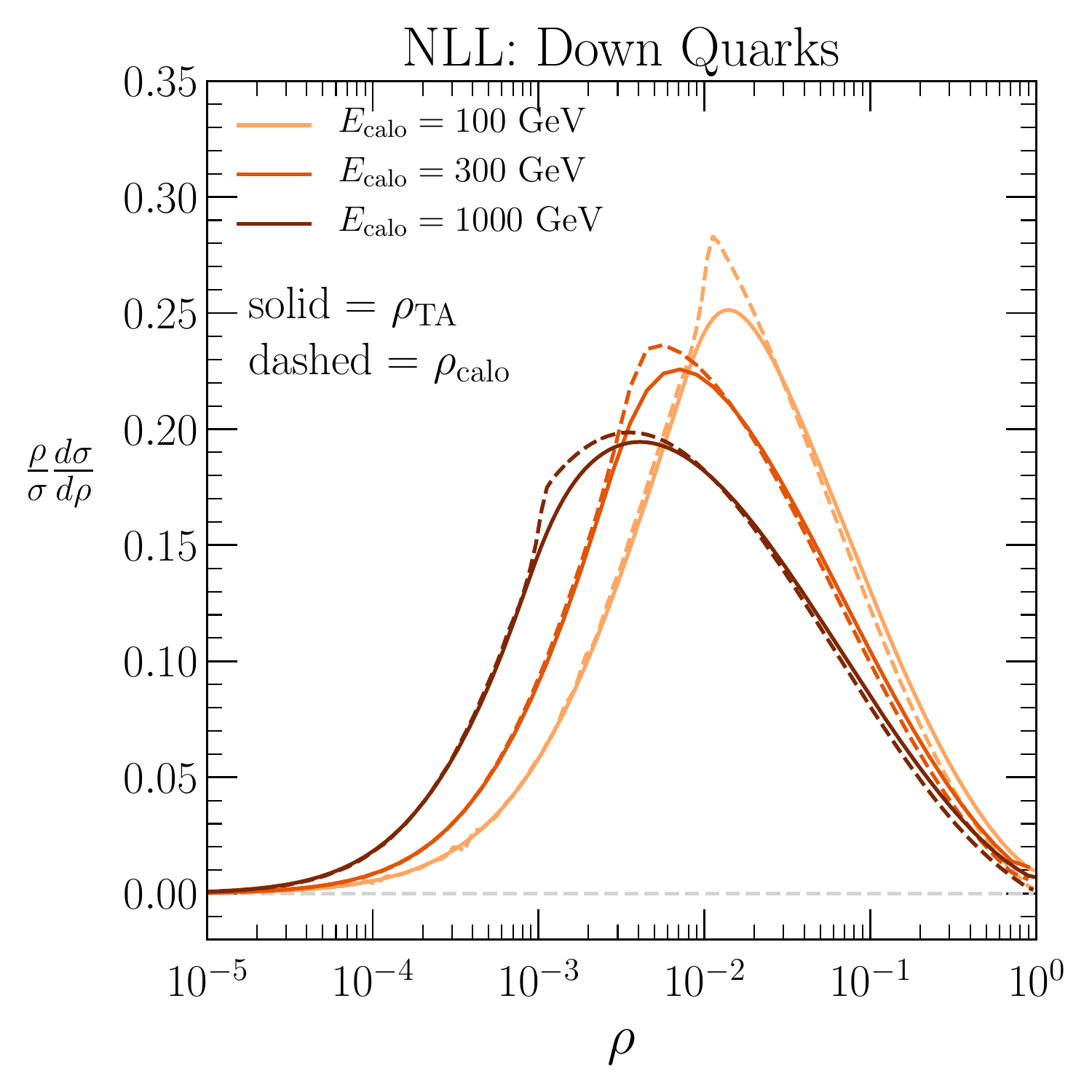}
		\label{fig:resummed-quark-nogroom}
	}

	\subfloat[]{
		\includegraphics[width=0.45\textwidth]{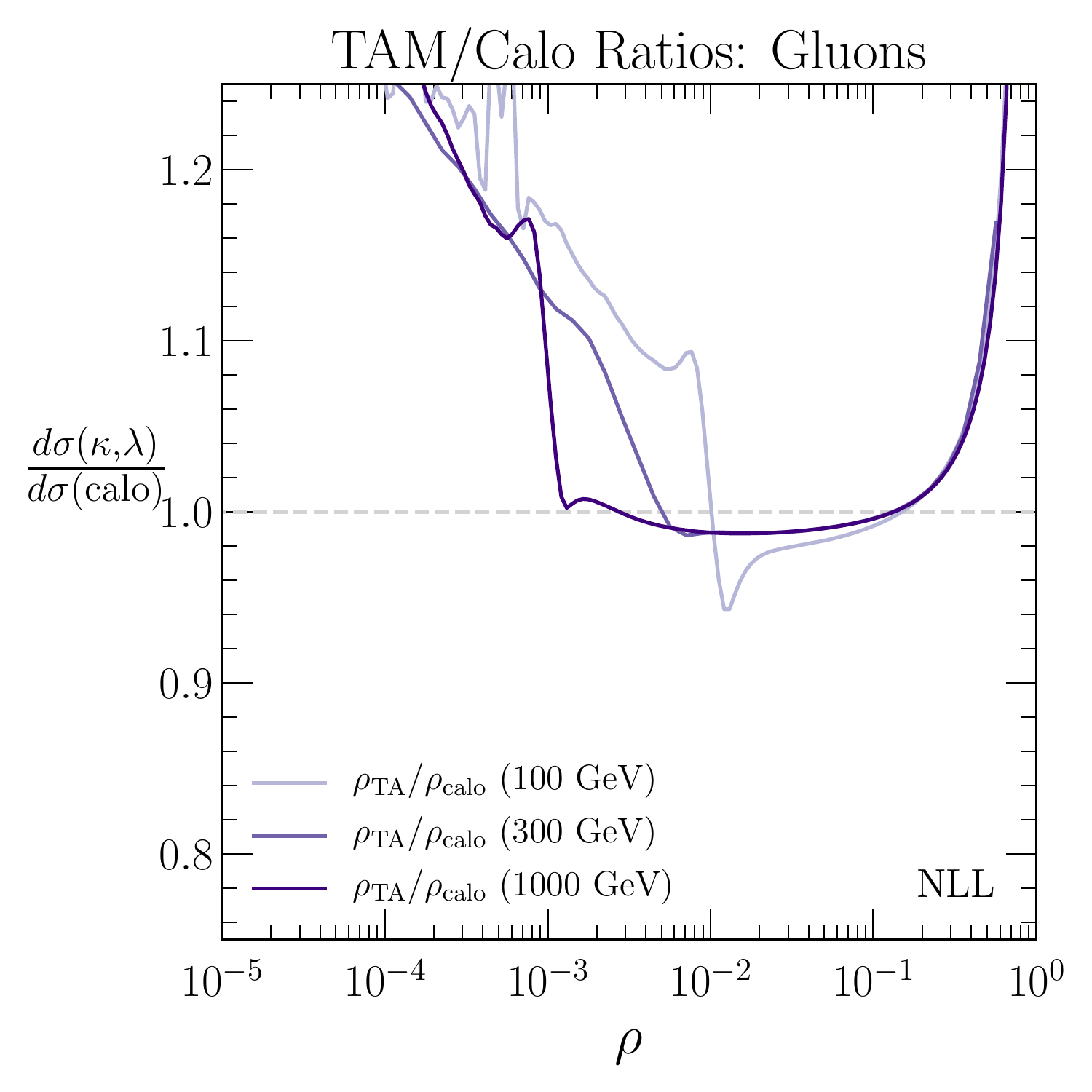}
		\label{fig:resummed-gluon-nogroom-ratio}
	}
	\subfloat[]{
		\includegraphics[width=0.45\textwidth]{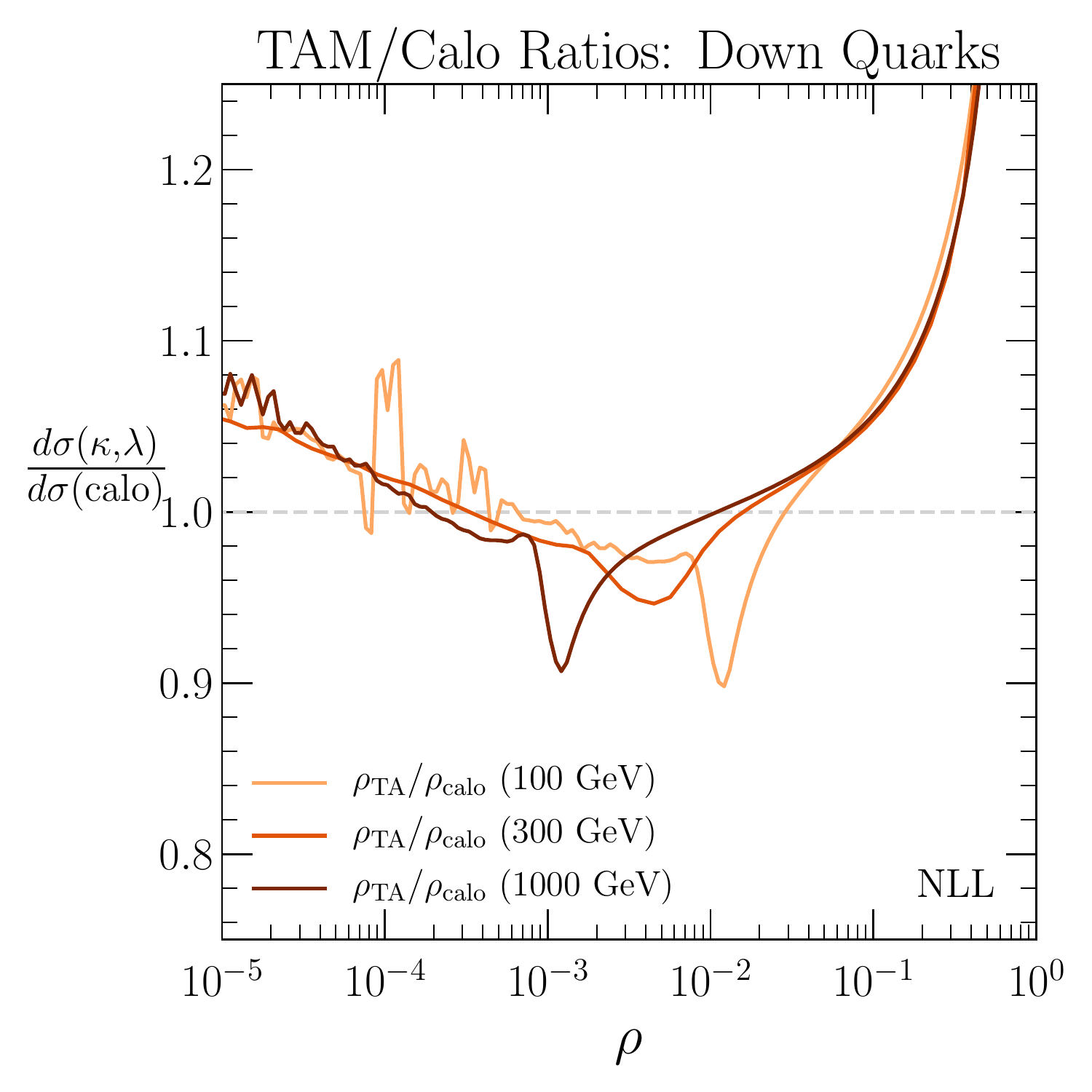}
		\label{fig:resummed-quark-nogroom-ratio}
	}
	\caption{Top row:  Resummed distributions at NLL for rescaled jet mass $\rho_{\rm calo}$ (dashed curves) and rescaled track-assisted mass $\rho_{\mbox{\tiny TA}}$ (solid curves) with $R=1$ for (a) gluon-initiated jets and (b) down-quark-initiated jets.  Shown are three different values of the jet energy, $E_{\rm calo}=$ 100 GeV, 300 GeV, and 1000 GeV. Bottom row:  Ratios of $\rho_{\rm TA}/\rho_{\rm calo}$ at the same jet energies to highlight the differences in the distributions.}
	\label{fig:resummed-nogroom}
\end{figure}

The differential distribution $\frac{\rho}{\sigma}\frac{\text{d}\sigma}{\text{d}\rho}$ is obtained by numerically taking the $\rho$ derivative of \Eq{eq:track-cumulative}.
These are plotted in \Fig{fig:resummed-nogroom} for gluon-initiated jets and for down-quark-initiated jets, with $R=1$ and $E_{\rm calo} = $ 100 GeV, 300 GeV, and 1000 GeV. 
The only dependence on the quark flavor comes from the track functions, as illustrated in \Fig{fig:track-functions-a}.
The track-assisted mass distributions are similar to the jet mass distributions for both quarks and gluons, but the similarity is much stronger for gluon jets, for reasons which we now explain.


\subsection{Insights at Fixed Coupling}
\label{sec:insights}


We can apply the same analysis from \Ref{Chang:2013iba} for track thrust to achieve some insight into the close correspondence between track-assisted mass and jet mass.
We proceed by examining \Eq{eq:track-cumulative} in the fixed-coupling approximation.
The appropriate scale to fix the coupling is the characteristic scale of the jet, which for $e^+e^-$ collisions is just $\mu = E_{\rm calo}R$. 
In this approximation, to NLL order in the observable $\rho$, the radiator becomes
\begin{equation}
\label{eq:radiator-fixed-coupling}
R_{\mbox{\tiny TA}}(\rho,x_j) \overset{\mbox{\tiny F.C.}}{=} \frac{\alpha_sC_i}{\pi}\int_{x_j\rho}^1 \text{d}x_k\, T_g(x_k)\, \bigg[\frac{1}{2}\ln^2(\tfrac{1}{\rho}) + \ln(\tfrac{1}{\rho})\left(\ln\left(\frac{x_k}{x_j}\right)+B_i\right)\bigg].
\end{equation}
The factor $B_i$ is $-\frac{3}{4}$ for quark-initiated jets and $-\frac{11}{12} + \frac{T_Fn_f}{3C_A}$ for gluon-initiated jets.%
\footnote{The flavor-changing process $g \to q \bar{q}$ is captured by the $B_i$ term.  Note that, to NLL accuracy, the integral over $T_g(x_k)$ drops out of the term proportional to $B_i$, which justifies ignoring the flavor change of the track function in our calculation.}
The color factors $C_i$ are $C_F = \tfrac{4}{3}$ for quarks and $C_A = 3$ for gluons.

In choosing to keep only terms with logarithms of $\rho$, we have neglected logarithms of $x_k$ and $x_j$, which become large near the track function endpoint $x\approx 0$. 
As discussed in \Sec{sec:calculation-definitions}, in the $x\rightarrow 0$ limit where logs of $x$ become large, the integrand $T_k(x)\ln(x) \rightarrow x^{\lambda_k}\ln(x) \rightarrow 0$. 
Thus, the track functions themselves suppress logs of $x_j$ and $x_k$, and resumming only  logarithms of $\rho$ is justified.

Working to NLL order, from \Eq{eq:radiator-fixed-coupling} we can compute
\begin{equation}
\label{eq:R-prime}
R_{\mbox{\tiny TA}}'(\rho,x_j) = \frac{\alpha_sC_i}{\pi} f^{g,0}(x_j\rho,1)\ln(\tfrac{1}{\rho}) \,,
\end{equation}
defined in terms of the partial logarithmic moments of the track functions
\begin{equation}
\label{eq:partial-log-moment}
f^{i,n}(a,b) = \Theta(\min(b,1)-a) \int_a^b \text{d}x\,T_i(x)\ln^n(x) \,, \ \ \ \ \ f^{i,n} \equiv f^{i,n}(0,1)\,.
\end{equation}
The zero-th moment is just the normalization of the track frunction, $f^{i,0} \equiv f^{i,0}(0,1) = 1$.\footnote{The $\Theta$-function and $\min$ are necessary since the integral bounds are set by the $\Theta$ functions of the observable (and the soft-drop $\Theta$ functions in \Sec{sec:soft-drop}).}
Since $x_j \le 1$ and we are working in the region where $\rho \ll 1$, we can expand the integral around $x_j\rho \approx 0$, keeping only the leading term. 
This amounts to extending the lower endpoint of the integral over $x_k$ in the radiator to 0.\footnote{For $\rho = 0.1$, the region $x_k < \rho$ accounts for only $\approx 1.2\%$ of the probability weight for down-quark track functions and $\approx 0.01\%$ for gluon track functions at 300 GeV. } 
With this approximation, $R' = R'(\rho)$ is independent of both track fractions, yielding
\begin{equation}
\label{eq:approx-fixed-coupling-NLL-radiator}
R_{\mbox{\tiny TA}}(\rho,x_j) = \frac{\alpha_sC_i}{\pi} \left[ \frac{1}{2}\ln^2(\tfrac{1}{\rho}) + \ln(\tfrac{1}{\rho})\left(f^{g,1} - \ln(x_j) + B_i\right)\right]\, ,\ \ \ \ R_{\mbox{\tiny TA}}'(\rho) = \frac{\alpha_sC_i}{\pi}\ln(\tfrac{1}{\rho})\,.
\end{equation}

The explicit expression for the cumulative distribution in this approximation is
\begin{align}
\begin{split}
\label{eq:approx-fixed-coupling-NLL-cumulative}
\Sigma_{\mbox{\tiny TA}}(\rho) &= \frac{e^{-\gamma_ER_{\mbox{\tiny TA}}'}}{\Gamma\left(1+R_{\mbox{\tiny TA}}'\right)} \exp\left\{-\frac{\alpha_sC_i}{\pi}\left(\frac{1}{2} \ln^2(\tfrac{1}{\rho})+\ln(\tfrac{1}{\rho})B_i\right)\right\}\\
&\hspace{1cm} \times \exp\left(-\frac{\alpha_s C_i}{\pi} \ln(\tfrac{1}{\rho}) f^{g,1}\right) \times \int_0^1 \text{d}x_j\, T_j(x_j)\exp\left\{\frac{\alpha_sC_i}{\pi}\ln(\tfrac{1}{\rho})\ln(x_j)\right\}.
\end{split}
\end{align}
We can further simplify the factor involving the down-quark track function with an exponential approximation \cite{Chang:2013iba},
\begin{equation}
\label{eq:exp-approx}
\int_0^1 \text{d}x_j\, T_j(x_j)\exp\left\{\frac{\alpha_sC_i}{\pi}\ln(\tfrac{1}{\rho})\ln(x_j)\right\} \approx \exp\left\{\frac{\alpha_sC_i}{\pi}\ln(\tfrac{1}{\rho}) f^{j,1}  \right\}\,.
\end{equation}
Replacing $\approx$ with $\geq$, this is just Jensen's inequality applied to expectation values of $\ln(x_j)$.
This approximation holds to within $10\%$ for all parton flavors over the range of energies considered in this paper down to $\rho \approx 10^{-6}$. 
It improves substantially with increasing jet energy and increasing $\rho$. 
This provides an approximate expression at NLL order with fixed coupling for the cumulative distribution,
\begin{align}\label{eq:cumulative-fixed}
\begin{split}
\Sigma_{\mbox{\tiny TA}}(\rho) \simeq \Sigma_{\rm calo}(\rho)\times \exp\left\{-\frac{\alpha_sC_i}{\pi}\ln(\tfrac{1}{\rho}) (f^{g,1}-f^{j,1})  \right\}.
\end{split}
\end{align}
The gluon and down-quark first logarithmic moments are $(f^{g,1},f^{q,1}) \approx (-0.53,-0.62)$ at 100 GeV, $(-0.52,-0.59)$ at 300 GeV, and $(-0.51,-0.58)$ at 1000 GeV. 

The result above demonstrates why track-assisted mass is numerically so close to ordinary jet mass, through the approximate cancellation between these two terms. 
While the cancellation is much more precise for gluon jets, $f^{j,1} = f^{g,1}$, than for quark jets with $f^{j,1} = f^{q,1}$, it is still not exact due to the approximations that we have just described. 
This is why the agreement in \Fig{fig:resummed-nogroom} between track-assisted mass and jet mass is closer for gluon jets than for quark jets, but still not perfect. 
Note that setting $f^{g,1} = f^{q,1} = 0$ recovers the fixed-coupling approximation for ordinary jet mass.


\subsection{Fixed-Order Corrections}
\label{sec:calculation-matching-fo}


The resummed calculation in \Sec{sec:calculation-resummed} only holds in the regime $\rho \ll 1$, where terms proportional to $\log^2\rho$ and $\log\rho$ dominate and terms which are powers of $\rho$ or constants can be neglected. 
Producing a distribution that is correct over the full kinematic range of $\rho$ requires matching this all-orders distribution to a fixed-order calculation for a specific process. 
We choose the process $e^+e^-\rightarrow \gamma^*/Z \rightarrow q\bar{q}g$ to match to the resummed quark-jet calculation and $e^+e^-\rightarrow H \rightarrow ggg (q\bar{q}g)$ for the gluon-jet one. 
Due to the Higgs coupling to gluons only through a quark loop, the process $e^+e^-\rightarrow H \rightarrow ggg (q\bar{q}g)$ is already $\mathcal{O}(\alpha_s^2)$ in the Standard Model. 
To make this computation more tractable, we compute the matrix elements for this process in the $m_t\rightarrow \infty$ limit, as discussed in \App{app:details-me}.
Since the differential distribution $\frac{\text{d}\sigma}{\text{d}\rho}$ diverges as $\rho \rightarrow 0$ for both processes, we placed a cutoff on the observable, and performed the numerical calculation for $\rho > 10^{-6}$. 
This allows us to neglect virtual terms which only contribute to a delta function at $\rho = 0$, and also removes contributions from the part of phase space where the matrix elements diverge.
Since we are ultimately interested in calculating the statistic in \Eq{eq:dist_kappa_lambda}, we fix the overall normalization of our distributions to 1.

For a three-parton final state, we can rewrite the observable \Eq{eq:obs-ee-parton-def} using the parton energy fractions $y_i = \frac{2E_i}{Q}$, which satisfy $y_1+y_2+y_3 = 2$.
If partons 1 and 3 are clustered in the same jet, this leads to a new formula for the observable,
\begin{equation}\label{eq:coordobs}
\hat{\rho}_{\mbox{\tiny TA}} = \left(\frac{4x_1x_3(1-y_2)}{(x_1 y_1 + x_3 (2-y_1-y_2))^2}\right) \frac{1}{R^2}\,.
\end{equation}
At this perturbative order, the differential cross section for track-assisted mass is (ignoring contributions proportional to a delta function at $\rho=0$),
\begin{equation}
\label{eq:fixed-order-cross-section}
\frac{\text{d}\sigma}{\text{d}\rho} = \int \text{d}y_1\, \text{d}y_2\, \frac{\text{d}\hat{\sigma}}{\text{d}y_1\text{d}y_2} \int \text{d}x_1\, \text{d}x_3\, T_1(x_1)\,T_3(x_3)\, \Theta (R-\theta_{13})\delta(\hat{\rho}_{\mbox{\tiny TA}}-\rho)\,,
\end{equation}
with $\hat{\rho}_{\mbox{\tiny TA}}$ given now by \Eq{eq:coordobs}. 
The $\Theta$ function requires that two partons be in the same jet. 
Explicit expressions for the partonic cross sections $\text{d}\hat{\sigma}/\text{d}y_1\text{d}y_2$ are given in \App{app:details-me}. 
Since track-assisted mass is not IRC safe, the cross section must be convolved with the track functions $T_1(x_1)$ and $T_3(x_3)$ for the two partons in the jet.




To match our all-orders result to the fixed-order calculation, we use the modified log-$R$ matching scheme \cite{Catani:1992ua}
\begin{equation}
\label{eq:matching-logR}
\ln\left(\Sigma_{\mbox{\tiny NLL+LO}}(\rho)\right) = \ln(\Sigma_{\mbox{\tiny NLL}}(\rho)) + \Sigma_{\mbox{\tiny LO}}(\rho) - \Sigma_{\mbox{\tiny NLL,}\alpha}(\rho)  \,.
\end{equation}
The cumulative distribution $\Sigma_{\mbox{\tiny NLL,}\alpha_s}$ is obtained by expanding the NLL cumulative distribution in powers of $\alpha_s$ and taking the $\mathcal{O}(\alpha_s)$ piece,
\begin{align}
\label{eq:cumulative-NLL-alpha-approx}
\Sigma(\rho)_{\mbox{\tiny NLL}} = 1 - \frac{\alpha_sC_i}{\pi}&\int_0^1\text{d}x_j\,T_j(x_j)\, \big[\tfrac{1}{2}f^{g,0}(x_j\rho,1)\ln^2(\tfrac{1}{\rho})\\ \nonumber
& + \ln(\tfrac{1}{\rho})f^{g,0}(x_j\rho,1)\left(B_i-\ln(x_j)\right) + f^{g,1}(x_j\rho,1)\big] + \mathcal{O}(\alpha_s^2)\,.
\end{align}
The cumulative cross section $\Sigma_{\mbox{\tiny LO}}$ is the integral of the corresponding differential cross section in \Eq{eq:fixed-order-cross-section}:
\begin{equation}
\Sigma_{\mbox{\tiny LO}}(\rho) = 1 - \int_\rho^1 \text{d}\rho'\, \frac{1}{\sigma_0}\frac{\text{d}\sigma_{\mbox{\tiny LO}}}{\text{d}\rho'} \,.
\end{equation}
The normalization factors for the quark and gluon fixed-order distributions are the Born-level cross sections
\begin{equation}
\sigma_0^q = \sum_{\rm colors}\sum_{f = u,\ldots b} \frac{4\pi \alpha^2}{3s}Q_f^2\,, \ \ \ \ \ \sigma_0^g = \frac{s^2}{8\pi} \left(\frac{m_eA}{v_{\rm EW}(s - m_H^2)}\right)^2 \,.
\end{equation}
Here $Q_f$ is the fermion electric charge, $v_{\rm EW}$ is the Higgs vacuum expectation value, and $A$ is the effective coupling constant in the $m_t \rightarrow \infty$ effective theory~\cite{Inami:1982xt,Dawson:1990zj,Spira:1995rr} in which the matrix elements for $e^+e^-\rightarrow H \rightarrow ggg (q\bar{q}g)$ were calculated (see \App{app:details-matching-comp}).

Ideally, we would like our matched NLL+LO distribution to vanish at the upper kinematic endpoint of the LO parton-level process, which is 
\begin{equation}
\label{eq:rho-max}
\rho_{\mbox{\scriptsize max,\tiny LO}} = \frac{1-\cos(R)}{2R^2} \approx 0.23 \ (R=1)\,.
\end{equation}
In the modified log-$R$ framework, we can approximate this condition with the replacement
\begin{equation}
\label{eq:shift}
\frac{1}{\rho} \rightarrow \frac{1}{\rho} - \frac{1}{\rho_{\mbox{\scriptsize max,\tiny LO}}} + e^{-B_i}\,,
\end{equation}
where $e^{-B_i}$ is the endpoint of the NLL resummed calorimeter mass distribution \cite{Marzani:2017mva,Marzani:2017kqd}.
Because we are evaluating the radiator numerically and not making subleading adjustments to enforce that \Eq{eq:shift} exactly matches the endpoints, there will be small residuals beyond the LO endpoint in the distributions below.

\begin{figure}[t]
	\centering
	\subfloat[]{
		\includegraphics[width=0.45\textwidth]{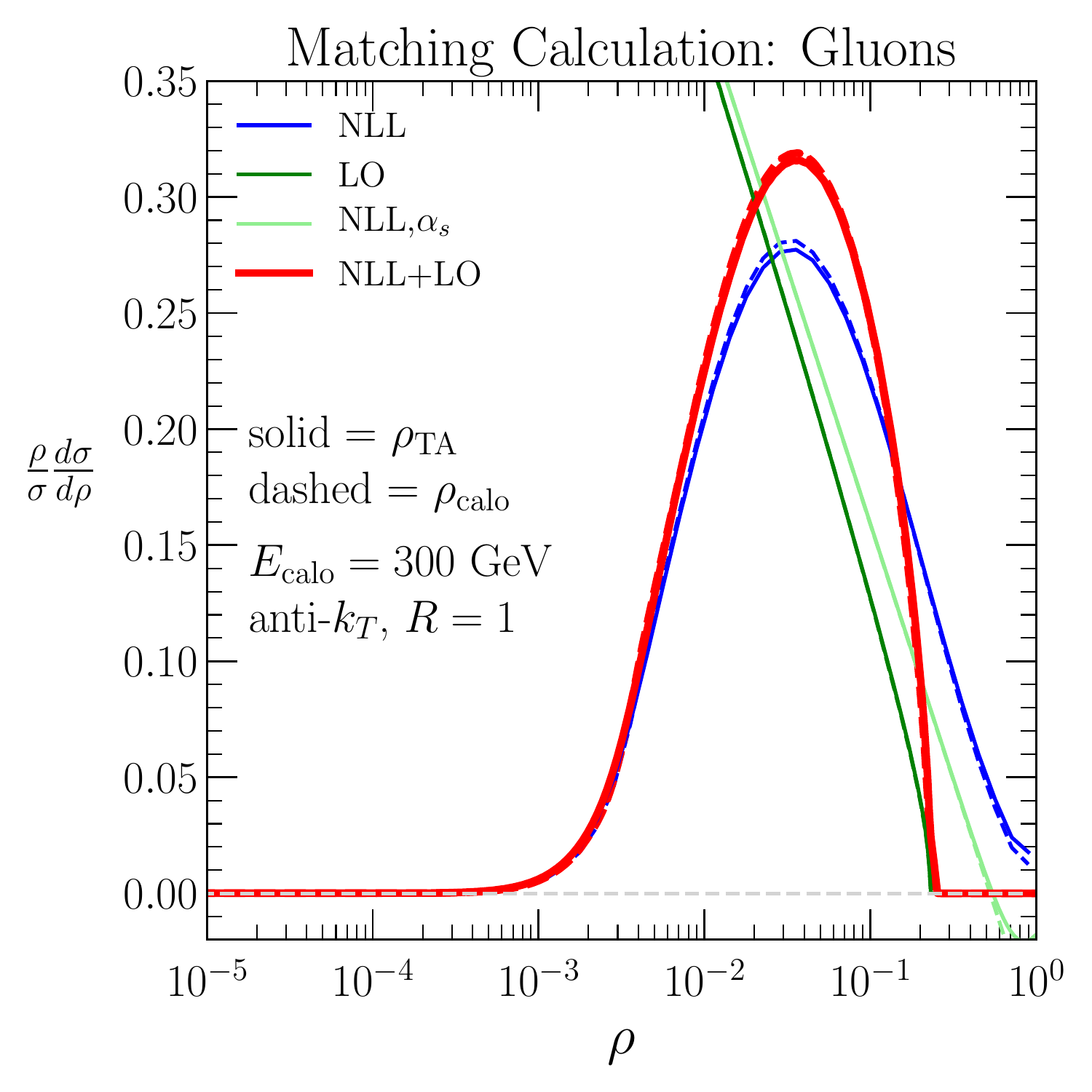}
		\label{fig:calculation-gluon-nogroom}
	}
	\subfloat[]{
		\includegraphics[width=0.45\textwidth]{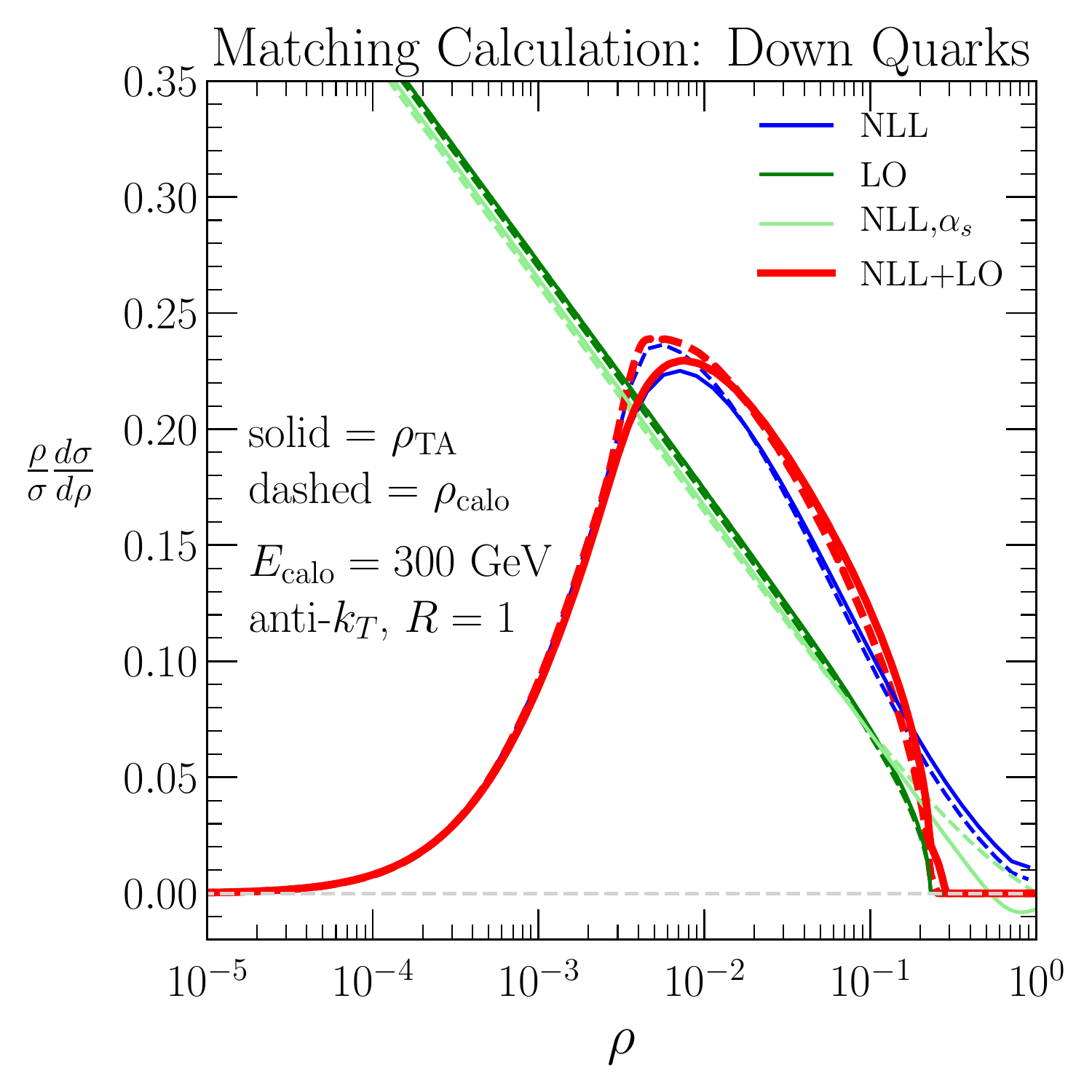}
		\label{fig:calculation-quark-nogroom}
	}

	\subfloat[]{
		\includegraphics[width=0.45\textwidth]{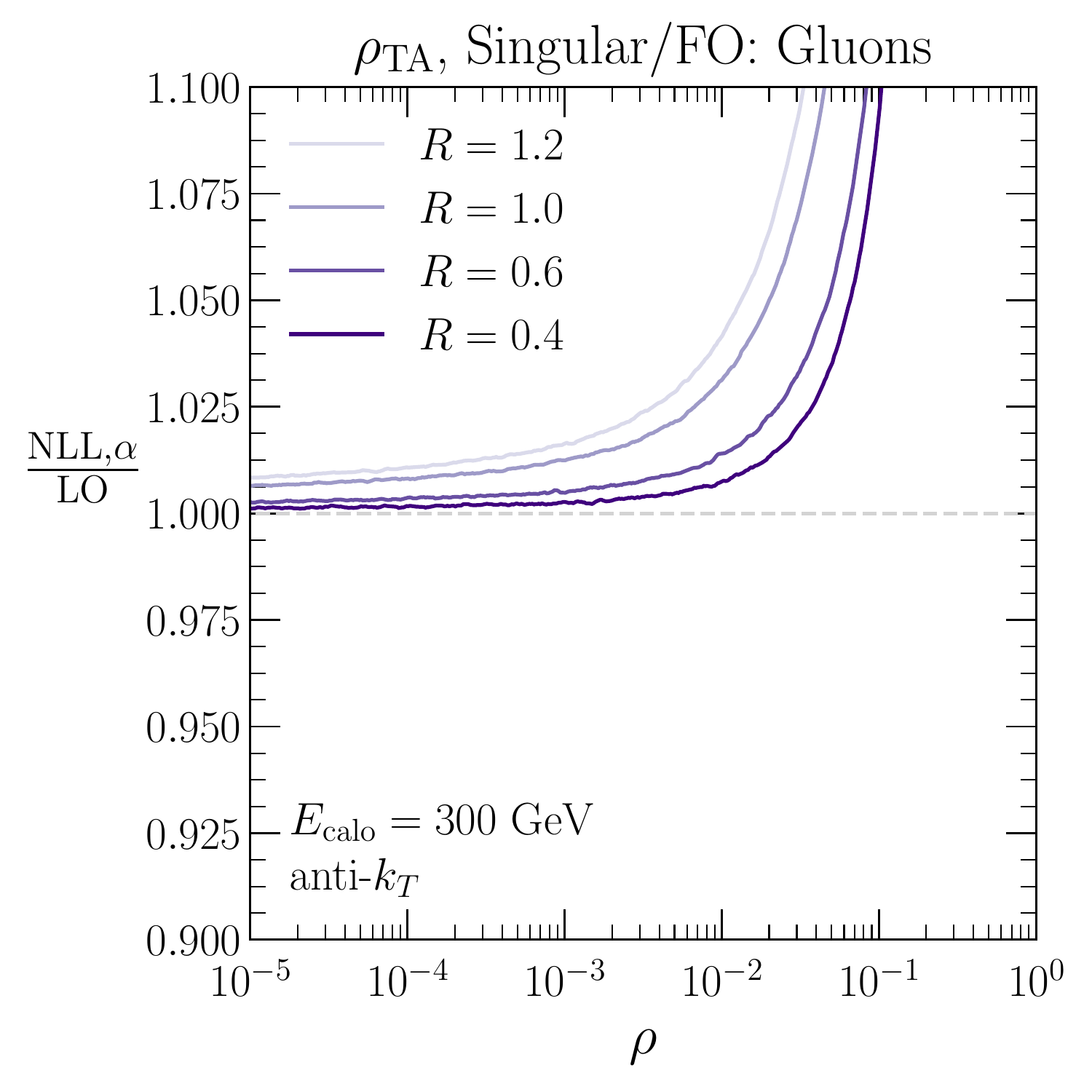}
		\label{fig:ratios-gluon}
	}
	\subfloat[]{
		\includegraphics[width=0.45\textwidth]{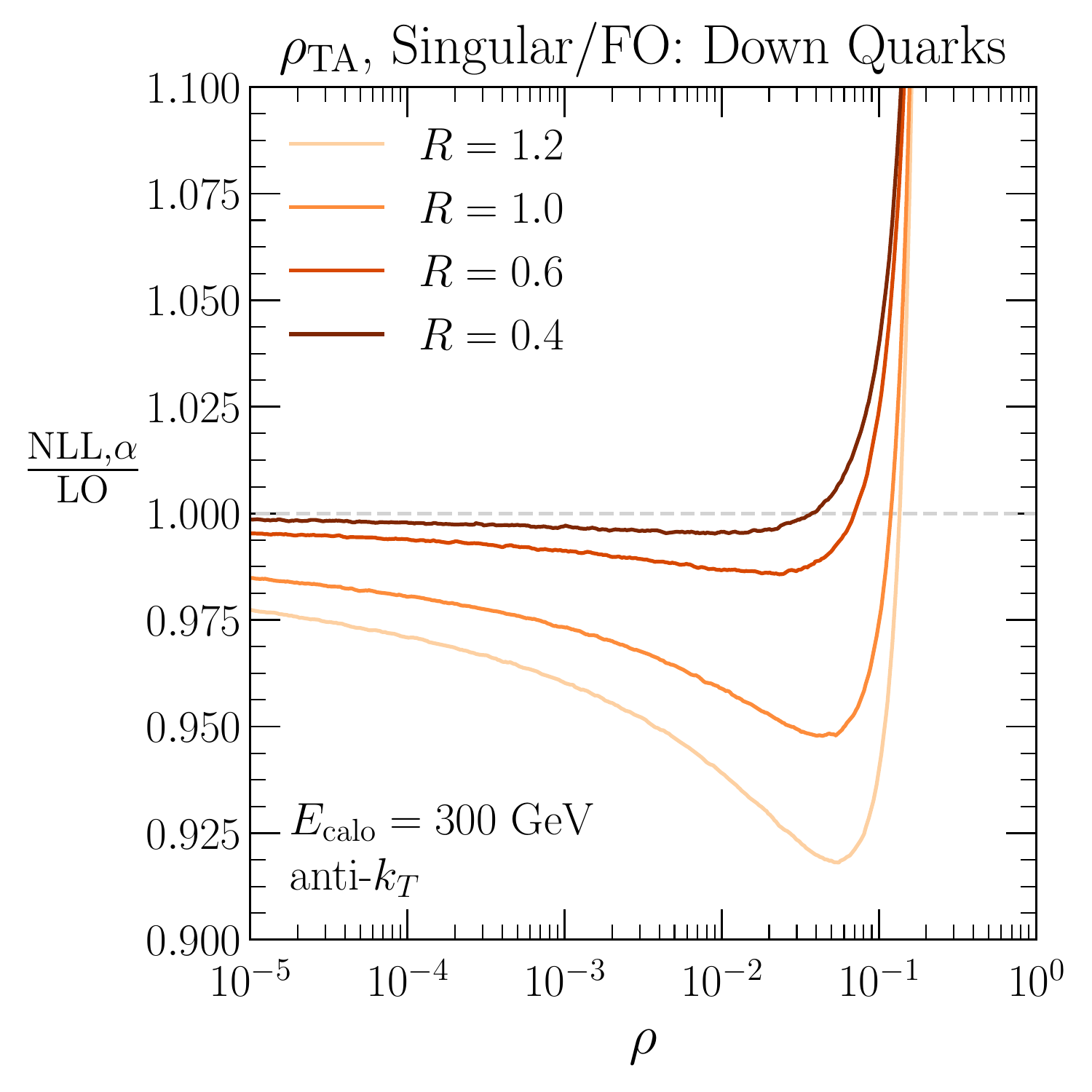}
		\label{fig:ratios-quark}
	}
	\caption{\label{fig:calculation-nogroom} Top row: Components of the matching calculation for (a) gluons and (b) down quarks. The full matched NLL+LO result is in red, the LO fixed-order distribution is in dark green, and the $\mathcal{O}(\alpha_s)$ piece of the fixed-order expansion of the NLL resummed distribution is in light green. The NLL distribution is plotted in blue. Bottom row: The corresponding ratios of the $\mathcal{O}(\alpha_s)$ piece of the NLL distribution over the LO differential distribution, for various values of jet radius $R$ to highlight the $R \log R$ residual.}
\end{figure}

In \Fig{fig:calculation-nogroom}, we illustrate the matching procedure described above. 
In the top row, we plot the gluon and down-quark NLL differential distributions in blue, and LO distributions in dark green. 
The light green curves (NLL,$\alpha_s$) are the differential distributions obtained from the $\rho$ derivative of the $\mathcal{O}(\alpha_s)$ piece of \Eq{eq:cumulative-NLL-alpha-approx}. 
As expected, these $\alpha_s$-expanded NLL distributions match the LO distributions in the $\rho\rightarrow 0$ limit, where the fixed-order result is dominated by large logs of $\rho$. 
With the log-$R$ prescription, this gives us the NLL+LO matched distributions (red curves).

The bottom row of \Fig{fig:calculation-nogroom} shows the ratio
\begin{equation}
\frac{1}{\sigma} \frac{\text{d}\sigma_{\mbox{\tiny NLL},\alpha_s}}{\text{d}\rho}\bigg/ \frac{1}{\sigma_0} \frac{\text{d}\sigma_{\mbox{\tiny LO}}}{\text{d}\rho} \,,
\end{equation}
for $M_{\rm calo}$ with $E_{\rm CM} = 300$ GeV and several values of the jet radius $R$. 
We might naively expect this ratio to approach one as $\rho \rightarrow 0$, since double and single logs of $\rho$ should dominate finite terms in this limit.
Instead, the difference between this ratio and unity scales as $R\log R $.
This is a power-suppressed effect in the small $R$ limit, as expected since we took the collinear limit of the observable in \Eq{eq:observable-sc-limit}.
When we go to the soft-drop groomed distributions in \Sec{sec:sd-calculation}, this residual will be noticeably smaller since soft drop grooms away wide-angle contributions that contribute to this power correction.


\subsection{Extension to Generalized Track-Assisted Mass}
\label{sec:calculation-gtam}

With the full NLL+LO machinery in place, the extension to GTAM is straightforward for arbitrary values of $\kappa$ and $\lambda$.
To compute the NLL-resummed GTAM distribution, we first need to rewrite the observable value from \Eq{eq:obs-ee-parton-def} for a parton-level splitting $i\rightarrow jk$ as
\begin{equation}
\label{eq:gtam-rs-obs}
\hat{\rho}_{\mbox{\tiny TA}}^{(\kappa,\lambda)} \simeq \frac{x_kz \theta^2}{x_j^{2\kappa-1}\langle x_i\rangle^{2\lambda}R^2}  \, , \qquad \qquad \langle x_i\rangle = \int_0^1 \text{d}x_i\, x_i \, T_i(x_i,\mu)\,.
\end{equation}
The track function in \Eq{eq:gtam-rs-obs} corresponds to the parton initiating the jet, and the expectation value of $x_i$ gives the ensemble-averaged charged energy fraction for that parton.
In terms of the NLL calculation, this just requires the replacement $R(\rho,x_j) \rightarrow R(\rho, x_j^{2\kappa-1}\langle x_i\rangle^{2\lambda})$ in \Eq{eq:track-cumulative}.
Similarly, to carry out the fixed-order calculation, we replace \Eq{eq:coordobs} with 
\begin{equation}
\label{eq:gtam-fo-obs}
\hat{\rho}_{\mbox{\tiny TA}}^{(\kappa,\lambda)} = \left( \frac{4x_1x_3(1-y_2)(2-y_2)^{2\kappa-2}}{(x_1 y_1 + x_3 (2-y_1-y_2))^{2\kappa}}\right)\frac{1}{\langle x_1\rangle^{2\lambda}R^2}\,.
\end{equation}
The only change to the log-$R$ matching scheme from \Sec{sec:calculation-matching-fo} is
\begin{equation}
\label{eq:gtam-matching-dist}
\Sigma_{\mbox{\tiny NLL,}\alpha} = - \int_0^1 \text{d}x_j\, T_j(x_j,\mu)\, R(\rho,x_j) \quad \Longrightarrow \quad \Sigma_{\mbox{\tiny NLL,}\alpha} = - \int_0^1 \text{d}x_j\, T_j(x_j,\mu)\, R(\rho,x_j^{2\kappa-1}\langle x_i\rangle^{2\lambda})\,.
\end{equation}

\begin{figure}[t]
	\centering
	\subfloat[]{
		\includegraphics[width=0.4\textwidth]{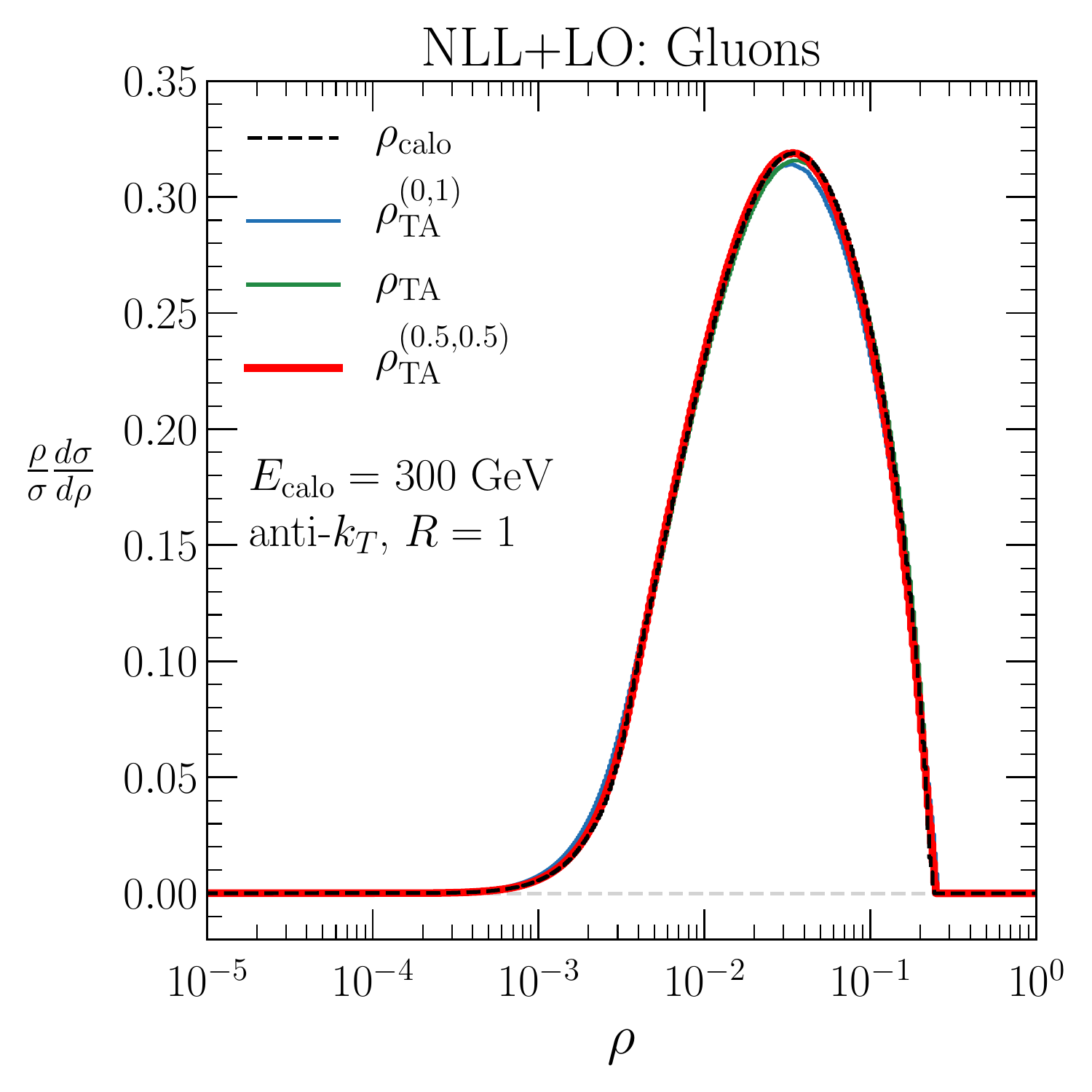}
		\label{fig:gtam-gluon-matched}
	}
	\subfloat[]{
		\includegraphics[width=0.4\textwidth]{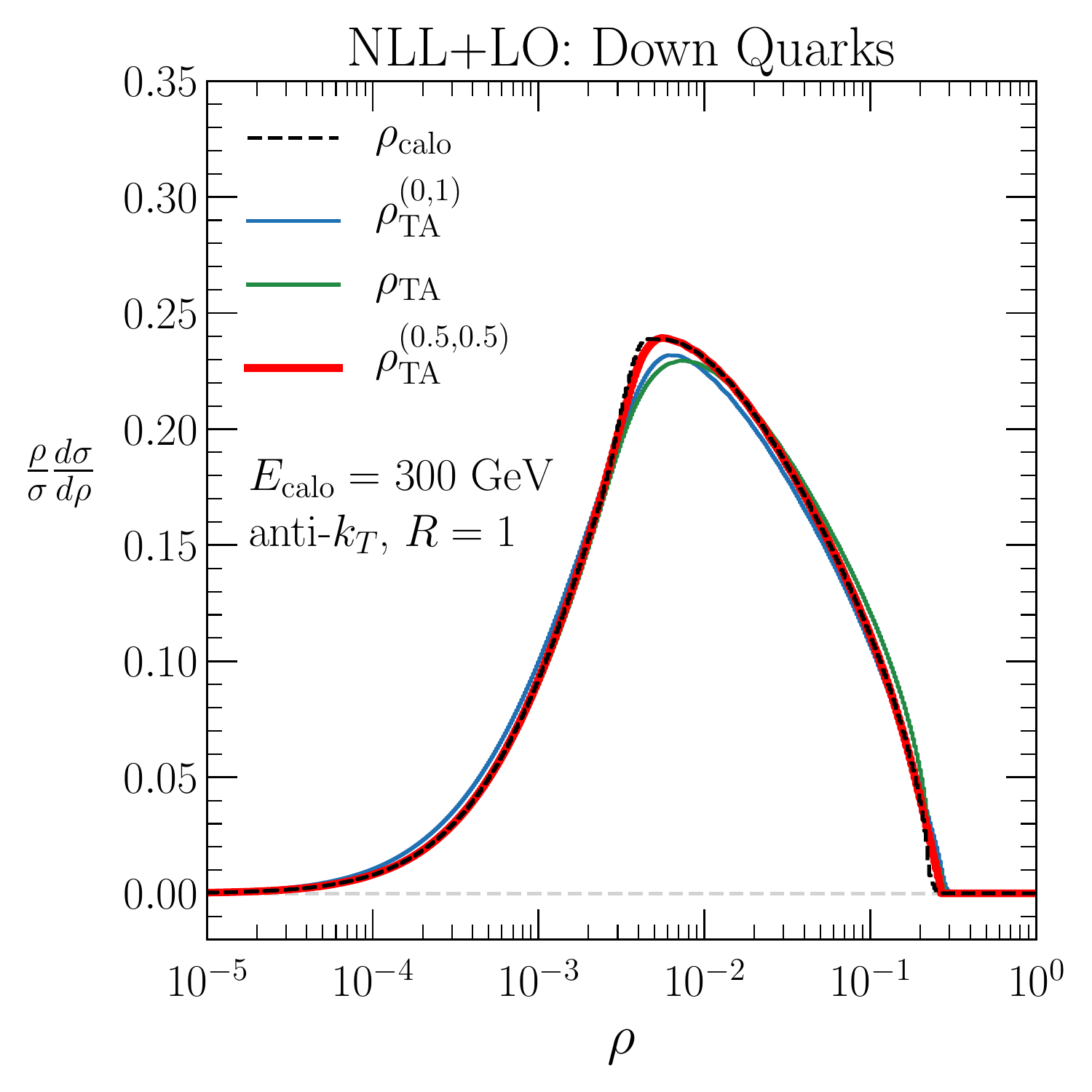}
		\label{fig:gtam-quark-matched}
	}
	
	\subfloat[]{
		\includegraphics[width=0.4\textwidth]{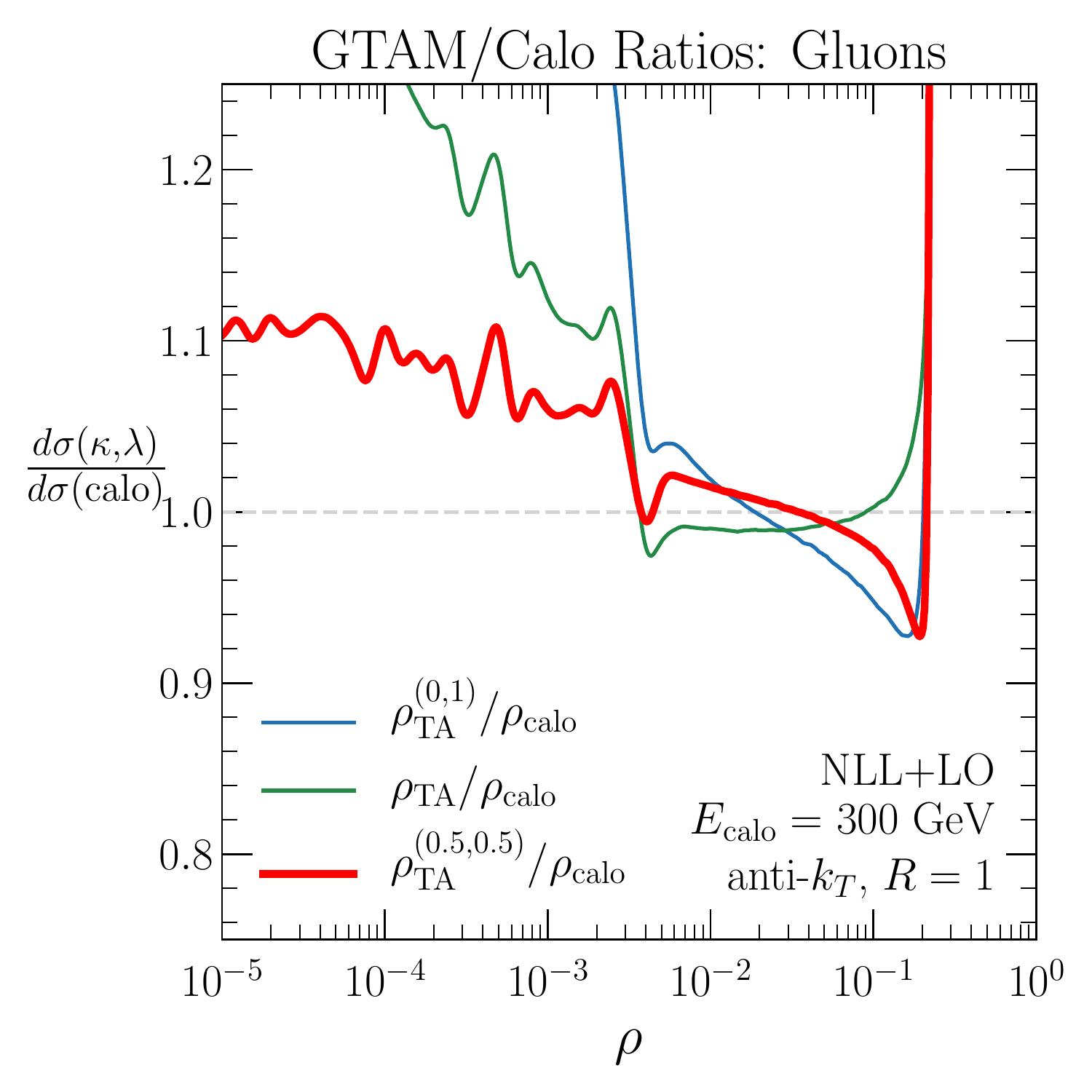}
		\label{fig:gtam-gluon-ratios}
	}
	\subfloat[]{
		\includegraphics[width=0.4\textwidth]{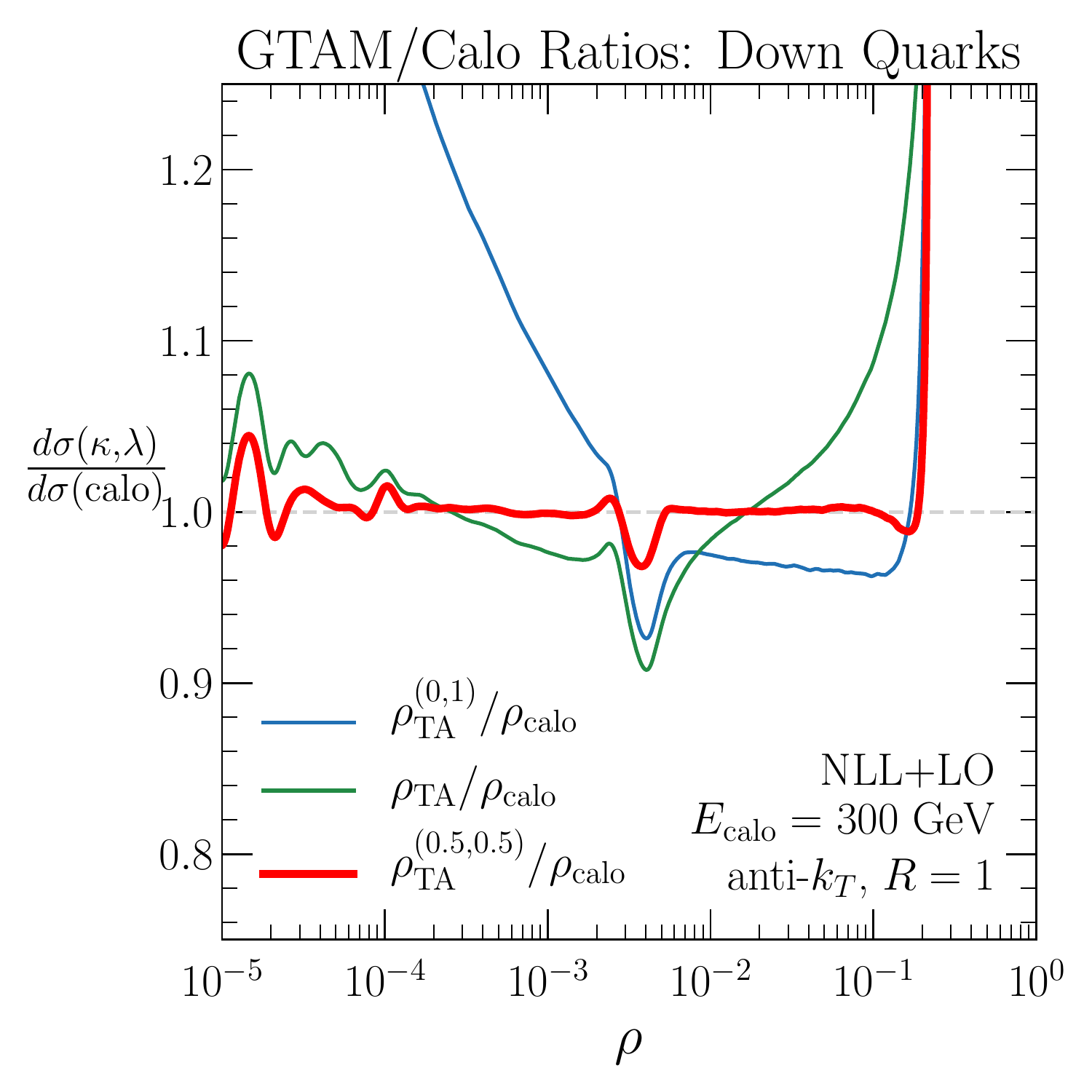}
		\label{fig:gtam-quark-ratios}
	}
	\caption{Top row: NLL+LO calculations of the GTAM distribution for (a) gluon jets and (b) down-quark jets, compared to ordinary jet mass.  Bottom row: corresponding ratio between GTAM and jet mass. The kink at $\rho \simeq \text{few} \times 10^{-3}$ is due to freezing the coupling $\alpha_s$ at $\mu_{\rm NP}$.}
	\label{fig:gtam-matched}
\end{figure}

In the top row of \Fig{fig:gtam-matched}, we show gluon and down-quark NLL+LO matched distributions for $\rho_{\rm track}$, $\rho_{\rm TA}$, and our recommended GTAM observable $\rho_{\rm TA}^{(0.5,0.5)}$.
To make the comparison between GTAM and jet mass more clear, we plot the ratio of $\rho_{\rm TA}^{(\kappa,\lambda)}$ over $\rho_{\rm calo}$ for these same observables in the bottom row of \Fig{fig:gtam-matched}.
As expected from the \textsc{Vincia} study, the $(\kappa,\lambda) = (0.5,0.5)$ distribution more closely tracks the $\rho_{\rm calo}$ shape in the peak region, though a precise comparison needs to include the non-perturbative effects from \Sec{sec:calculation-np}.

Following \Sec{sec:insights}, we can do a fixed-coupling analysis to estimate the optimal value of $\kappa$ and $\lambda$.  
The only change in the argument comes from the replacement for $\hat{\rho}_{\rm TA} \rightarrow \hat{\rho}_{\rm TA}^{(\kappa,\lambda)}$ in \Eq{eq:gtam-rs-obs}. 
Making exactly the same approximations as in \Sec{sec:insights}, we obtain an expression analogous to \Eq{eq:cumulative-fixed} for a jet of flavor $j$: 
\begin{align}\label{eq:cumulative-fixed-gtam}
\begin{split}
\Sigma_{\mbox{\tiny TA}}^{(\kappa,\lambda)}(\rho) \simeq \Sigma_{\rm calo}(\rho)\times \exp\left\{-\frac{\alpha_sC_i}{\pi}\ln(\tfrac{1}{\rho}) \big(f^{g,1}-(2\kappa-1)f^{j,1} -2\lambda \ln \langle x_i\rangle \big) \right\}.
\end{split}
\end{align}
Thus, we obtain a simple linear relation to describe the values of $\kappa$ and $\lambda$ which should give the closest agreement between $\rho_{\rm calo}$ and $\rho_{\rm TA}^{(\kappa,\lambda)}$:
\begin{equation}
\label{eq:gtam-rs-fixed-coupling-min}
f^{g,1}-(2\kappa-1)f^{j,1} -2\lambda \ln \langle x_i\rangle = 0\,.
\end{equation}
Using the values $f^{g,1} = -0.52$, $f^{q,1} = -0.59$, and $\langle x_i\rangle = 0.62$, which approximately hold for quark or gluon jets at $p_T = 300$ GeV, gluon jets prefer $\kappa \approx 1.00 - 0.92 \, \lambda$ while quark jets prefer $\kappa \approx 0.94 - 0.81 \, \lambda$.
Notably, both lines are consistent with the naive expectation of $\kappa = 1- \lambda$.
These preferred lines intersect at $(\kappa, \lambda) = (0.50, 0.54)$, which is in surprisingly good agreement with our \textsc{Vincia} findings for $pp$ collisions in \Fig{fig:heatmap-pp}, though this will not be the case for our more detailed numerical study in \Sec{sec:calculation-best-fit}.


\subsection{Non-Perturbative Corrections}
\label{sec:calculation-np}

For low values of $\rho$, or equivalently low values of $M$, perturbative all-orders contributions to the cross section are dominated by non-perturbative effects. 
In particular, an analytic calculation which does not include non-perturbative information will not correctly predict the location of the peak of the jet mass distribution. 
As described in \App{app:details-resummed}, the appropriate scale $\mu$ to evaluate the coupling in the radiator functions \Eqs{eq:radiator-calorimeter}{eq:track-radiator} is the momentum transfer of the splitting, $\mu = E_{\rm calo}z\theta$. 
For low $\rho$ values, the lower bounds on the integrals over $z$ and $\theta$ are very small, and $\mu$ can enter the non-perturbative regime. 
One way to handle this problem is to freeze the coupling at a scale $\mu_{\rm NP} \simeq \Lambda_{\rm QCD}$. 
Non-perturbative effects below this scale can often be handled by convolution with a non-perturbative shape function~\cite{Manohar:1994kq,Dokshitzer:1995zt,Korchemsky:1999kt,Korchemsky:2000kp,Salam:2001bd,Lee:2006nr,Hoang:2007vb,Mateu:2012nk,Stewart:2014nna}.

These non-perturbative effects will occur with a characteristic scale $E_{\rm NP} \simeq \Lambda_{\rm QCD}$, and will therefore be suppressed in a hard interaction with scale $Q$ by powers of $\Lambda_{\rm QCD}/Q$. 
The quantity with the appropriate dimensions is
\begin{equation}
\label{eq:tau-defn}
\tau_n = \frac{M_n^2}{E_n} \sim \tau_n^{\rm PT} + \tau_n^{\rm NP}\,,
\end{equation}
where $n = $ calo or track.
We can write the differential cross section including non-perturbative corrections as
\begin{equation}
\label{eq:shape-fcn-convolution}
\left(\frac{\text{d}\sigma}{\text{d}\rho}\right) = \int_0^{Q\rho} \text{d}\tau \,  F_{\rm NP}(\tau) \, \left(\frac{\text{d}\sigma(\rho -\tau/Q)}{\text{d}\rho}\right)_{\rm PT}\,,
\end{equation}
where $Q$ is the scale of the hard interaction. 
Following \Ref{Stewart:2014nna}, we use a shape function with the parametrized form 
\begin{equation}
\label{eq:shape-functions}
F_{\rm NP}(\tau_n) = \frac{4\tau_n}{\Omega_n^2}\, e^{-\frac{2\tau_n}{\Omega_n}}\,, \ \ n=\mbox{calo, track}\,,
\end{equation}
which is normalized to one, falls off exponentially for large $\tau_n$, and goes linearly to zero at small $\tau_n$. 

To perform the convolution in \Eq{eq:shape-fcn-convolution}, we need to identify the appropriate energy scale $Q$ to divide $\tau_n$. 
For $\tau_{\rm calo}$, the relevant scale is $Q_{\rm calo} = E_{\rm calo}R^2$. 
For $\tau_{\rm track}$, we can write
\begin{equation}
\label{eq:track-np-scaling}
\rho_{\mbox{\tiny TA}}^{(\kappa,\lambda)} =  \frac{M_{\rm track}^2}{E_{\rm calo}^2R^2} \left(\frac{E_{\rm calo}}{E_{\rm track}}\right)^{2\kappa} \left\langle \frac{E_{\rm calo}}{E_{\rm track}} \right\rangle^{2\lambda}=  \left(\frac{\tau_{\rm track}}{E_{\rm calo}R^2}\right)\left(x_i\right)^{1-2\kappa} \left\langle x_i \right\rangle^{-2\lambda}\,.
\end{equation}
In the limit that the track function $T_i$ of the initiating parton is narrow, we can replace the track fraction $x_i$ in \Eq{eq:track-np-scaling} with its average value.%
\footnote{Alternately, one could try to convolve with a track function. The perturbative cross section in \Eq{eq:shape-fcn-convolution} is the NLL+LO matched distribution, so it is not straightforward to assign this track fraction to one of track functions already used in the calculation.  Note that the mismatch from replacing $x_i$ with $\langle x_i \rangle$ dominantly affects the non-perturbative region.  From \Fig{fig:differential-delta}, we see that the primary benefit of using GTAM with $\kappa \simeq 1- \lambda \approx 0.5$ comes from the resummed and fixed-order regions, where $x_i$ versus $\langle x_i \rangle$ is treated properly in our calculation.}
This gives us the hard scale by which non-perturbative effects $\tau_{\rm track}$ are suppressed
\begin{equation}
Q_{\rm track} = E_{\rm calo}R^2\langle x_i\rangle^{2\kappa + 2\lambda -1} = Q_{\rm calo} \langle x_i\rangle^{2\kappa + 2\lambda -1}\,.
\end{equation}

Following the reasoning of \Ref{Chang:2013iba}, we expect that $\Omega_{\rm track} \approx \langle x_g \rangle \Omega_{\rm calo}$. 
This should hold in the limit that the matrix element which defines the non-perturbative parameter $\Omega_{\rm track}$ in the operator product expansion is dominated by a single gluon emission, and in the limit that the track function of the initiating parton $T_i$ is narrow.
We will take the track fraction $\langle x_i \rangle = \langle x_g \rangle = 0.6$ in these relations. 
The gluon and quark shift parameters should be related by approximate Casimir scaling, $\Omega_n^g/\Omega_n^q \approx C_A/C_F$, as expected for observables which are additive in the soft-collinear limit \cite{Larkoski:2013eya,Gras:2017jty}.

\begin{figure}[t]
	\centering
	\subfloat[]{
		\includegraphics[width=0.45\textwidth]{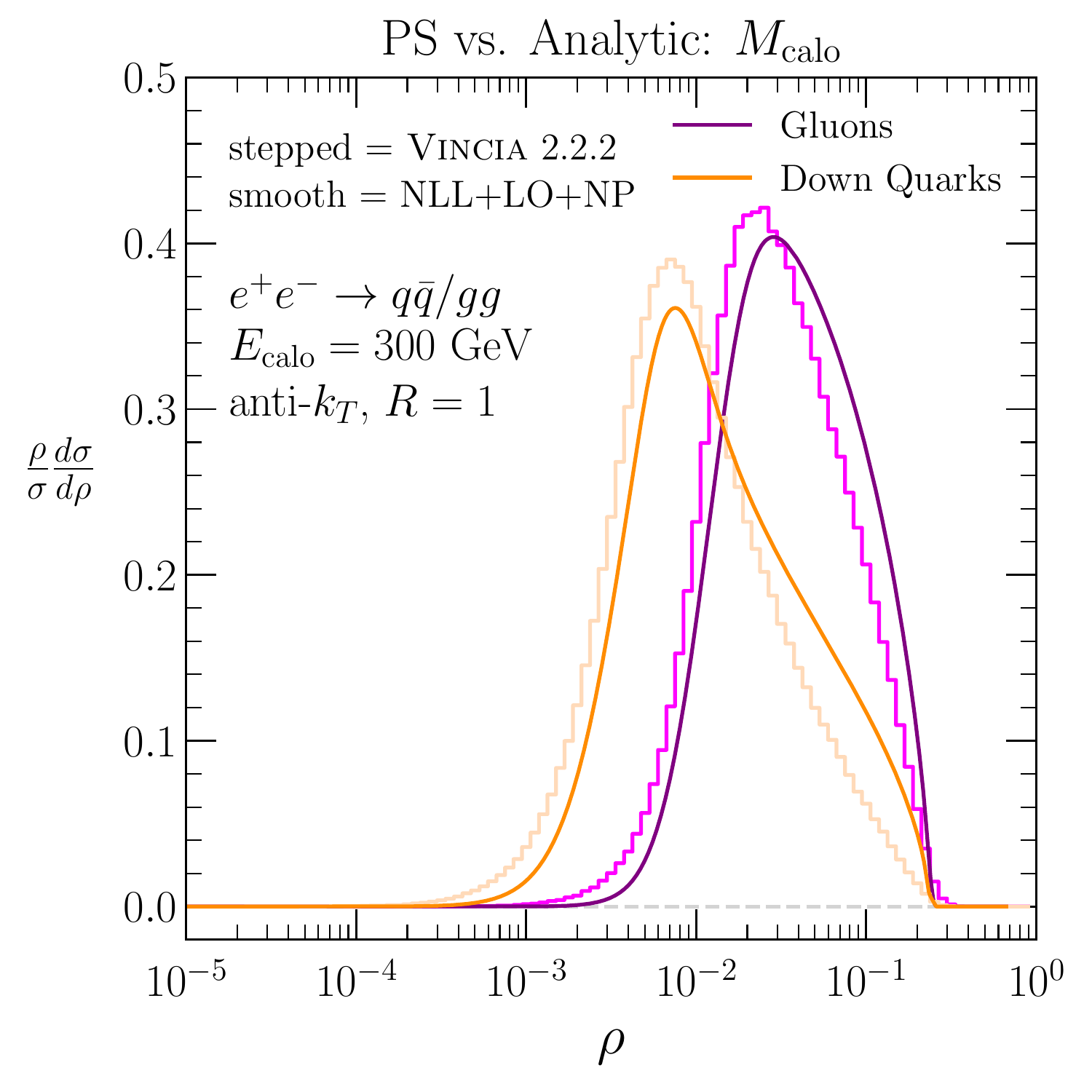}
		\label{fig:np-a}
	}
	\subfloat[]{
		\includegraphics[width=0.45\textwidth]{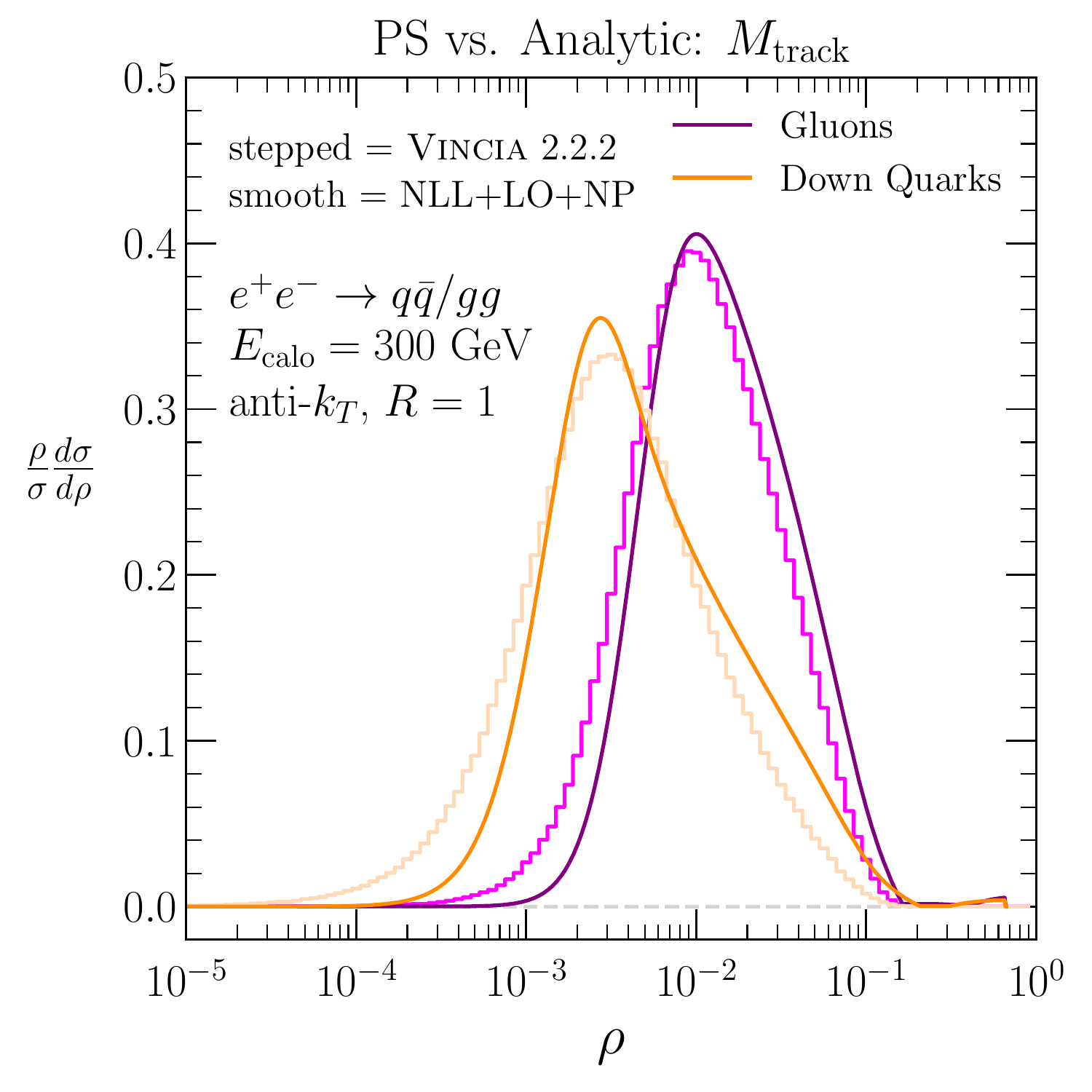}
		\label{fig:np-b}
	}
	
	\subfloat[]{
		\includegraphics[width=0.45\textwidth]{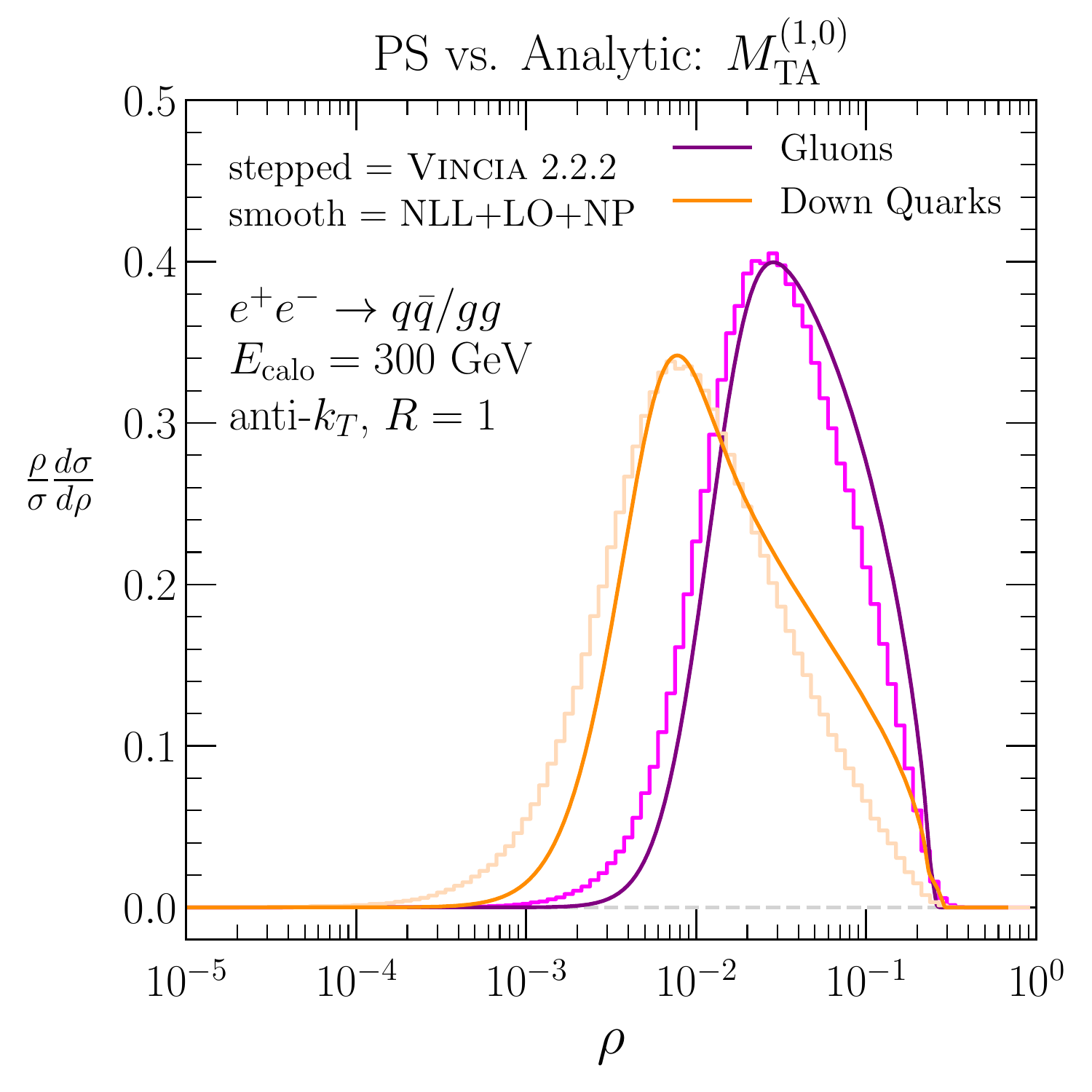}
		\label{fig:np-c}
	}
	\subfloat[]{
		\includegraphics[width=0.45\textwidth]{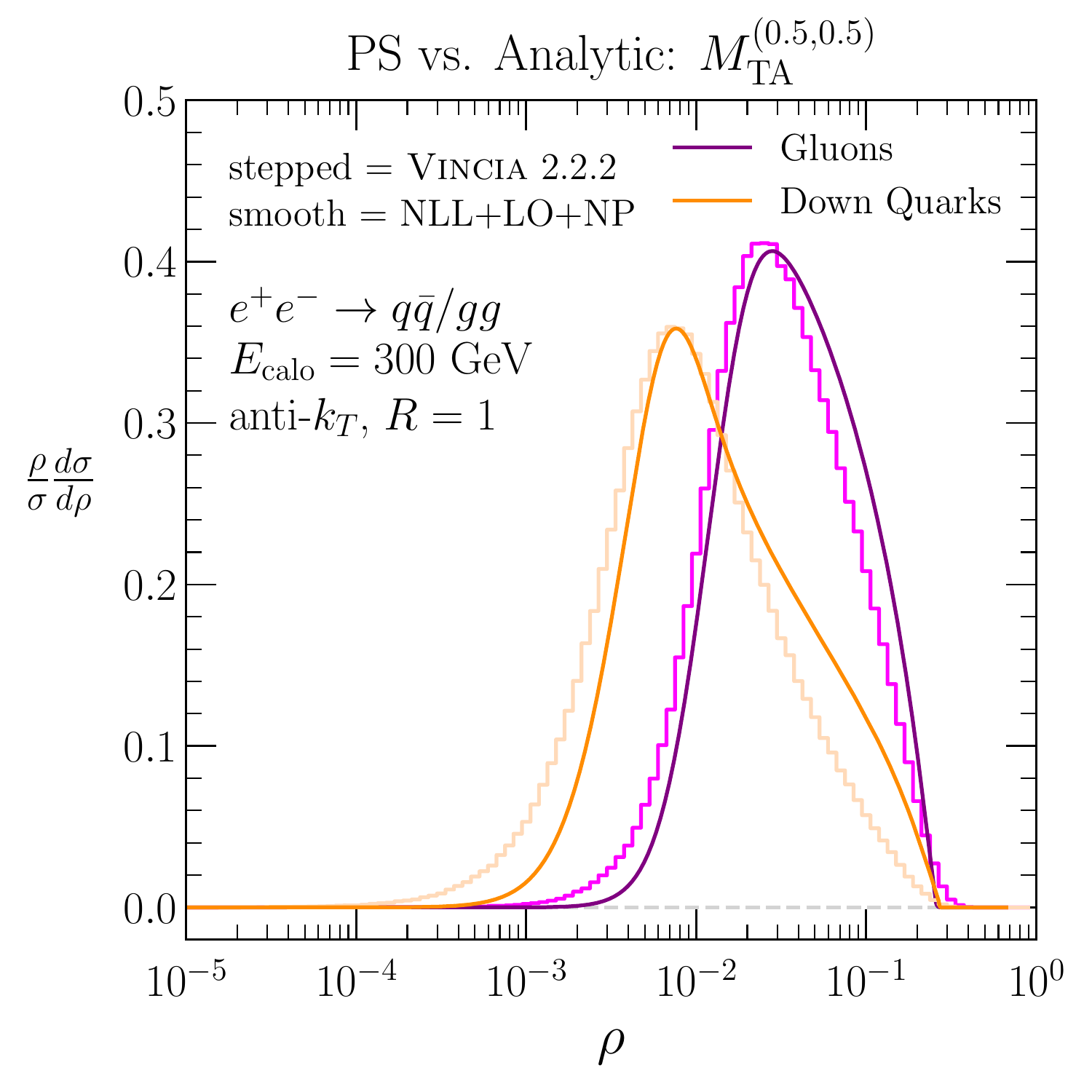}
		\label{fig:np-d}
	}
	\caption{GTAM distributions computed from $\textsc{Vincia}$ (stepped histograms) and NLL+LO with non-perturbative corrections (smooth curves). Shown are (a) calorimeter jet mass, (b) track mass, (c) track-assisted mass, and (d) GTAM with $\kappa=0.5$ and $\lambda = 0.5$.  The high-side discrepancy between our analytic calculation and \textsc{Vincia} is already present in the log-$R$ matching procedure, even before convolution with the non-perturbative shape function.}
	\label{fig:np-shape-fcn} 
\end{figure}

\begin{figure}[t]
	\centering
	\subfloat[]{
		\includegraphics[width=0.45\textwidth]{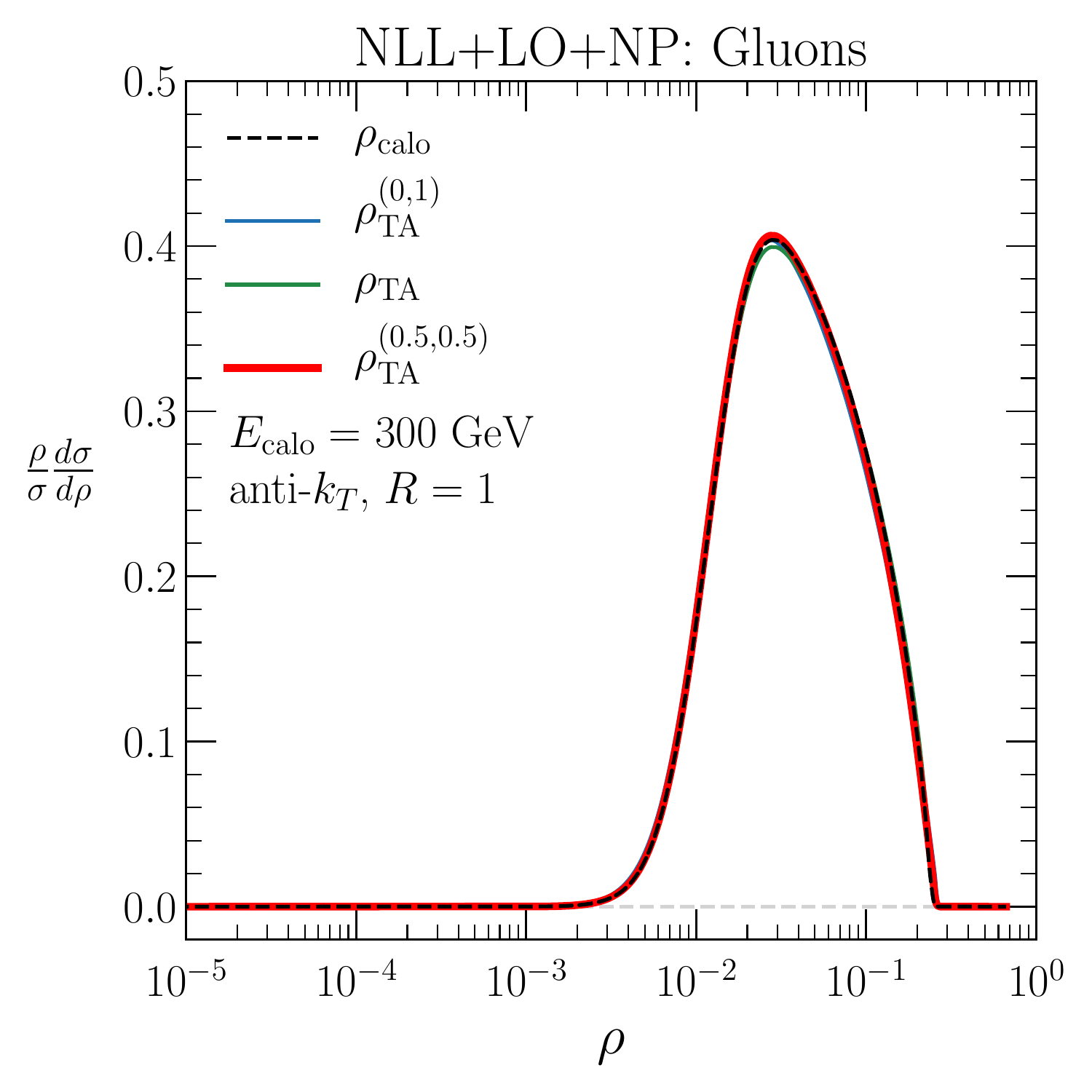}
		\label{fig:diff-delt-a}
	}
	\subfloat[]{
		\includegraphics[width=0.45\textwidth]{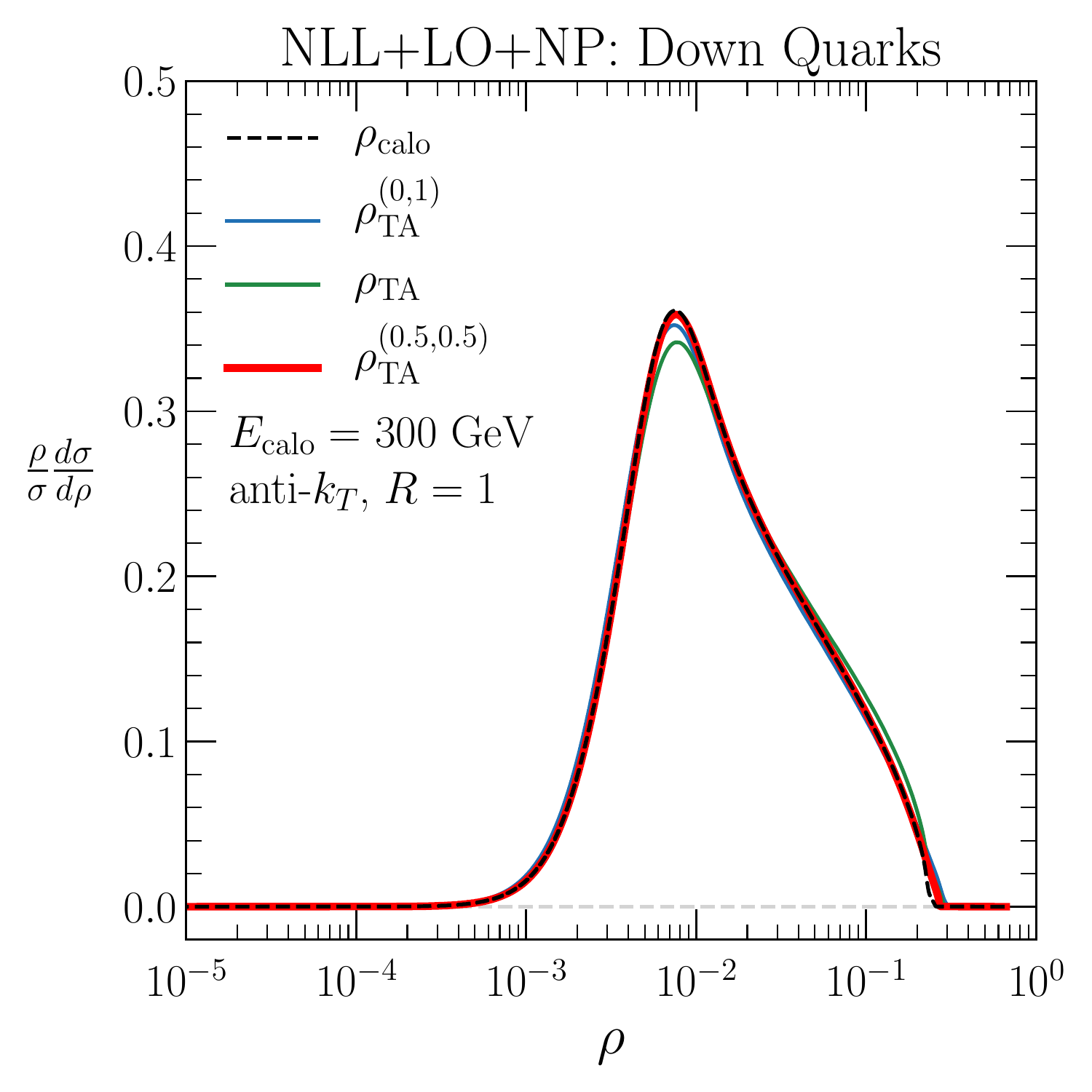}
		\label{fig:diff-delt-b}
	}
	
	\subfloat[]{
		\includegraphics[width=0.45\textwidth]{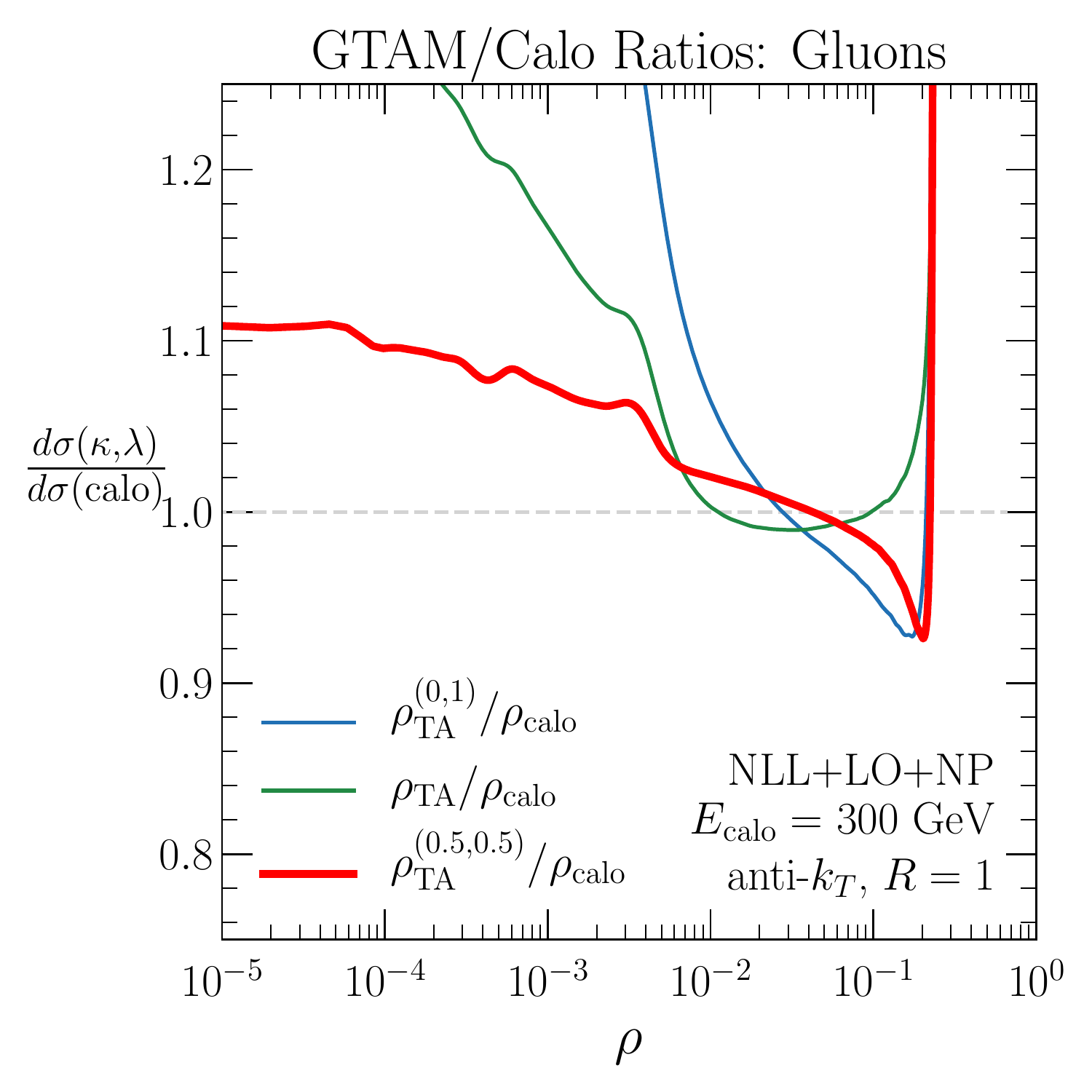}
		\label{fig:diff-delt-c}
	}
	\subfloat[]{
		\includegraphics[width=0.45\textwidth]{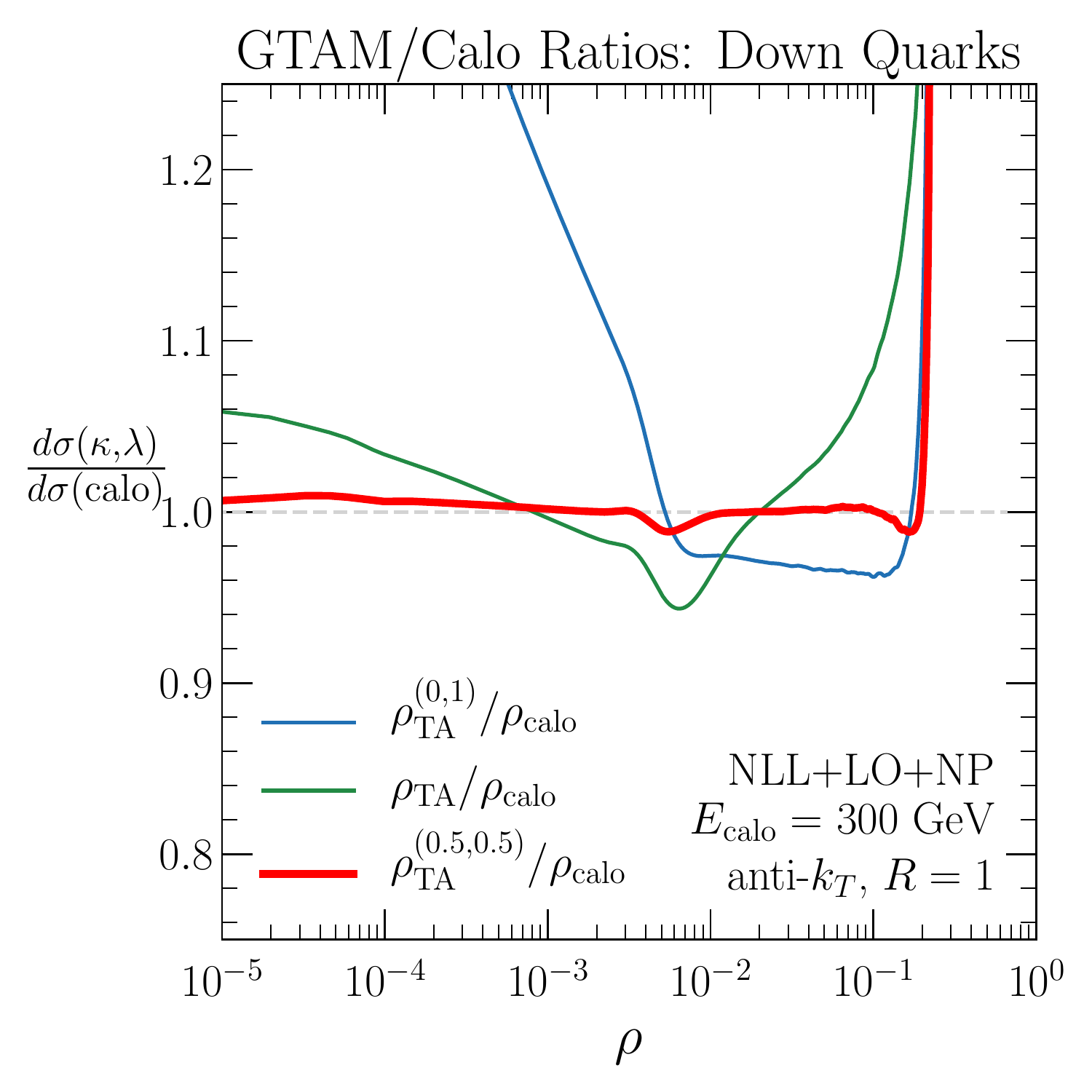}
		\label{fig:diff-delt-d}
	}
	\caption{Same as \Fig{fig:gtam-matched}, but including the non-perturbative corrections.}
	\label{fig:finaldistributions} 
\end{figure}

As shown in \Fig{fig:np-shape-fcn}, we find a reasonable match between the analytic calculations and $\textsc{Vincia}$ with the functional form of \Eq{eq:shape-functions} and the non-perturbative parameters
\begin{align}
\begin{split}
\Omega_{\rm calo}^q &= 0.40\ \mbox{GeV}\,,\ \ \Omega_{\rm calo}^g = 1.00\ \mbox{GeV}\,, \\
\Omega_{\rm track}^q &= 0.25\ \mbox{GeV}\,,\ \ \Omega_{\rm track}^g = 0.60\ \mbox{GeV}\,,
\end{split}
\end{align}
which obey the relations just described.
These parameters, which we found by fitting to predictions from a parton shower, must ultimately be extracted from fits to experimental data.
For completeness, our final NLL+LO+NP predictions are shown in \Fig{fig:finaldistributions}.

\subsection{Best Fit Parameters}
\label{sec:calculation-best-fit}

\begin{figure}[t]
	\centering
	\subfloat[]{
		\includegraphics[width=0.45\textwidth]{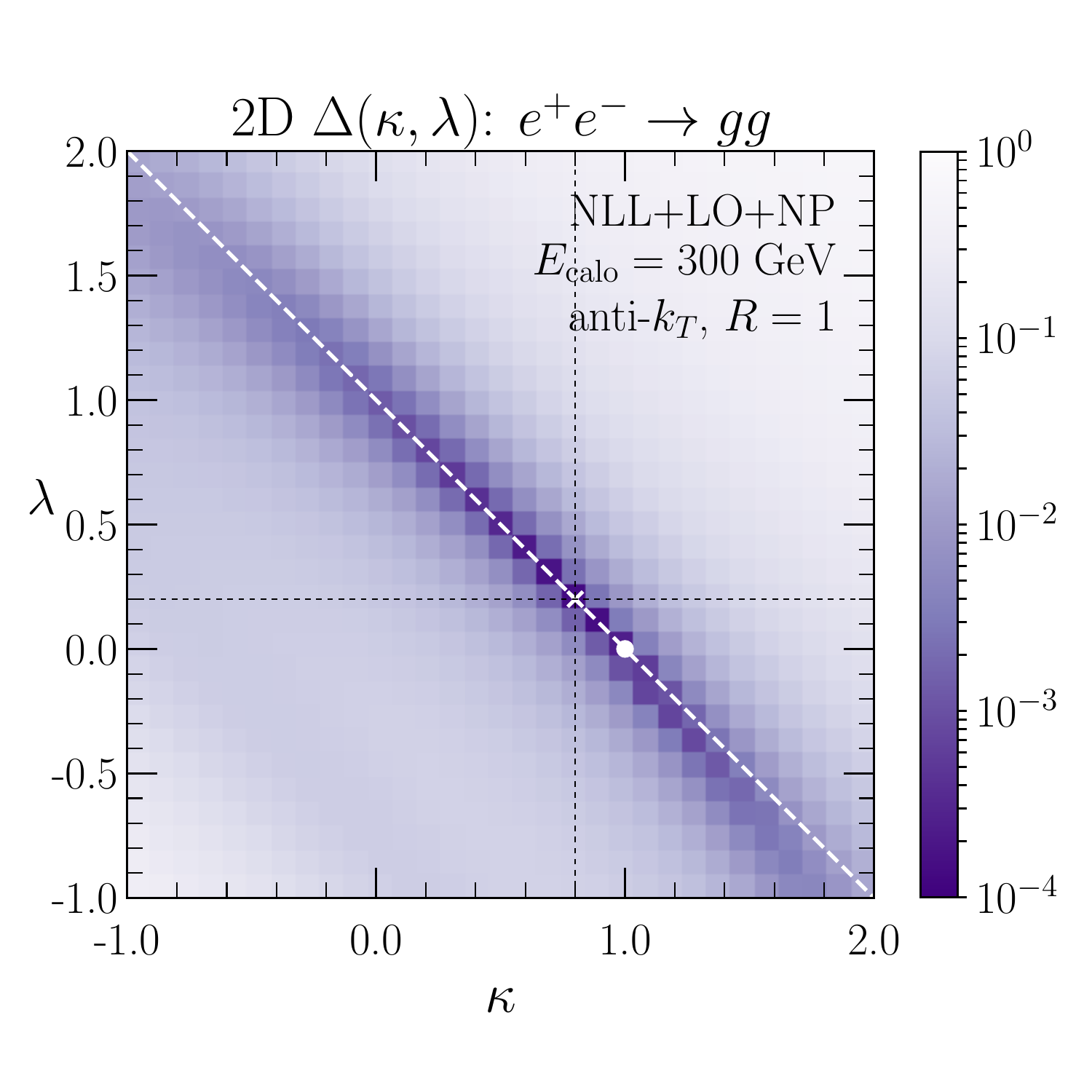}
		\label{fig:gtam-gluon-heatmap}
	}
	\subfloat[]{
		\includegraphics[width=0.45\textwidth]{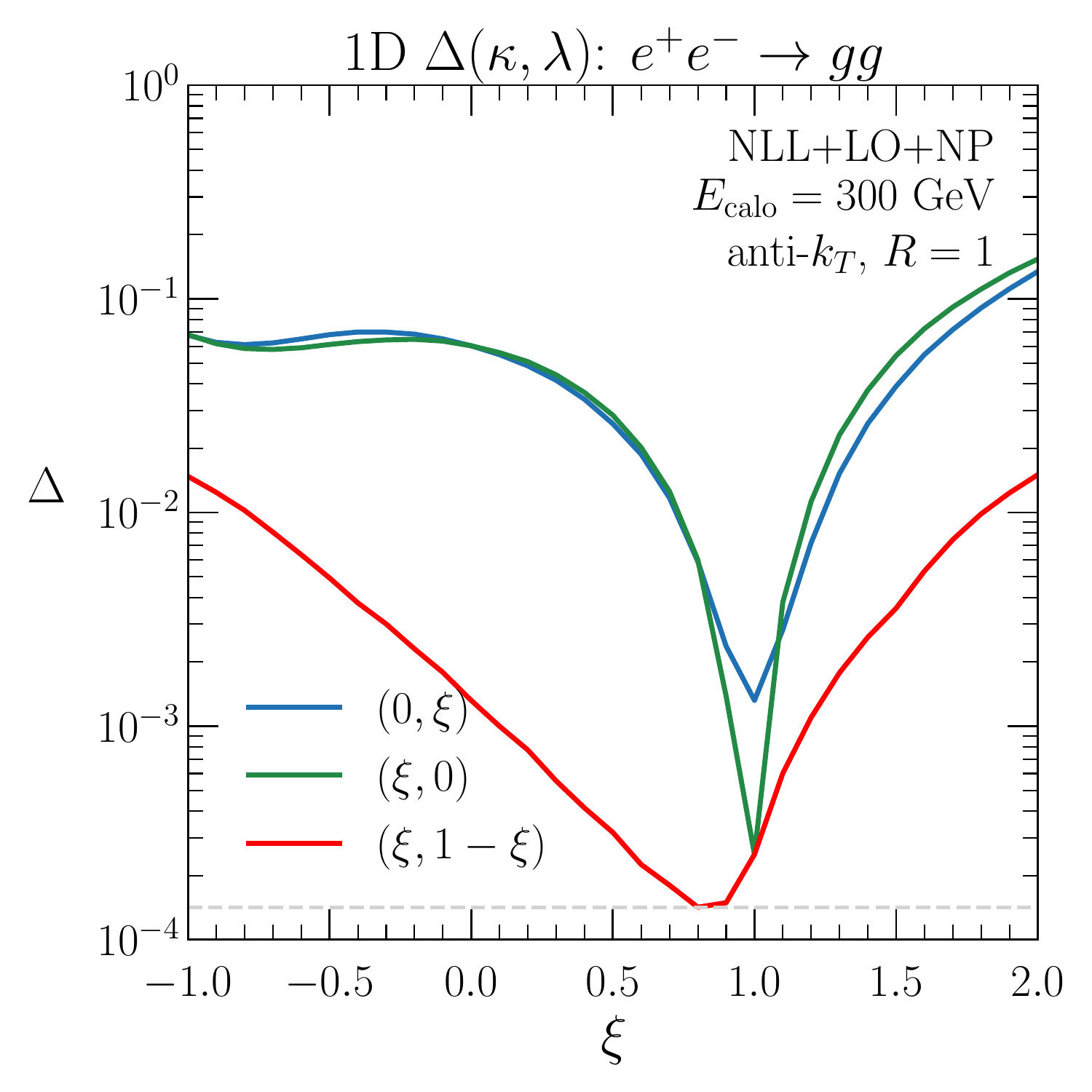}
		\label{fig:gtam-gluon-curves}
	}
	
	\subfloat[]{
		\includegraphics[width=0.45\textwidth]{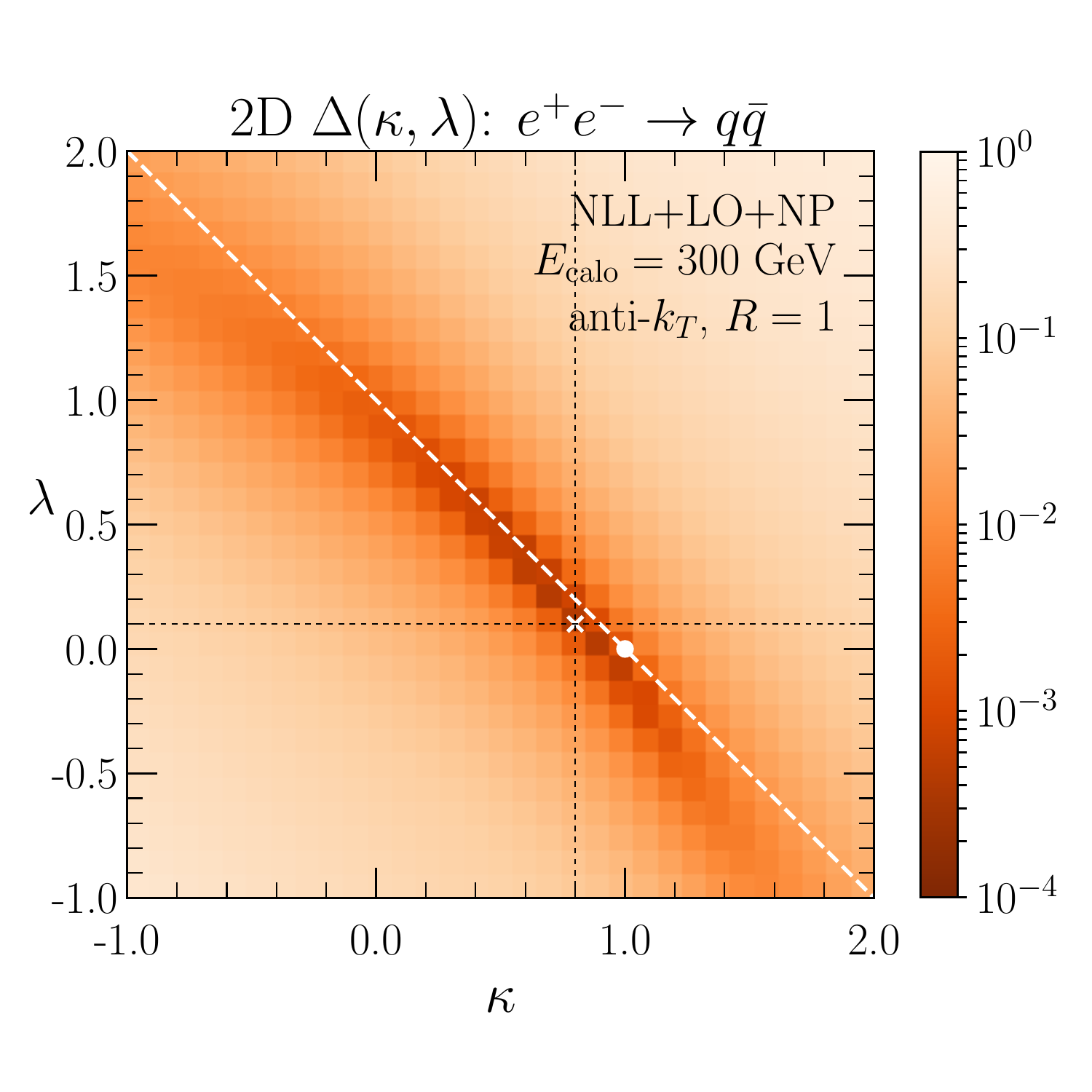}
		\label{fig:gtam-quark-heatmap}
	}
	\subfloat[]{
		\includegraphics[width=0.45\textwidth]{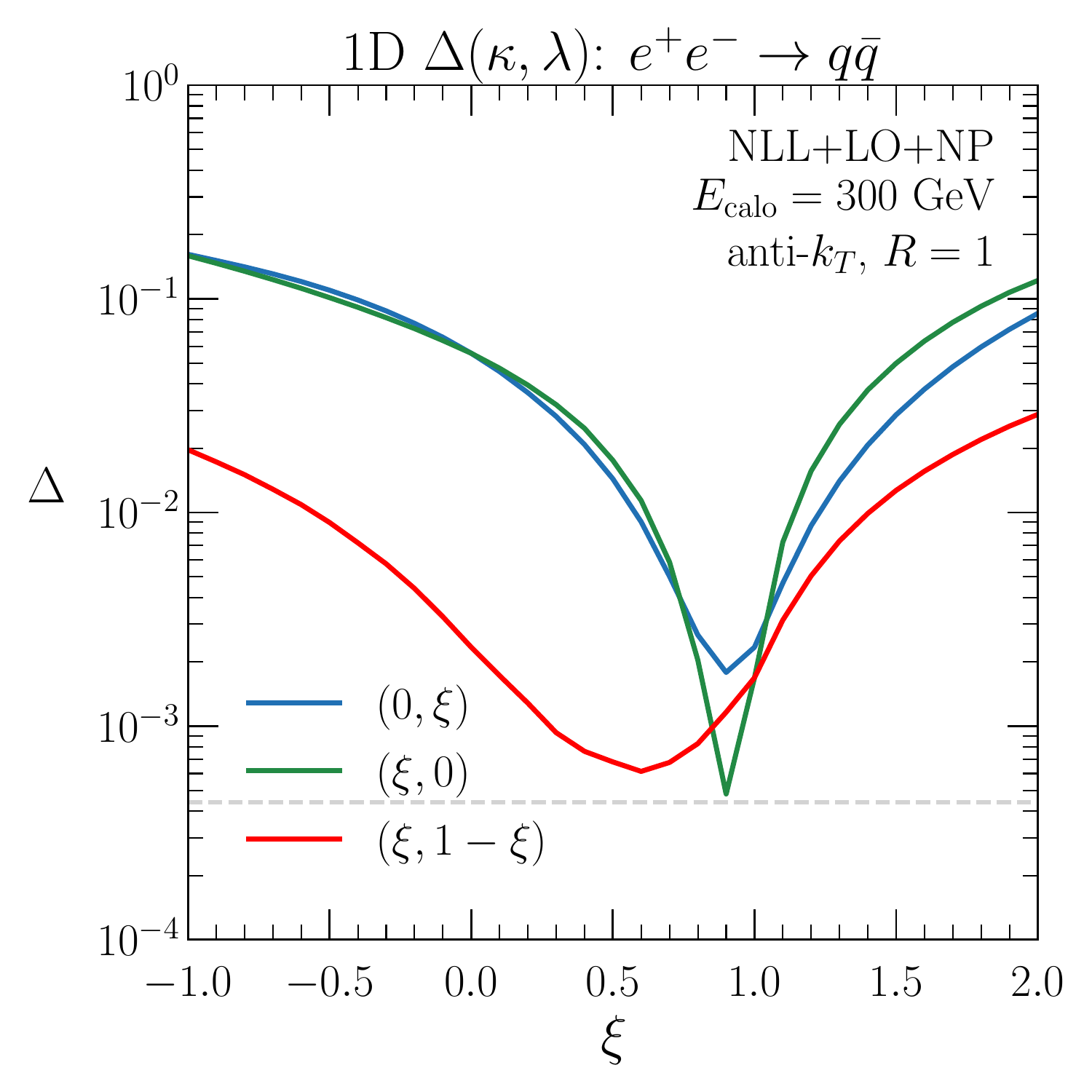}
		\label{fig:gtam-quark-curves}
	}
	\caption{\label{fig:gtam-delta}Left column:  Distribution of $\Delta(\kappa,\lambda)$ computed using the analytic NLL+LO distributions convolved with the shape function $F_{\rm NP}$ for gluons (top) and down quarks (bottom). The white cross marks the best-fit value, which is $(\kappa,\lambda) = (0.8,0.2)$ for gluons and $(0.8,0.1)$ for down quarks. The white dot marks track-assisted mass, $\rho_{\mbox{\tiny TA}} = \rho_{\mbox{\tiny TA}}^{(1,0)}$. Right column:  One-dimensional slices of the distributions on the left. The slight offset of the minimum values in (c) from the line $\lambda = 1-\kappa$ manifests in the green $(\xi,0)$ curve dipping below the red $(\xi,1-\xi)$ curve in (d) before intersecting again at $\xi = 1$. }
\end{figure}

With the complete NLL+LO+NP calculation of the GTAM distributions, we can now compute $\Delta(\kappa,\lambda)$ from first principles. 
This statistic is plotted in the $(\kappa,\lambda)$ plane for gluons (\Fig{fig:gtam-gluon-heatmap}) and down quarks (\Fig{fig:gtam-quark-heatmap}), with one-dimensional slices in \Figs{fig:gtam-gluon-curves}{fig:gtam-quark-curves}. 
Note that we have not attempted to estimate theoretical uncertainties on these distributions.%
\footnote{One way to estimate these uncertainties is by varying the renormalization scale in the jet mass distributions (see, e.g., \Ref{Chang:2013iba}).  This, however, is likely to yield an underestimate (overestimate) of the uncertainty in the $\Delta(\kappa,\lambda)$ distribution if one treats the renormalization scale variations as being fully correlated (uncorrelated).}

Compared to the \textsc{Vincia} $pp$ result in \Fig{fig:heatmap-pp}, the analytic calculation of $\Delta(\kappa,\lambda)$ has the same qualitative features, with a broad minimum at $\kappa + \lambda = 1$.
The precise value of the minimum is different, though, with the NLL+LO+NP result predicting a minimum at $(\kappa=0.8,\lambda = 0.2)$ for gluons and $(\kappa=0.8,\lambda = 0.1)$ for down quarks, as compared to $(\kappa=0.55,\lambda=0.48)$ for the \textsc{Vincia} $pp\rightarrow $ dijets study. 
This is likely due to the fact that the overall degree of agreement between $M_{\rm calo}$ and $M_{\rm TA}^{(\kappa,\lambda)}$ is closer in the analytic calculation than in \textsc{Vincia}, so ensemble-averaged information plays a less effective role.
We checked that the inclusion of $F_\text{NP}$ can change $\Delta(\kappa,\lambda)$ by upwards of 50\% relative to NLL+LO alone, though the optimal values of $\kappa$ and $\lambda$ are relatively insensitive to non-perturbative corrections.
The one exception is for the quark case, where without non-perturbative corrections, the global minimum moves to be along the $\kappa + \lambda = 1$ line at $(\kappa=0.6,\lambda = 0.4)$.


\section{Impact of Soft-Drop Grooming}
\label{sec:soft-drop}

In this section, we study track-assisted mass used in parallel with soft-drop grooming \cite{Larkoski:2014wba}. 
We first discuss different approaches to track-assisted grooming in \Sec{sec:ta_grooming}. 
In \Sec{sec:soft-drop-parton-shower}, we perform a \textsc{Vincia} study on groomed jet GTAM at the LHC.
In \Sec{sec:sd-calculation}, we perform an NLL+LO calculation for $e^+e^-$ collisions.
The qualitative lessons from the groomed case mirror the ungroomed case, so our discussion here will be relatively brief.

\subsection{Track-Assisted Grooming}
\label{sec:ta_grooming}

To implement soft-drop grooming on an ordinary jet, the jet is first reclustered into a Cambridge/Aachen tree \cite{Dokshitzer:1997in,Wobisch:1998wt}.
Then the soft-drop condition,
\begin{equation}
\label{eq:sd-groom}
\frac{\min(p_{T1},p_{T2})}{p_{T1}+p_{T2}} > z_{\rm cut} \left(\frac{\Delta R_{12}}{R_0}\right)^\beta\,,
\end{equation}
is applied at each splitting in the tree beginning with the widest. 
If a splitting fails \Eq{eq:sd-groom}, then the softer branch is removed from the jet and declustering continues to the next splitting. 
When a splitting passes \Eq{eq:sd-groom}, both subjets are kept, and the grooming procedure stops.
The groomed jet mass has been been measured at ATLAS~\cite{Aaboud:2017qwh} and CMS~\cite{CMS:2017tdn}, showing good agreement with resummed calculations~\cite{Frye:2016okc,Frye:2016aiz, Marzani:2017mva,Marzani:2017kqd}. 

We aim to find a track-assisted proxy for groomed jet mass. 
To do this, we groom the jet first, and then compute GTAM using the groomed jet constituents. 
In particular, the factors of $p_{T,\rm calo}$ we use in the definition of the groomed GTAM observable are the $p_T$ of the groomed jet. 
This approach ensures that the charged particles used to compute $M_{\rm TA}^{(\kappa,\lambda)}$ are the same as the charged particles in the soft-drop groomed jet whose mass we are trying to reproduce.

The drawback to grooming before restricting to only charged particles is that the grooming procedure itself requires angular information. 
This problem is less serious than the angular resolution problem when computing the mass $M_{\rm calo}$, in part because the C/A clustering tree mainly relies on the relative angular order between particles, and is therefore less sensitive to the angular resolution.
That said, whether or not an emission passes \Eq{eq:sd-groom} does depend on absolute $\Delta R_{12}$ information.
For this reason, we explore an alternative approach in \App{app:alternate-sd-implementation}, where we restrict to charged particles before grooming, though this makes it difficult to define $p_{T,\rm calo}$ for the reweighting factors.

We will only show results for $\beta \ge 0$, where $\beta = 0$ corresponds to the modified Mass Drop Tagger~\cite{Dasgupta:2013ihk}.
For $\beta < 0$, the soft-drop algorithm acts like a tagger and completely grooms away all jets without two hard, well-separated prongs. 
This leads to a bimodal jet mass distribution with a large population near zero mass, a peak closer to the endpoint, and a gap between the two. 
Our analytic analysis does not give sensible results if the spike near zero mass is included, since $\Delta$ is determined primarily by jets with masses of only a few GeV. 
Thus, to avoid any confusion about this issue, we will not show any $\beta<0$ results in this paper.\footnote{One way to get a sensible result for $\beta < 0$ is to place a cut on the groomed jet mass of $M_{\rm calo} > 3$ GeV.  The fixed-order calculation can then be used to compute the part of the $\beta < 0$ distribution above this small mass cut.}


\subsection{Parton Shower Study}
\label{sec:soft-drop-parton-shower}

\begin{figure}[t]
	\centering
	\subfloat[]{
		\includegraphics[width=0.45\textwidth]{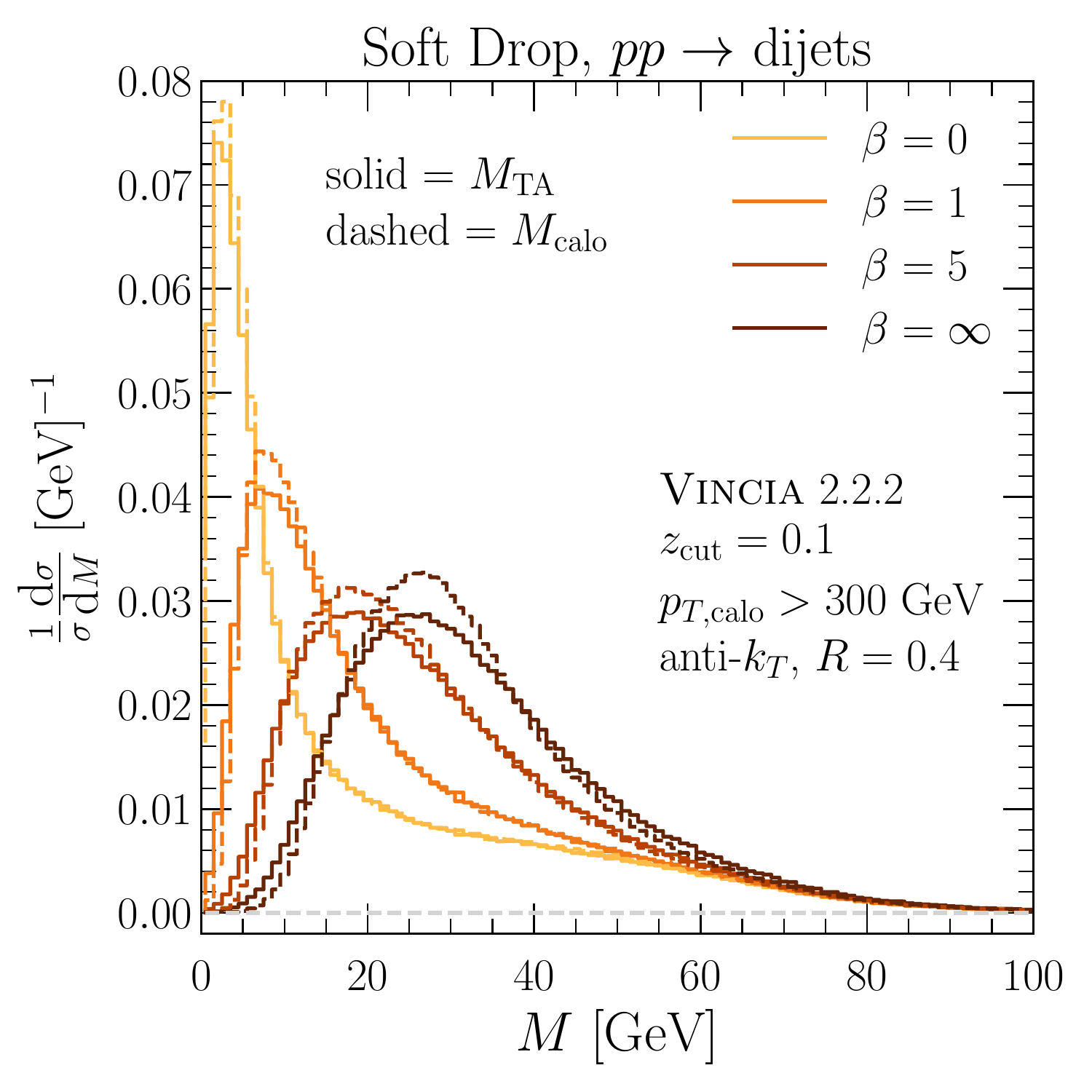}
		\label{fig:sd-distributions-beta}
	}
	\subfloat[]{
		\includegraphics[width=0.45\textwidth]{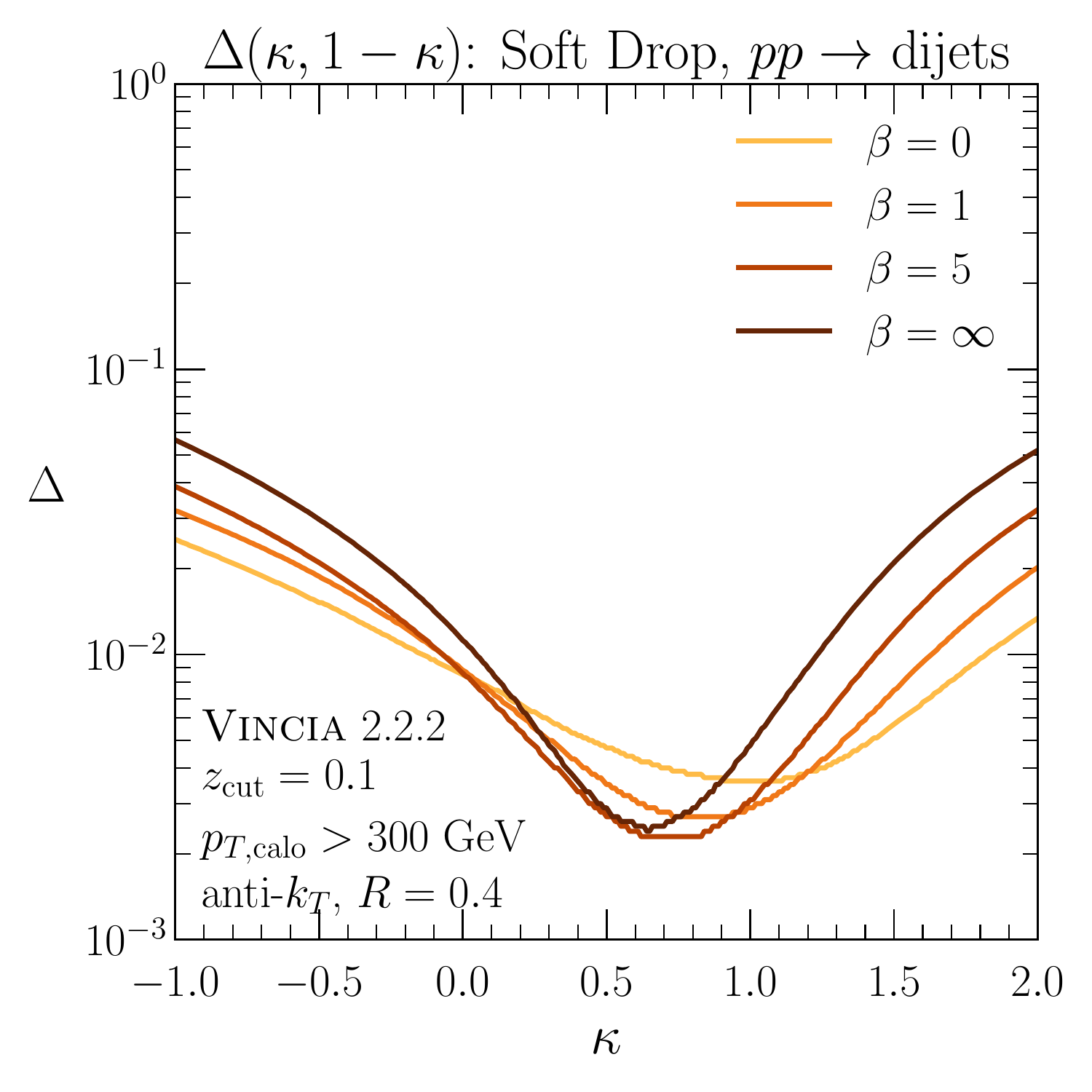}
		\label{fig:sd-distance-beta}
	}
	\caption{(a) Groomed jet mass and track-assisted mass distributions in \textsc{Vincia} for $z_{\rm cut} = 0.1$ and various values of $\beta$.  The $\beta = \infty$ curve corresponds to the ungroomed distribution. (b):  The statistic $\Delta(\kappa,1-\kappa)$ for the same values of $\beta$.}
	\label{fig:sd-beta}
\end{figure}

For our groomed parton shower study, we use the same \textsc{Vincia} event samples as \Sec{sec:MCexplore-results}, taking an ensemble of jets for which the ungroomed jet has $p_{T,\rm calo}>300$ GeV. 
\Fig{fig:sd-distributions-beta} shows distributions of groomed track-assisted mass and groomed jet mass with $z_{\rm cut}= 0.1$ for a range of $\beta$ values. 
We see that groomed track-assisted mass continues to accurately reproduce the groomed jet mass distribution for a wide range of soft-drop parameters.

When incorporating soft-drop grooming, we want to compute $\Delta(\kappa,\lambda)$ where both $M_{\rm TA}^{(\kappa,\lambda)}$ and $M_{\rm calo}$ are groomed using the same soft-drop parameters $\beta$ and $z_{\rm cut}$.  
We checked the full two-dimensional distribution of $\Delta(\kappa,\lambda)$, and found that, as in \Fig{fig:heatmap-pp}, the best-fit values obeyed $\kappa + \lambda = 1$. 
Since this conclusion is unchanged from \Sec{sec:MCexplore}, we fix $\lambda = 1-\kappa$ and sweep $\kappa$ in this section. 
\Fig{fig:sd-distance-beta} shows $\Delta(\kappa,1-\kappa)$ as a function of $\kappa$ for the same grooming parameters.
We can see that lower values of $\beta$, which correspond to more aggressive grooming, lead to best-fit parameters with a higher $\kappa$ value.

Though not shown, we checked that this trend can also be seen by fixing $\beta$ and scanning a range of $z_{\rm cut}$ values, with higher $z_{\rm cut}$ leading to more aggressive grooming and higher best-fit $\kappa$. 
As we argued in \Sec{sec:MCexplore}, the ensemble-averaged reweighting factor controlled by $\lambda$ corrects for fluctuations in the angular distribution of neutral particles. 
Since grooming removes soft, wide-angle radiation, we expect the impact of these angular fluctuations to be smaller for jets with more aggressive grooming parameters. 
As in \Sec{sec:MCexplore-results}, we checked that underlying event has a modest effect on the $\Delta(\kappa,\lambda)$ distributions, with almost no sensitivity to underlying event using the most aggressive grooming parameters.


\subsection{Analytic Calculation with Grooming}
\label{sec:sd-calculation}


We can easily adapt the calculations in \Sec{sec:calculation} to handle soft-drop grooming.
One of the major benefits of soft-drop grooming from the theoretical perspective is that it removes radiation corresponding to NGLs, which are difficult to calculate and were neglected in \Sec{sec:calculation}.

Starting from the NLL resummation, the radiator function in \Eq{eq:track-radiator} describes the probability of an emission with $\hat{\rho}(z,\theta) > \rho$ and therefore involves an integral over the allowed phase space for this emission. 
Grooming restricts this allowed phase space, so the soft-drop condition in \Eq{eq:sd-groom} must be included in the radiator, which becomes
\begin{equation}
\label{eq:sd-radiator}
R_{\mbox{\tiny TA}}(\rho,x_j) = \int_0^1 \text{d}x_k\, T_g(x_k,\mu) \int_0^1 \text{d}z\, P_i(z)\int_0^R \frac{\text{d}\theta}{\theta}\, \frac{\alpha_s(E_{\rm calo}z\theta)}{\pi} \Theta\left(\hat{\rho}_{\mbox{\tiny TA}}-\rho\right) \Theta\left( z-z_{\rm cut} \left(\tfrac{\theta}{R}\right)^\beta \right)\, .
\end{equation}
The function $\hat{\rho}$ is unchanged from \Eq{eq:obs-ee-parton-def}.\footnote{When one branch of the splitting fails the soft-drop condition, we have simply groomed away the softer particle, without accounting for flavor-changing effects present at finite $z_{\rm cut}$.  A precise calculation taking this effect into account would have a radiator with matrix structure in flavor space, as described in \Refs{Dasgupta:2013ihk,Marzani:2017mva}. 
}
The track fractions do not appear in the grooming $\Theta$-function because grooming is applied to the full jet, as described above; see \App{app:alternate-sd-implementation} for an alternate prescription where the grooming is applied to the track-only jet.

Truncating to NLL order and making the fixed-coupling approximation as in \Sec{sec:insights}, we can again see the cancellation of the track function logarithmic moments. 
For $\beta > 0$, the radiator in the fixed-coupling approximation becomes (an analogous expression holds for $\beta = 0$, see \App{app:details-resummed})
\begin{align}
\label{eq:sd-radiator-nll}
R_{\mbox{\tiny TA}}(\rho,x_j) &\overset{\mbox{\tiny F.C.}}{=} \frac{\alpha_sC_i}{\pi} \bigg\{ \tfrac{1}{2}\ln^2(\tfrac{1}{\rho})\bigg[\frac{\beta}{2+\beta} f^{g,0}\left(y^*,1\right) + f^{g,0}\left(x_j\rho,y^*\right)\bigg]\\ \nonumber
&\hspace{1cm} + \ln(\tfrac{1}{\rho})\bigg[B_i f^{g,0}\left(x_j\rho,1\right) -\ln(x_j)\bigg( \frac{\beta}{2+\beta} f^{g,0}\left(y^*,1\right)+ f^{g,0}\left(x_j\rho,y^*\right)   \bigg) \\ \nonumber
&\hspace{1cm} +\frac{\beta}{2+\beta}f^{g,1}(y^*,1) + f^{g,1}\left(x_j\rho,y^*\right)   + \frac{2}{2+\beta} f^{g,0}\left(y^*,1\right) \ln\left(\tfrac{1}{z_{\rm cut}}\right)  \bigg]\bigg\}\,,
\end{align}
where $y^* = \min\big(\frac{x_j\rho}{z_{\rm cut}},1\big)$. 
The terms proportional to $f^{g,n}(y^*,1)$ account for the regions of $x_j,x_k$ phase space where grooming is active, characterized by the boundary $\frac{x_kz_{\rm cut}}{x_j} < \rho$. 
The region where grooming is inactive, $\frac{x_kz_{\rm cut}}{x_j} > \rho$, contributes the terms proportional to $f^{g,n}(x_j\rho,y^*)$.

Now setting $x_j\rho$ in the integral endpoints to zero and making the exponential approximation in \Eq{eq:exp-approx} for the integral over $x_j$, we obtain the approximate cumulative distribution:%
\footnote{For $\beta = 0$, the $x_j\rho \to 0$ endpoint is cut off by $z_{\rm cut}$, but this expression is still valid in the approximation where one keeps logarithms of $z_{\rm cut}$ but drops powers of  $z_{\rm cut}$.}
\begin{align}
\label{eq:sd-cumulative-fixed-coupling}
\begin{split}
\Sigma_{\mbox{\tiny TA}}(\rho) &\simeq \exp\bigg\{-\tfrac{\alpha_s C_i}{\pi}\bigg[\tfrac{\beta}{2(2+\beta)}\ln^2(\tfrac{1}{\rho}) + \ln(\tfrac{1}{\rho}) \bigg(\tfrac{\beta}{2+\beta} \gamma_E + B_i +\tfrac{2}{2+\beta}\ln(\tfrac{1}{z_{\rm cut}})   \bigg) \bigg] \bigg\}\\
&\hspace{1cm} \times \Gamma\left(1+ \tfrac{\alpha_sC_i}{\pi}\tfrac{\beta}{2+\beta} \ln(\tfrac{1}{\rho})\right)^{-1} \times \exp\left\{-\tfrac{\alpha_sC_i}{\pi}\tfrac{\beta}{2+\beta} (f^{g,1}-f^{j,1})\ln(\tfrac{1}{\rho}) \right\}\,.
\end{split}
\end{align}
We see that the cancellation between logarithmic moments of the track functions also occurs in the region of phase space where the grooming is active. 
From this fixed-coupling approximation, we can again predict that in the full calculation, the track-assisted mass will be more similar to the jet mass for gluon jets than for quark jets because of the more complete cancellation. 
We emphasize that to obtain results which are formally at least NLL order, the full radiator \Eq{eq:sd-radiator} with running $\alpha_s$ and the accompanying expressions for $R'$ in \App{app:details-resummed} are required. 

The LO cross section for track-assisted mass measured on soft-drop groomed jets only requires a $\Theta$ function to implement the grooming,
\begin{equation}
\label{eq:fixed-order-cross-section-sd}
\frac{\text{d}\sigma}{\text{d}\rho} = \int \text{d}y_1\, \text{d}y_2\, \frac{\text{d}\hat{\sigma}}{\text{d}y_1\text{d}y_2} \int \text{d}x_1\, \text{d}x_3\, T_1(x_1)\,T_3(x_3)\,  \Theta(R-\theta_{13})\, \delta (\hat{\rho}_{\mbox{\tiny TA}}-\rho) \Theta \left(z -z_{\rm cut} \big(\tfrac{\theta_{13}}{R} \big)^\beta \right)\,.
\end{equation}
The matching procedure is then exactly the same as in \Sec{sec:calculation-matching-fo}. 
It is important that soft-drop grooming with $\beta\ge 0$ does not alter the total jet production rate, so the normalization of the fixed-order cross sections is the same as for the ungroomed distribution. 
The required expressions for the $\mathcal{O}(\alpha_s)$ piece of the expansion of the resummed distribution, $\tfrac{1}{\sigma}\tfrac{d\sigma_{\mbox{\tiny NLL,}\alpha}}{d\rho}$, are given in \App{app:details-resummed}.

\begin{figure}[t]
	\centering
	\subfloat[]{
		\includegraphics[width=0.45\textwidth]{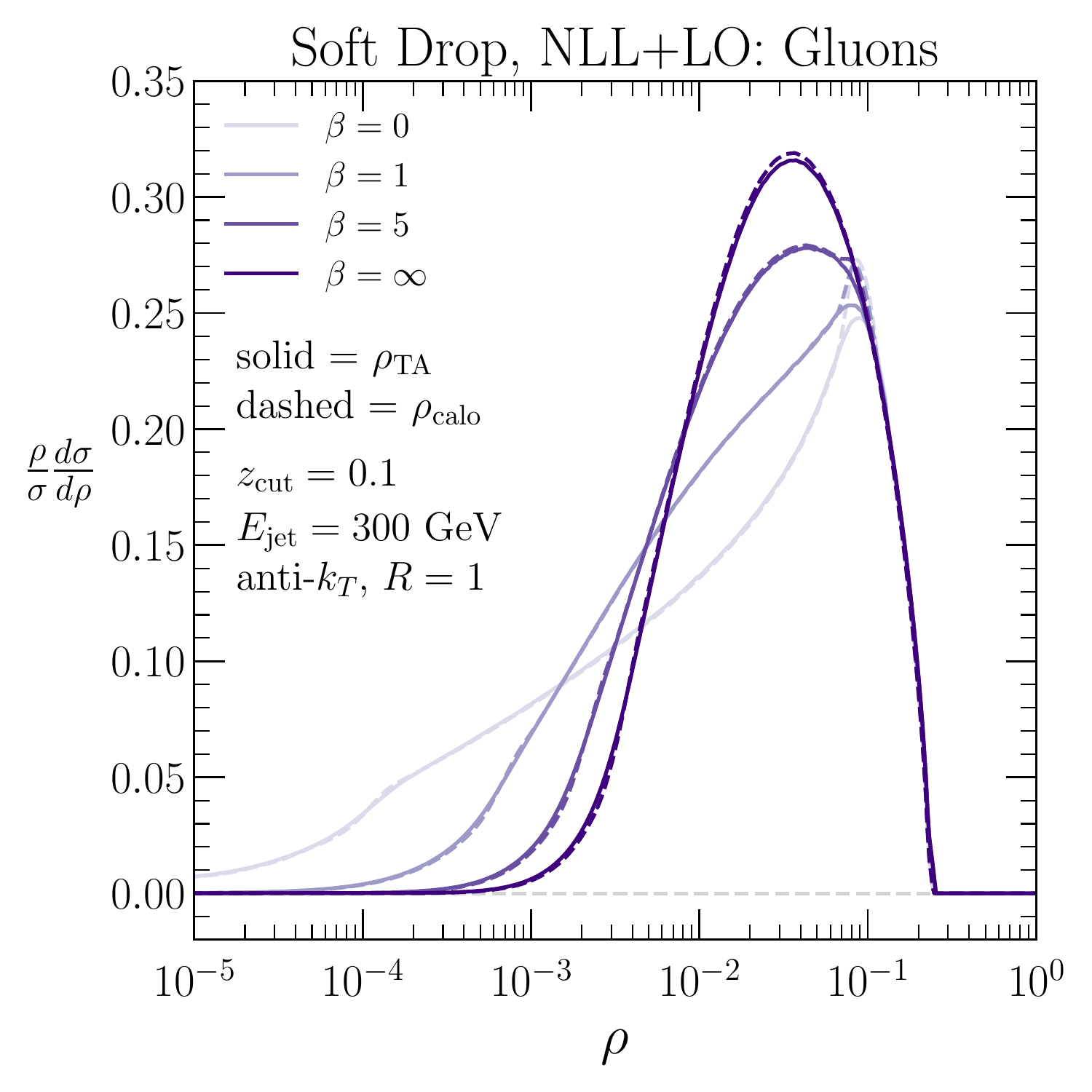}
		\label{fig:sd-matched-gluon}
	}
	\subfloat[]{
		\includegraphics[width=0.45\textwidth]{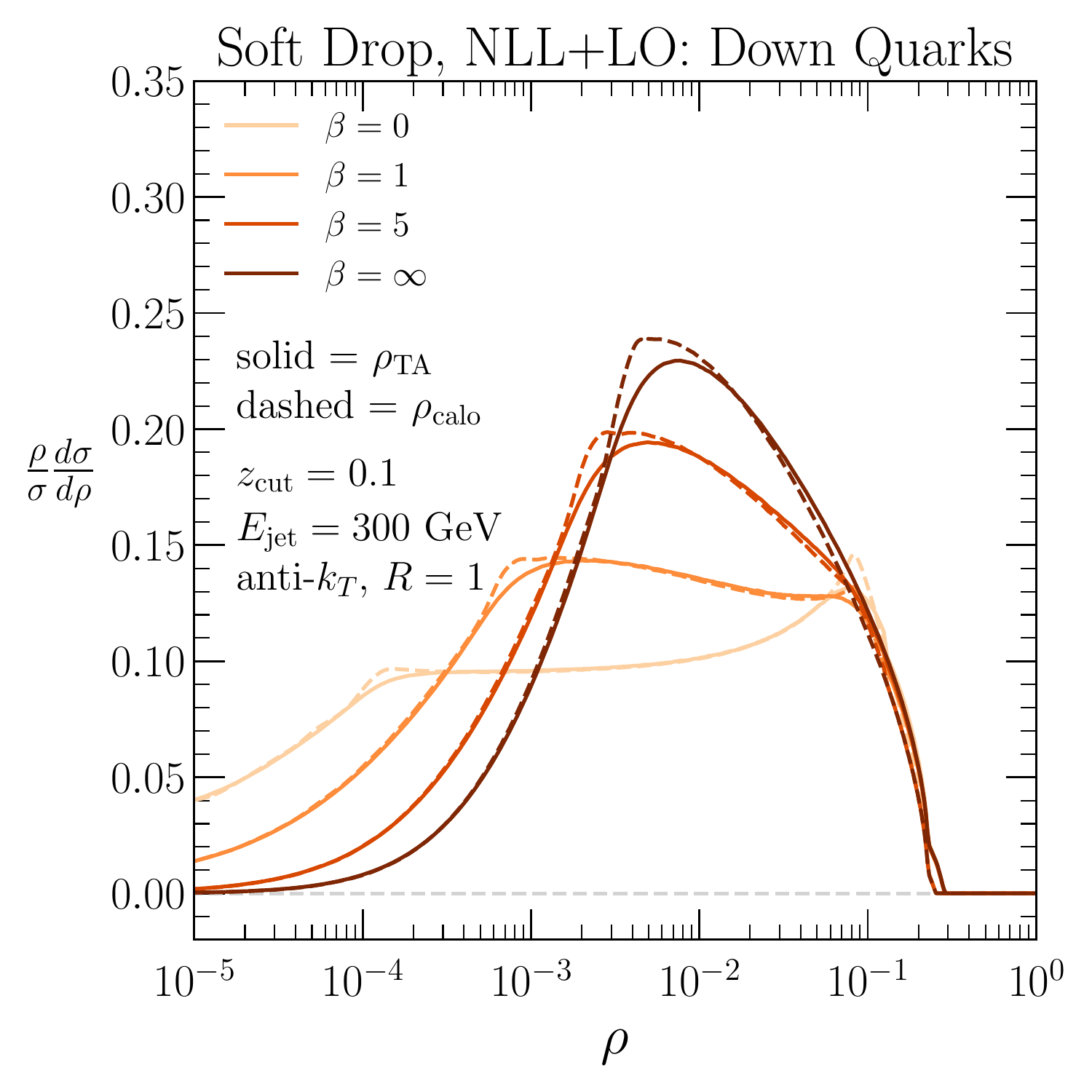}
		\label{fig:sd-matched-quark}
	}
	\caption{ Matched distributions for rescaled jet mass $\rho_{\rm calo}$ (solid curves) and rescaled track-assisted mass $\rho_{\mbox{\tiny TA}}$ (dashed curves) after soft-drop grooming for (a) gluon jets and (b) down-quark jets.  Shown are soft-dropped grooming parameters $z_{\rm cut}=0.1$ and $\beta = \{0,1,5,\infty\}$.}
	\label{fig:sd-matched}
\end{figure}

Plots of the NLL+LO $\rho_{\rm calo}$ and $\rho_{\mbox{\tiny TA}}$ distributions are shown in \Fig{fig:sd-matched} for gluon jets and down-quark jets. 
These distributions have grooming parameters $z_{\rm cut}=0.1$ and $\beta = \{0,1,5,\infty\}$, where the $\beta=\infty$ case is the same as the ungroomed observable.
As expected from the $\textsc{Vincia}$ study, we find that the close relationship between track-assisted mass and jet mass holds over a broad range of parameters. 
This gives a theoretical prediction of the groomed track-assisted mass which can be compared directly to experimental data, without necessitating an unfolding of the track-assisted mass measurement to ordinary jet mass.

Non-perturbative emissions which contribute to the distribution of $\rho \sim z\theta^2$ must be either soft or collinear. 
Since Soft Drop changes the relative proportions of soft and collinear radiation, the appropriate non-perturbative parameter $\Omega$ as described \Sec{sec:calculation-np} would depend on the grooming parameters $z_{\rm cut}$ and $\beta$ \cite{Manohar:1994kq,Waalewijn:2012sv,Larkoski:2013paa,Gras:2017jty}. 
We leave a study of these non-perturbative corrections to future work. 

\begin{figure}[t]
	\centering
	\subfloat[]{
		\includegraphics[width=0.45\textwidth]{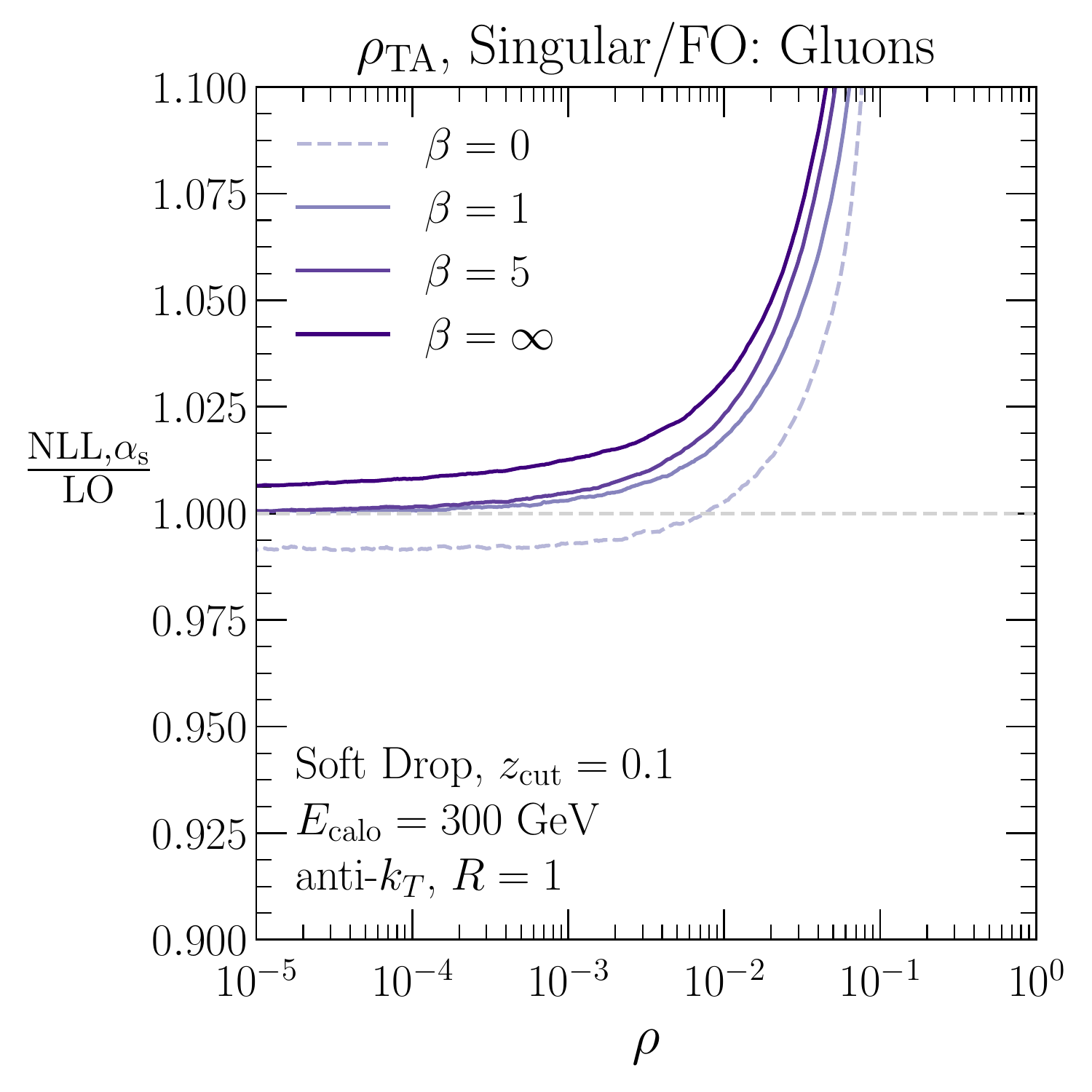}
		\label{fig:ratios-softdrop-gluon}
	}
	\subfloat[]{
		\includegraphics[width=0.45\textwidth]{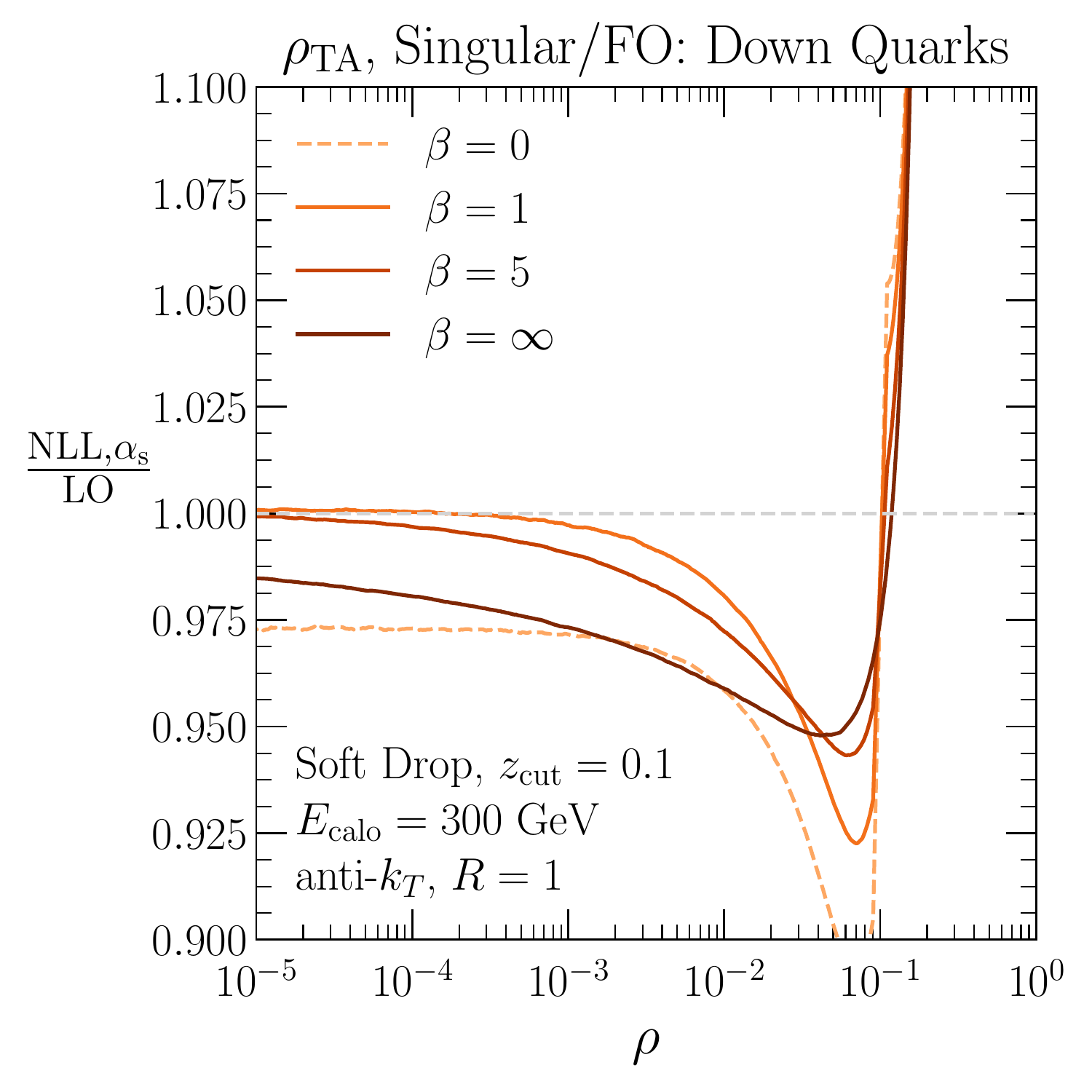}
		\label{fig:ratios-softdrop-quark}
	}
	\caption{Ratios of the soft-drop groomed $\mathcal{O}(\alpha_s)$ piece of the expanded NLL distributions over the LO $\rho_{\mbox{\tiny TA}}$ distributions. The $\beta=0$ case is dashed since this calculation is formally only LL order.}
	\label{fig:sd-ratios} 
\end{figure}

In \Fig{fig:sd-ratios}, we again plot the ratio of the $\mathcal{O}(\alpha_s)$ piece of the resummed distribution over the fixed-order distribution, this time for groomed jets with $R=1$, $z_{\rm cut} = 0.1$, and $\beta = \{0,1,5,\infty\}$. 
For $\beta = \infty$, we recover the $R=1$ result from \Fig{fig:calculation-nogroom}.
With $\beta = 1$ or 5, the two distributions have a much closer match, since the grooming is removing wide-angle radiation that was contributing to the $R \log R$ power correction.
The $\beta = 0$ distribution again exhibits a mismatch as $\rho \rightarrow 0$, due to constant terms in the radiator which are beyond the order of this calculation.
These terms have an effect of the same magnitude as the single logarithmic terms in the $\beta = \infty$ distribution because there are no double-logarithmic terms in the resummed distribution of $\beta = 0$ groomed jet mass \cite{Dasgupta:2013ihk}, so the $\beta = 0$ calculation is formally LL instead of NLL.


\section{Conclusions}
\label{sec:conclusion}

The different granularity of calorimetry and tracking at the LHC experiments makes it worthwhile to explore proxies for key jet observables like jet mass that can better exploit the fine angular resolution available for charged particle tracks. 
The original track-assisted mass, as defined by the ATLAS collaboration in \Eq{eq:tam-definition}, is one example of such a proxy, which trades fluctuations in the charged-to-neutral mass fraction for improved track mass resolution.
In this paper, we showed that the generalized version of track-assisted mass in \Eq{eq:gtam-intro} can act as an even better proxy for jet mass, by balancing the fluctuations in charged-to-neutral mass fraction against ensemble-averaged information.
In the spirit of \Ref{Coleman:2017fiq}, this same GTAM philosophy could be applied to situations where both tracking and electromagnetic calorimeter information is used to determine jet mass, but hadronic calorimeter information is only used to determine jet $p_T$.
One could even imagine more exotic proxies for jet mass, such as ones that weight higher energy particles more than lower energy ones.

As a step towards a comparison with experimental results, we performed a NLL resummed calculation of the distribution of \Eq{eq:obs-ee-def}, the rescaled (squared) track-assisted mass, and its generalizations, in $e^+e^-$ collisions. 
Since track-assisted mass is not an IRC-safe observable, we used the track function formalism to regulate its collinear divergences and regain calculational control. 
This computation offers some insight into the close correspondence between track-assisted mass and jet mass, as well as the values of the best-fit parameters for GTAM, in terms of the similarity of logarithmic moments of the track functions. 
We matched our track-assisted mass distributions to the fixed-order processes $e^+e^- \rightarrow q\bar{q}g$ and $e^+e^-\rightarrow H\rightarrow ggg(q\bar{q}g)$ using the log-$R$ matching scheme, and implemented non-perturbative corrections using a shape function.
We found reasonable agreement between our NLL+LO+NP calculation and distributions obtained from \textsc{Vincia}.
We also explored the impact of soft-drop grooming on track-based observables, where we found that groomed GTAM was effective as a proxy for groomed jet-mass.
Though we did not compute contributions from NGLs in this work, these contributions are happily removed by the soft-drop grooming procedure. 
It would be interesting to extend our calculations to alternate grooming procedures like recursive soft drop~\cite{Dreyer:2018tjj}.

Future work on this topic would quantify the expected gain in sensitivity to highly collimated decay products at the LHC for generalized track-assisted mass measurements as compared to track-assisted mass and ordinary jet mass using detector simulations. 
Furthermore, in order to make a quantitative comparison to experimental data, a more precise theoretical calculation is required. 
Such a calculation would include contributions from NGLs in the ungroomed case, and flavor-changing finite $z_{\rm cut}$ effects for the soft-drop groomed case. 
To be competitive with other state-of-the-art jet substructure predictions, the calculation also needs to be pushed to NNLL, including contributions from the RG evolution of the track functions. 
The fixed-order corrections have been included only to LO, which is not sufficient for a precision measurement. 
Lastly, it would be interesting to adapt the machinery developed in \Ref{Larkoski:2014pca} to study the double-differential distribution of $M_{\rm calo}$ and $M_{\rm TA}^{(\kappa,\lambda)}$ in order to assess their jet-by-jet correspondence. 
The use of generalized track-assisted mass as a benchmark jet observable at hadron colliders, with comparisons to precision theoretical calculations, offers the possibility to improve the sensitivity and flexibility of experimental measurements. 
This will be of vital importance in the high-energy limit at the LHC and at future colliders.

\section*{Acknowledgements}

We thank David Miller and Benjamin Nachman for discussions about track-assisted mass at ATLAS and for comments on our manuscript.
We also thank Massimiliano Procura, Wouter Waalewijn, and our anonymous referee for additional helpful feedback.
This work was supported by the Office of High Energy Physics of the U.S. Department of Energy (DOE) under grant DE-SC-0012567.

\appendix


\section{Pure Quark and Gluon Ensembles}
\label{app:pure-quark-gluon}


\begin{figure}[p]
	\centering
	\subfloat[]{
		\includegraphics[width=0.45\textwidth]{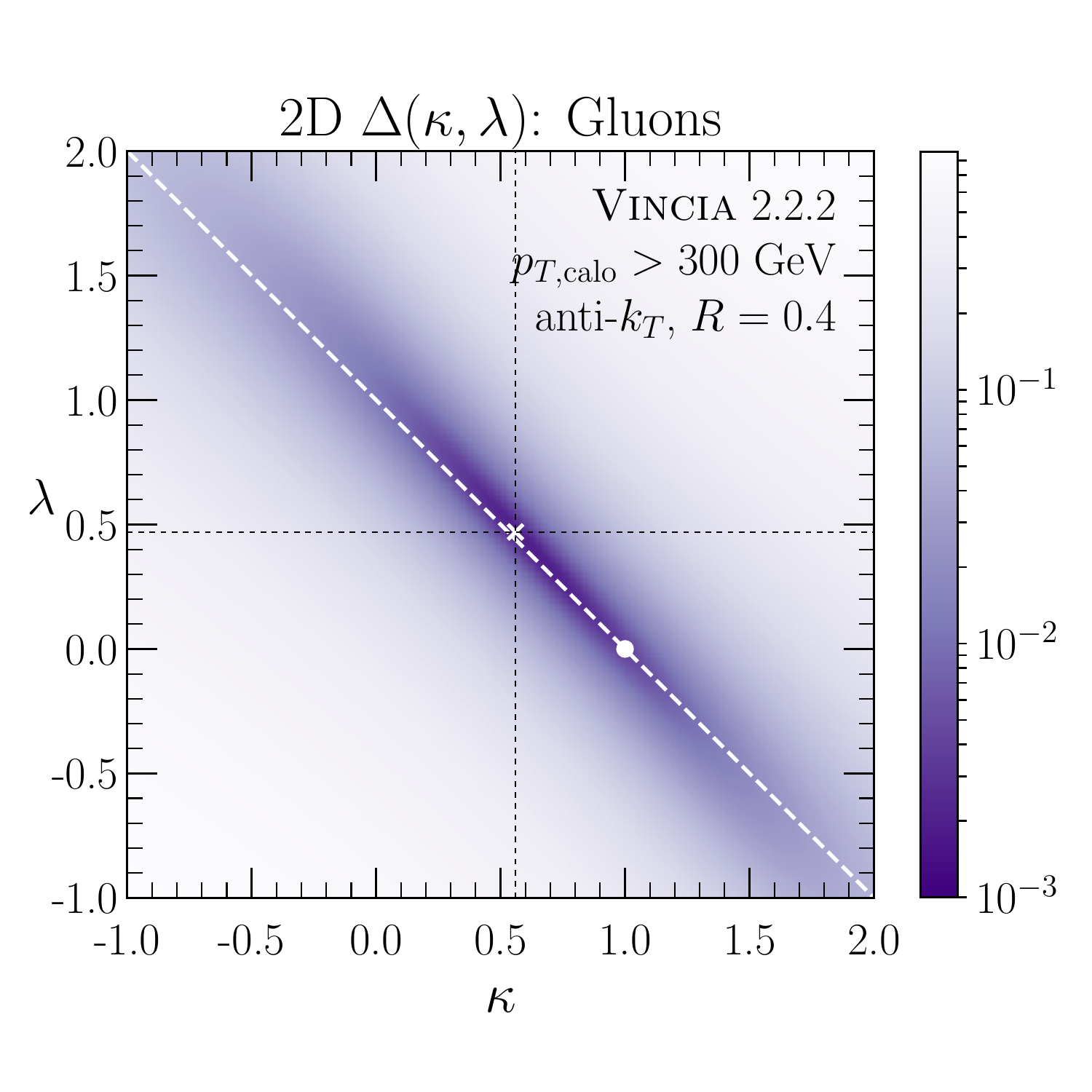}
		\label{fig:pure-qg-distance-a}
	}
	\subfloat[]{
		\includegraphics[width=0.45\textwidth]{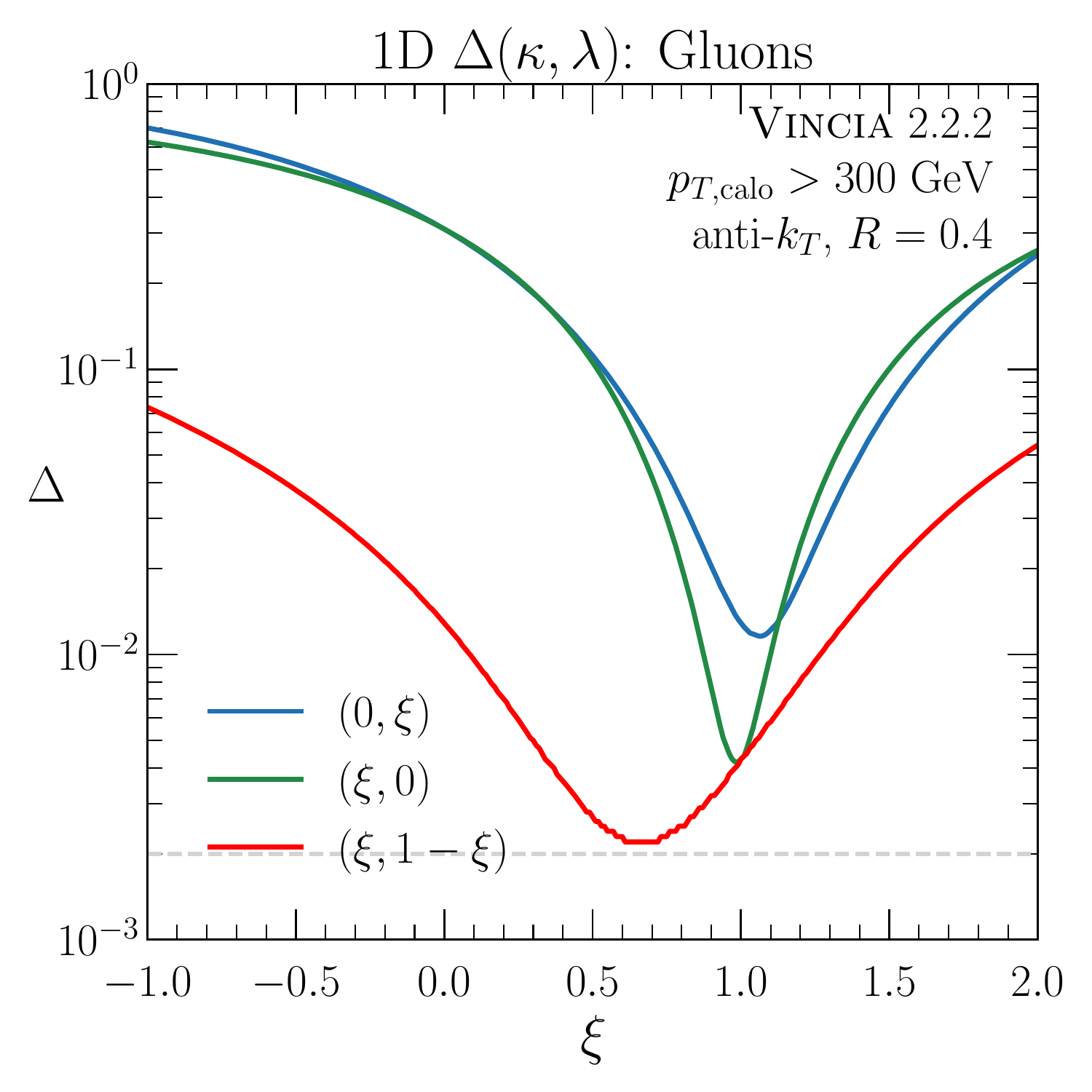}
		\label{fig:pure-qg-distance-b}
	}
	
	\subfloat[]{
		\includegraphics[width=0.45\textwidth]{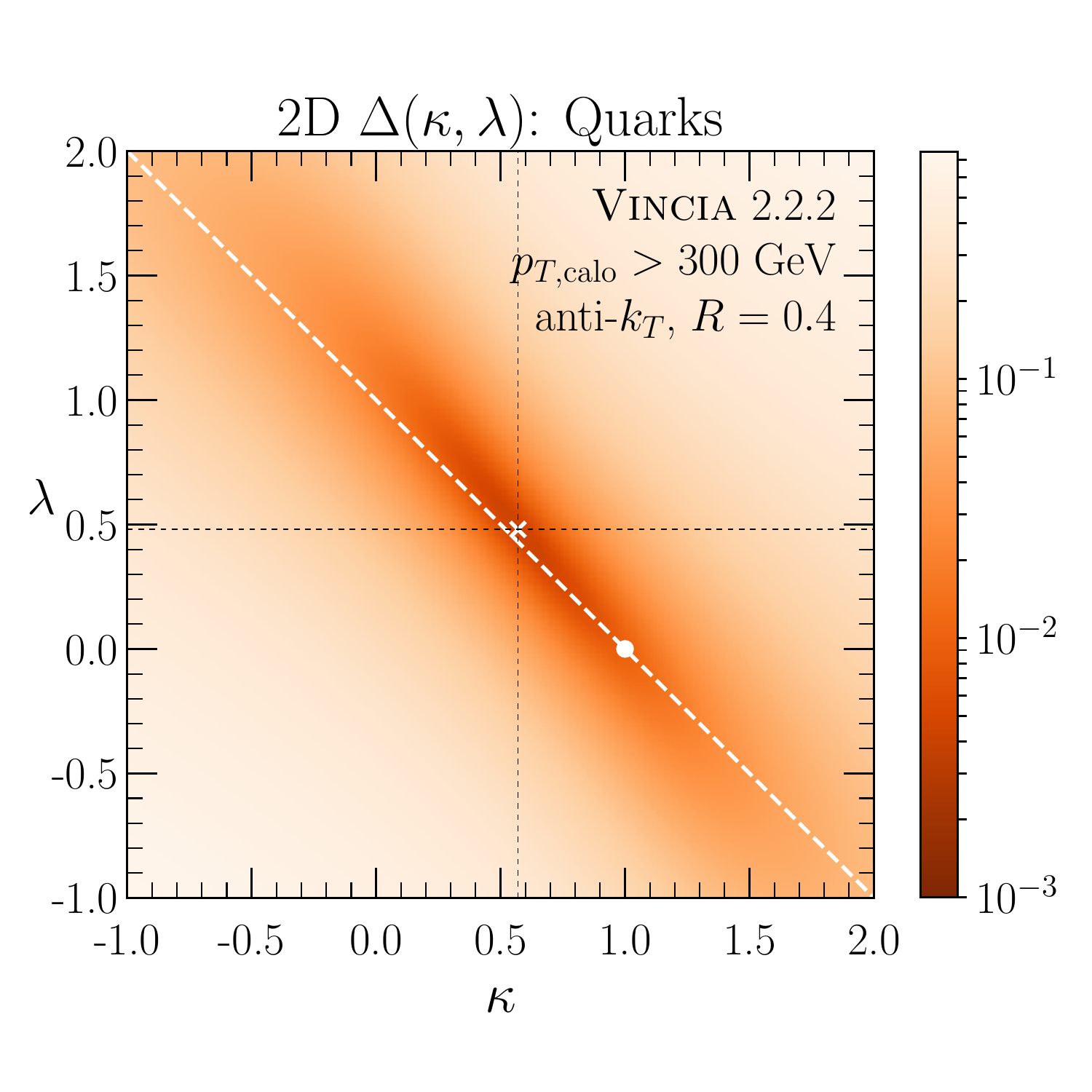}
		\label{fig:pure-qg-distance-c}
	}
	\subfloat[]{
		\includegraphics[width=0.45\textwidth]{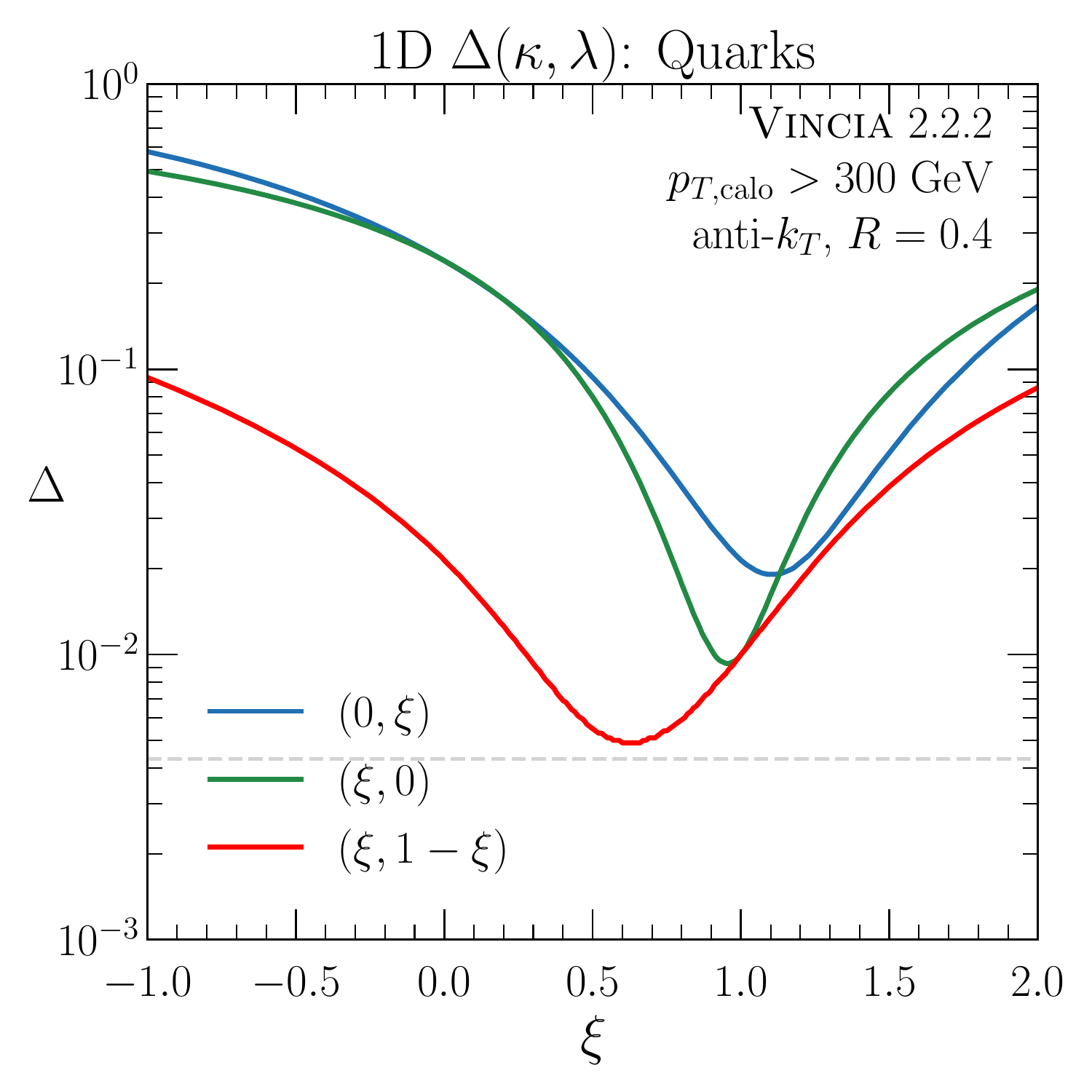}
		\label{fig:pure-qg-distance-d}	
	}
	\caption{\label{fig:pure-qg-distance}Distribution of $\Delta(\kappa,\lambda)$ for the processes $pp\rightarrow gg$ (top) and $pp\rightarrow q\bar{q}$ (bottom), to be compared to \Fig{fig:heatmap}. The full two-dimensional distributions are on the left, and slices of these distributions are shown on the right.}
\end{figure}

The analysis of GTAM in \Sec{sec:MCexplore} was carried out using $pp\rightarrow $ dijet events generated with $\textsc{Vincia}$ 2.2.2. 
It is conceivable that the flavor content of the jet could change the best-fit parameters and the degree of the correspondence between jet mass and GTAM. 
To investigate this possibility, we repeated the analysis using ensembles of purely quark-initiated and purely gluon-initiated jets, as labeled by the $\textsc{Vincia}$ hard process.

The best-fit GTAM parameters $(\kappa_{\rm best},\lambda_{\rm best})$ turn out to be rather insensitive to the species of parton initiating the jet. 
As shown in \Fig{fig:pure-qg-distance}, the best-fit values $(\kappa_{\rm best},\lambda_{\rm best})$ were found to be (0.56, 0.47) for gluons and (0.57,0.48) for quarks, as compared to $(0.59,0.44)$ for a mixture of quark and gluon jets. 
GTAM is a closer match to jet mass by about a factor of two (as measured by $\Delta$) for gluon jets than for quark jets. 
This is to be expected, since the variance of the gluon track function is smaller than that of the quark track function ($\Delta T_g = 0.15$ and $\Delta T_q = 0.2$), so the track fraction reweighting factors with exponent $\lambda$ will smear the GTAM distribution less for gluons than for quarks. 
This also matches the conclusion from the analytic approximations in \Sec{sec:calculation-resummed} where gluon jets produced a more complete cancellation of track function effects on the resummed GTAM distribution.


\section{Details of Resummed Calculation}
\label{app:details-resummed}


In order to obtain a finite differential distribution for the observable $\rho \equiv M^2/(E_{\rm calo}R)^2$ in the region $\rho \ll 1$, where the fixed-order perturbative expansion breaks down due to large logarithms of $\rho$, it is necessary to calculate contributions proportional to $\alpha_s^n \log^{2n-1}(\rho)$ (LL) to all orders in perturbation theory. 
Quantitative agreement with experimental data requires resummation of terms up to at least NLL order. 

Running coupling effects are taken into account with the two-loop $\beta$-function. 
The appropriate scheme for the coupling in the resummed cumulative distribution \Eq{eq:track-cumulative} is the CMW scheme \cite{Catani:1990rr}, which is related to the $\overline{\rm MS}$ scheme by
\begin{equation}
\alpha_s^{\mbox{\tiny CMW}} = \alpha_s^{\overline{\mbox{\tiny MS}}}\left(1+\left[C_A\left(\tfrac{67}{18}-\tfrac{\pi^2}{6}\right) - \tfrac{5}{9}n_f\right]\frac{\alpha_s^{\overline{\mbox{\tiny MS}}}}{2\pi}\right)\,.
\end{equation}
The scale at which $\alpha_s$ is evaluated in \Eqs{eq:radiator-calorimeter}{eq:track-radiator} is $\mu = E_{\rm calo}z\theta$ ($p_{T,\rm jet}z\theta$ for $pp$ collisions), the transverse momentum of the radiated particle in the soft-collinear limit. 
This scale enters the non-perturbative regime for low enough values of $\rho$. 
To handle this effect, we freeze the coupling at $\alpha_s(1 \mbox{ GeV}) = 0.42$, which is the result of using the two-loop $\beta$-function to run from $\alpha_s(M_Z) = 0.1182$. 
In the derivative of the radiator, all terms are one logarithmic order lower than in the radiator itself, so to this order we can evaluate the coupling in $R'$ at the hard scale $E_{\rm calo}R$. 

As described in \Sec{sec:calculation-resummed}, the running of the track function contributes only at NNLL order. 
Additionally, the first moment of the track functions is extremely scale insensitive, see \Fig{fig:track-functions-b}, with a fractional change of only 0.04$\%$ for gluon or quark-singlet (average over quark and anti-quark species) jets when evolved from 10 GeV to $10^6$ GeV.
The fractional change of the second moment is about $2\%$ over this same scale evolution.  
The track functions are fixed at the hard scale $\mu = E_{\rm calo}R$ in the calculation of $R$ and $R'$.
A higher-order calculation could include this effect using the nonlinear DGLAP-like evolution equation described in \cite{Chang:2013rca,Elder:2017bkd}. 

The radiators for quark and gluon jets also include the (real) reduced splitting functions $P_{q}$ and $P_{g}$ respectively:
\begin{align}
P_{q}(z) = P_{gq}(z) &= C_F\left[\frac{1+(1-z)^2}{z}\right]\,, \\
P_{g}(z) = \frac{1}{2}p_{gg}(z) + n_fp_{qg}(z) &= C_A \left[\frac{2(1-z)}{z} + z(1-z) \right] + n_fT_F(z^2 + (1-z)^2)\,.
\end{align}
The splitting functions do not have a virtual part, and no plus function regularization is required, since the observable value $\rho$ cuts off the singularities in both the $z$ and $\theta$ integrals. 
The $z\leftrightarrow 1-z$ symmetry in the $g\rightarrow gg$ splitting was exploited to write $p_{gg}$ in a form which is singular only at $z=0$.

We give the full expressions needed for the cumulative track-assisted mass distributions, including soft-drop grooming with $\beta \ge 0$.
Although they are already available in the literature \cite{Larkoski:2014wba}, we give the equivalent expressions for ordinary jet mass and soft-drop groomed mass as well for comparison. 
In the fixed-coupling approximation (but without the endpoint approximations in \Eq{eq:approx-fixed-coupling-NLL-radiator}), the calorimeter and track radiators are given to NLL order by\footnote{The cases $\beta > 0$ and $\beta = 0$ are distinct at NLL order because we are keeping finite $z_{\rm cut}$ terms. For this reason, the $\beta\rightarrow 0$ limit of \Eq{eq:radiator-exact-betagt0} does not recover \Eq{eq:radiator-exact-betaeq0}. The ungroomed $(\beta=\infty)$ case can be recovered from the $\beta > 0$ expression by taking either the $\beta\rightarrow \infty$ limit or the $z_{\rm cut} \rightarrow 0$ limit. For $x_j\rho>0$, this gives $f^{i,n}(y^*,1) \rightarrow 0$ and $f^{i,n}(x,y^*) \rightarrow f^{i,n}(x,1)$ for $z_{\rm cut} = x_j\rho$. It is easy to see that for any $\rho>0$, the ungroomed limits $z_{\rm cut}\rightarrow 0$ or $\beta\rightarrow \infty$ commute with removing the tracking procedure by setting all track fractions to one and all track functions to $\delta(1-x)$. }
\begin{align}
\label{eq:radiator-exact-betainfty}
(\beta = \infty)&:\\ \nonumber
R_{\rm calo}(\rho) &= \frac{\alpha_sC_i}{\pi} \bigg\{\tfrac{1}{2}\ln^2(\tfrac{1}{\rho}) + B_i\ln(\tfrac{1}{\rho})\bigg\}\,,\\ \nonumber
R_{\mbox{\tiny TA}}(\rho,x_j) &= \frac{\alpha_sC_i}{\pi} \bigg\{\tfrac{1}{2} \ln^2(\tfrac{1}{\rho})f^{g,0}(x_j\rho,1) + \ln(\tfrac{1}{\rho})\bigg[f^{g,0}\left(x_j\rho,1\right)\left(B_i - \ln(x_j)\right) +f^{g,1}\left(x_j\rho,1\right)\bigg]\bigg\}\,,\displaybreak[0]
\\
\label{eq:radiator-exact-betagt0}
(\beta > 0) &:\\ \nonumber
R_{\rm calo}(\rho,x_j) &= \frac{\alpha_sC_i}{\pi} \bigg\{\tfrac{1}{2}\ln^2(\tfrac{1}{\rho}) \bigg[\frac{\beta}{2+\beta} \Theta(z_{\rm cut}-\rho ) + \Theta(\rho - z_{\rm cut})\bigg]\\ \nonumber
&\hspace{1cm} + B_i\ln(\tfrac{1}{\rho}) + \frac{2}{2+\beta} \Theta(z_{\rm cut}-\rho)\ln\left(\tfrac{1}{\rho}\right)\ln\left(\tfrac{1}{z_{\rm cut}}\right)  \bigg]\bigg\}\,,\\ \nonumber
R_{\mbox{\tiny TA}}(\rho,x_j) &= \frac{\alpha_sC_i}{\pi} \bigg\{\tfrac{1}{2} \ln^2(\tfrac{1}{\rho}) \bigg[\frac{\beta}{2+\beta} f^{g,0}\left(y^*,1\right) + f^{g,0}\left(x_j\rho,y^*\right)\bigg]\\ \nonumber
&\hspace{1cm} + \ln(\tfrac{1}{\rho})\bigg[B_i f^{g,0}\left(x_j\rho,1\right) -\ln(x_j)\bigg( \frac{\beta}{2+\beta} f^{g,0}\left(y^*,1\right)+ f^{g,0}\left(x_j\rho,y^*\right)   \bigg) \\ \nonumber
&\hspace{1cm} +\frac{\beta}{2+\beta}f^{g,1}(y^*,1) + f^{g,1}\left(x_j\rho,y^*\right)   + \frac{2}{2+\beta} f^{g,0}\left(y^*,1\right) \ln\left(\tfrac{1}{z_{\rm cut}}\right)  \bigg]\bigg\}\,,\displaybreak[0]\\
\label{eq:radiator-exact-betaeq0}
(\beta = 0) &:\\ \nonumber
R_{\rm calo}(\rho,x_j) &= \frac{\alpha_sC_i}{\pi}\bigg\{\tfrac{1}{2}\Theta(\rho - z_{\rm cut})\ln^2(\tfrac{1}{\rho}) + B_i\ln(\tfrac{1}{\rho})\\ \nonumber
&\hspace{2.5cm} +\Theta(z_{\rm cut}-\rho)\ln(\tfrac{1}{\rho})\left[\tfrac{z_{\rm cut}}{4}(4-z_{\rm cut}) + \ln\left(\tfrac{1}{z_{\rm cut}}\right)\right] \bigg\}\,,\\ \nonumber
R_{\mbox{\tiny TA}}(\rho,x_j) &= \frac{\alpha_sC_i}{\pi}\bigg\{\tfrac{1}{2} \ln^2(\tfrac{1}{\rho}) f^{g,0}\left(x_j\rho,y^*\right) + \ln(\tfrac{1}{\rho}) \left[f^{g,0}\left(x_j\rho,y^*\right) (B_i - \ln(x_j)) + f^{g,1}\left(x_j\rho,y^*\right)\right] \\ \nonumber
&\hspace{2.5cm} +f^{g,0}\left(y^*,1\right)\ln(\tfrac{1}{\rho})\left[ B_i + \tfrac{z_{\rm cut}}{4}(4-z_{\rm cut}) + \ln\left(\tfrac{1}{z_{\rm cut}}\right)\right] \bigg\}\,,
\end{align}
where we have defined
\begin{equation}
y^* = \min \left(\frac{x_j\rho}{z_{\rm cut}},1\right)\,.
\end{equation}
Note that most of the $f^{i,n}(a,b)$ expressions above can be set to $f^{i,n}$ or 0 to NLL accuracy, but we leave the full expressions to help clarify the integration regions.

The radiator derivatives appearing in the multiple-emissions prefactor are easily read off from the $\ln^2\tfrac{1}{\rho}$ terms, and can be expressed for $\beta = \infty$ (ungroomed), $\beta > 0$, and $\beta = 0$ by
\begin{align}
\label{eq:rderiv-exact-sd}
\nonumber
R_{\rm calo}'(\rho) &= \frac{\alpha_sC_i}{\pi} \bigg[\ln(\tfrac{1}{\rho})\bigg(\frac{\beta}{2+\beta} \Theta(z_{\rm cut}-\rho) + \Theta(\rho-z_{\rm cut})\bigg) + \frac{2}{2+\beta}\Theta(z_{\rm cut}-\rho)\ln\left(\tfrac{1}{z_{\rm cut}}\right)\bigg]\,,\\
R_{\mbox{\tiny TA}}'(\rho,x_j) &= \frac{\alpha_sC_i}{\pi} \bigg[\ln(\tfrac{1}{\rho})\bigg(\frac{\beta}{2+\beta} f^{g,0}\left(y^*,1\right) + f^{g,0}\left(x_j\rho,y^*\right)\bigg) + \frac{2}{2+\beta}f^{g,0}(y^*,1)\ln\left(\tfrac{1}{z_{\rm cut}}\right)\bigg]\,.
\end{align}
The $\ln z_{\rm cut}$ term has been included in the multiple-emissions prefactor even though it is formally beyond NLL order in $\rho$. 
Near $\rho \simeq z_{\rm cut}$, the $\ln z_{\rm cut} $ term is just as important as the dominant $\log \rho$ terms which are being resummed and cannot be neglected. 
Using the radiators, \Eqs{eq:radiator-exact-betagt0}{eq:radiator-exact-betaeq0}, and their derivatives, \Eq{eq:rderiv-exact-sd}, in the general expression for the cumulative distribution of a track-assisted observable, \Eq{eq:track-cumulative}, gives the fixed-coupling expressions for the cumulative distributions of track-assisted mass for $\beta\ge 0$ values. 
To include the effects of the running coupling in our numerical results, we integrate the radiator \Eq{eq:sd-radiator} instead of using \Eqs{eq:radiator-exact-betagt0}{eq:radiator-exact-betaeq0}, but we used \Eq{eq:rderiv-exact-sd} for the multiple-emissions prefactor. 
As mentioned in \Sec{sec:calculation-gtam}, computing the GTAM distribution for parameters besides $\kappa = 1$ and $\lambda = 0$ simply requires making the replacement $x_j \rightarrow x_j^{2\kappa-1}\langle x_j\rangle^{2\lambda}$ in the appropriate radiator and its derivative.

In the matching calculation in \Sec{sec:calculation-matching-fo}, we needed the $\mathcal{O}(\alpha_s)$ piece of the fixed-order expansion of the differential NLL distribution. 
For completeness, we include these here:\footnote{The $\rho$-dependent endpoints do not contribute to this derivative. They are power suppressed, and therefore of the same order as other power-suppressed terms that were neglected in restricting to NLL order in the radiator. If terms of this order are to be included, they must all be computed for a consistent result.}
\begin{align}
\label{eq:dsigma-nll-alpha-beta-infty}
(\beta = \infty):&\\ \nonumber
\frac{1}{\sigma}\frac{\text{d}\sigma_{\mbox{\tiny NLL},\alpha}}{\text{d}\rho} &= \frac{\alpha_sC_i}{\pi} \int_0^1 \text{d}x_j\, T_j(x_j)\bigg\{\frac{1}{\rho} \bigg[ f^{g,0}(x_j\rho,1)\left(\ln\left(\tfrac{1}{\rho}\right) + B_i - \ln(x_j)\right) + f^{g,1}(x_j\rho,1)\bigg]\,,\displaybreak[0]\\ 
\label{eq:dsigma-nll-alpha-betagt0}
(\beta > 0):&\\ \nonumber
\frac{1}{\sigma}\frac{\text{d}\sigma_{\mbox{\tiny NLL},\alpha}}{\text{d}\rho} &=\frac{\alpha_sC_i}{\pi}  \int_0^1 \text{d}x_j\, T_j(x_j) \bigg\{ \frac{1}{\rho} \bigg[ f^{g,0}(x_j\rho,y^*)\left(\ln\left(\tfrac{1}{\rho}\right) + B_i - \ln(x_j)\right)\\ \nonumber
&\hspace{1cm} + f^{g,0}(y^*,1)\left(\tfrac{\beta}{2+\beta}\left(\ln\left(\tfrac{1}{\rho}\right) - \ln(x_j)\right) + \tfrac{2}{2+\beta}\ln\left(\tfrac{1}{z_{\rm cut}}\right) + B_i \right)\\ \nonumber
&\hspace{1cm} + f^{g,1}(x_j\rho,y^*) + \tfrac{\beta}{2+\beta}f^{g,1}(y^*,1) \bigg]\,, \displaybreak[0]\\
\label{eq:dsigma-nll-alpha-betaeq0}
(\beta=0):&\\ \nonumber
\frac{1}{\sigma}\frac{\text{d}\sigma_{\mbox{\tiny NLL},\alpha}}{\text{d}\rho} &=\frac{\alpha_sC_i}{\pi} \int_0^1 \text{d}x_j\, T_j(x_j)\bigg\{ \frac{1}{\rho}  \bigg[ f^{g,0}(x_j\rho,y^*)\left(\ln\left(\tfrac{1}{\rho}\right) + B_i - \ln(x_j)\right) + f^{g,1}(x_j\rho,y^*) \\ \nonumber
&\hspace{1cm}  + f^{g,0}(y^*,1)\left(B_i + \ln\left(\tfrac{1}{z_{\rm cut}}\right) - \tfrac{1}{4}z_{\rm cut}^2 + z_{\rm cut} \right)\bigg]\,.
\end{align}
Note that the coupling must be evaluated in the $\overline{\rm MS}$ scheme, not the CMW scheme, for the $\mathcal{O}(\alpha_s)$ piece of the NLL distribution, otherwise it cannot cancel the singular terms in the fixed-order distribution.


\section{Details of Fixed-Order Matching}
\label{app:details}


\subsection{Matrix Elements}
\label{app:details-me}

When computing a collinear-unsafe observable such as track-assisted mass, the cancellation of IR singularities guaranteed by the KLN theorem for sufficiently inclusive observables does not take place. 
The key to the track function (and more broadly the GFF) formalism is that the collinear singularities in the parton-level matrix element are absorbed into the track functions. 
Canceling the collinear singularities in the $e^+e^-\rightarrow q\bar{q}g$ and $e^+e^-\rightarrow H \rightarrow ggg(q\bar{q}g)$ matrix elements, which appear at $\mathcal{O}(\alpha_s)$, requires computing the track functions at parton level also to $\mathcal{O}(\alpha_s)$. 
After this cancellation, the partonic cross section $\frac{\text{d}\hat{\sigma}}{\text{d}y_1\text{d}y_2}$ in \Eq{eq:fixed-order-cross-section} is replaced by a non-singular matching coefficient, which depends on the renormalization scheme used to calculate the track functions. 
This procedure was demonstrated explicitly for quark track functions at $\mathcal{O}(\alpha_s)$ in \Ref{Chang:2013rca}. 
For the process $e^+e^-\rightarrow q\bar{q}g$, the matching coefficient is just the part of the parton-level cross section not proportional to $\delta(1-y_1)$ or $\delta(1-y_2)$, the points in phase space where the collinear singularities appear,
\begin{equation}
\frac{\text{d}^2\hat{\sigma}}{\text{d}y_1\text{d}y_2} = \sigma_0 \frac{\alpha_s(\mu) C_F}{\pi} \frac{\Theta(y_1+y_2-1)(y_1^2+y_2^2)}{2(1-y_1)(1-y_2)}\,,
\end{equation}
Here $\sigma_0$ is the Born cross section and $y_i = 2E_i/Q$, with $Q=\sqrt{s}$ the center-of-mass energy of the collision. 

Since we are computing the distribution of $\rho$ with a cutoff $\rho > 10^{-6}$, at LO we do not need to calculate virtual terms in the partonic cross sections in \Eq{eq:fixed-order-cross-section} and \Eq{eq:fixed-order-cross-section-sd} or to compute the $\mathcal{O}(\alpha_s)$ gluon track function. 
We only need the squared, spin-summed matrix element for the process $e^+e^-\rightarrow H \rightarrow ggg (gq\bar{q})$.
Since the Higgs couples to gluons through a quark loop, the lowest perturbative order at which this process can produce a non-zero jet mass is $\mathcal{O}(\alpha_s^2)$. 
In order to simplify the calculation, we work in the $m_t\rightarrow \infty$ limit, with an effective $Hgg$ coupling. 
The effective Lagrangian coupling the Higgs to gluons in this limit is~\cite{Inami:1982xt,Dawson:1990zj,Spira:1995rr}
\begin{equation}
\mathcal{L}_{eff} = \frac{\alpha_sA}{12\pi} G^{A}_{\mu \nu} G^{A,\mu \nu} \left( \frac{H}{v_{\rm EW}}\right)\,.
\end{equation}
where $v_{\rm EW}$ is the Higgs vacuum expectation value, $G^A_{\mu \nu}$ is the gluon field strength, and the $\mathcal{O}(\alpha_s^2)$ effective coupling constant $A=\frac{\alpha_s}{3\pi v_{\rm EW}}\left(1+\frac{11\alpha_s}{4\pi}\right)$.
The Feynman rules for the $Hgg$, $Hggg$, and $Hgggg$ vertices are proportional to the same tensors as the QCD gauge boson vertices. 
These are illustrated in \Fig{fig:feynman-rules}, where the relevant tensor structures are
\begin{align}
\mathcal{T}^{\mu \nu} &= g^{\mu \nu}p_1\cdot p_2 - p_1^\nu p_2^\mu \,,\\
\mathcal{T}^{\mu \nu \rho} &=  (p_1-p_2)^\rho g^{\mu \nu} + (p_2-p_3)^\mu g^{\nu\rho} + (p_3 - p_1)^\nu g^{\rho\mu}\,, \\ \nonumber
\mathcal{T}^{\mu \nu \rho\sigma}_{abcd} &= f^{abe}f^{cde}(g^{\mu \rho}g^{\nu\sigma} - g^{\mu \sigma}g^{\nu\rho}) + f^{ace}f^{bde}(g^{\mu \nu}g^{\rho\sigma} - g^{\mu \sigma}g^{\nu\rho})\\
& + f^{ade}f^{bce}(g^{\mu \nu}g^{\rho\sigma} - g^{\mu \rho}g^{\nu\sigma})\,.
\end{align}

\begin{figure}[t]
	\centering
	\subfloat[]{
		\begin{tikzpicture}[thick,scale=1.0]
			\draw[scalar] (0,0) -- (1,0);
			\draw[gluon,purple] (1,0) -- (2,1);
			\draw[gluon,purple] (1,0) -- (2,-1);
			\node[] at (2.2,1.2) {$p_1,\mu,a$};
			\node[] at (2.2,-1.2) {$p_2,\nu,b$};
			\draw[black,fill=black] (1,0) circle (.5ex);
			\node[] at (0,0.5) {$iA\delta^{ab}\mathcal{T}^{\mu\nu}$};
		\end{tikzpicture}
		\label{fig:Hgg}
	}
	\subfloat[]{
		\begin{tikzpicture}[thick,scale=1.0]
			\node[] at (-1,0) {};
			\draw[scalar] (0,0) -- (1,0);
			\draw[gluon,purple] (1,0) -- (1,1);
			\draw[gluon,purple] (1,0) -- (2,0);
			\draw[gluon,purple] (1,0) -- (1,-1);
			\node[] at (1,1.2) {$p_1,\mu,a$};
			\node[] at (2.5,0.4) {$p_2,\nu,b$};
			\node[] at (1,-1.2) {$p_3,\rho,c$};
			\draw[black,fill=black] (1,0) circle (.5ex);
			\node[] at (-0.5,0.5) {$-Ag_sf^{abc}\mathcal{T}^{\mu\nu\rho}$};
		\end{tikzpicture}
		\label{fig:Hggg}
	}
	\subfloat[]{
		\begin{tikzpicture}[thick,scale=1.0]
			\node[] at (-1.5,0) {};
			\draw[scalar] (-0.5,0) -- (1,0);
			\draw[gluon,purple] (1,0) -- (0,-1);
			\draw[gluon,purple] (1,0) -- (0,1);
			\draw[gluon,purple] (1,0) -- (2,1);
			\draw[gluon,purple] (1,0) -- (2,-1);
			\node[] at (0,1.2) {$p_1,\mu,a$};
			\node[] at (2.2,1.2) {$p_2,\nu,b$};
			\node[] at (0,-1.2) {$p_3,\rho,c$};
			\node[] at (2,-1.2) {$p_4,\sigma,d$};
			\draw[black,fill=black] (1,0) circle (.5ex);
			\node[] at (3,0) {$-iAg_s^2\mathcal{T}^{\mu\nu\rho\sigma}_{abcd}$};
		\end{tikzpicture}
		\label{fig:Hgggg}
	}
	\caption{\label{fig:feynman-rules}Feynman rules for $Hgg$, $Hggg$, and $Hgggg$ couplings in the $m_t\rightarrow \infty$ EFT, where all momenta are taken to be ingoing. }
\end{figure}
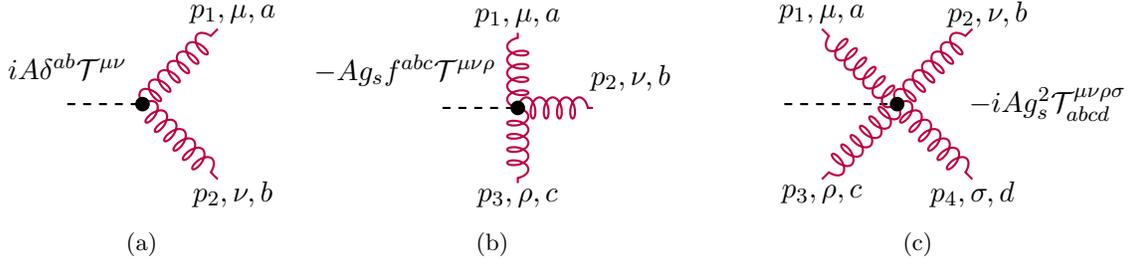

\begin{figure}[t]
	\centering
	\subfloat[]{
		\begin{tikzpicture}[thick,scale=1.0]
			\draw[particle] (-0.25,0.75) -- (0.5,0);
			\draw[antiparticle] (-0.25,-0.75) -- (0.5,0);			
			\draw[scalar] (0.5,0) -- (2,0);
			\draw[gluon,purple] (2,0) -- (3,1);
			\draw[gluon,purple] (2,0) -- (3,-1);
			\draw[gluon,purple] (3,1) -- (3.75,1.75);
			\draw[gluon,purple] (3,1) -- (3.75,0.25);
			\node[] at (4.25,1.75) {$k_1$};
			\node[] at (3.5,-1) {$k_2$};
			\node[] at (4.25,0.25) {$k_3$};
			\node[] at (1,-2) {};
			\node[] at (1,2) {};
		\end{tikzpicture}
		\label{fig:gluon-me-diag1}
	}
	\subfloat[]{
		\begin{tikzpicture}[thick,scale=1.0]
			\draw[particle] (-0.25,0.75) -- (0.5,0);
			\draw[antiparticle] (-0.25,-0.75) -- (0.5,0);
			\draw[scalar] (0.5,0) -- (2,0);
			\draw[gluon,purple] (2,0) -- (3,1);
			\draw[gluon,purple] (2,0) -- (3,-1);
			\draw[gluon,purple] (3,-1) -- (3.75,-0.25);
			\draw[gluon,purple] (3,-1) -- (3.75,-1.75);
			\node[] at (4.25,-0.25) {$k_3$};
			\node[] at (3.5,1) {$k_1$};
			\node[] at (4.25,-1.75) {$k_2$};
			\node[] at (1,-2) {};
			\node[] at (1,2) {};
		\end{tikzpicture}
		\label{fig:gluon-me-diag2}
	}
	\subfloat[]{
		\begin{tikzpicture}[thick,scale=1.0]
			\draw[particle] (-0.25,0.75) -- (0.5,0);
			\draw[antiparticle] (-0.25,-0.75) -- (0.5,0);
			\draw[scalar] (0.5,0) -- (2,0);
			\draw[gluon,purple] (2,0) -- (3.5,1);
			\draw[gluon,purple] (2,0) -- (3.5,0);
			\draw[gluon,purple] (2,0) -- (3.5,-1);
			\node[] at (4,1) {$k_1$};
			\node[] at (4,0) {$k_2$};
			\node[] at (4,-1) {$k_3$};
			\node[] at (1,-2) {};
			\node[] at (1,2) {};
		\end{tikzpicture}
		\label{fig:gluon-me-diag3}
	}
	
	\subfloat[]{
		\begin{tikzpicture}[thick,scale=1.0]
		\draw[particle] (-0.25,0.75) -- (0.5,0);
		\draw[antiparticle] (-0.25,-0.75) -- (0.5,0);
		\draw[scalar] (0.5,0) -- (2,0);
		\draw[gluon,purple] (2,0) -- (3,1);
		\draw[gluon,purple] (2,0) -- (3,-1);
		\draw[particle,orange] (3,1) -- (3.75,1.75);
		\draw[antiparticle,orange] (3,1) -- (3.75,0.25);
		\node[] at (4.25,1.75) {$k_1$};
		\node[] at (3.5,-1) {$k_2$};
		\node[] at (4.25,0.25) {$k_3$};
		\node[] at (1,-2) {};
		\node[] at (1,2) {};
		\end{tikzpicture}
		\label{fig:gluon-me-diag4}
	}
	\subfloat[]{
		\begin{tikzpicture}[thick,scale=1.0]
		\draw[particle] (-0.25,0.75) -- (0.5,0);
		\draw[antiparticle] (-0.25,-0.75) -- (0.5,0);
		\draw[scalar] (0.5,0) -- (2,0);
		\draw[gluon,purple] (2,0) -- (3,1);
		\draw[gluon,purple] (2,0) -- (3,-1);
		\draw[particle,orange] (3,-1) -- (3.75,-0.25);
		\draw[antiparticle,orange] (3,-1) -- (3.75,-1.75);
		\node[] at (4.25,-0.25) {$k_3$};
		\node[] at (3.5,1) {$k_1$};
		\node[] at (4.25,-1.75) {$k_2$};
		\node[] at (1,-2) {};
		\node[] at (1,2) {};
		\end{tikzpicture}
		\label{fig:gluon-me-diag5}
	}
	\caption{\label{fig:gluon-me-diags}Feynman diagrams for (a,b,c) $e^+e^- \rightarrow H \rightarrow ggg$ and (d,e) $e^+e^-\rightarrow H \rightarrow gq\bar{q}$ processes. }
\end{figure}
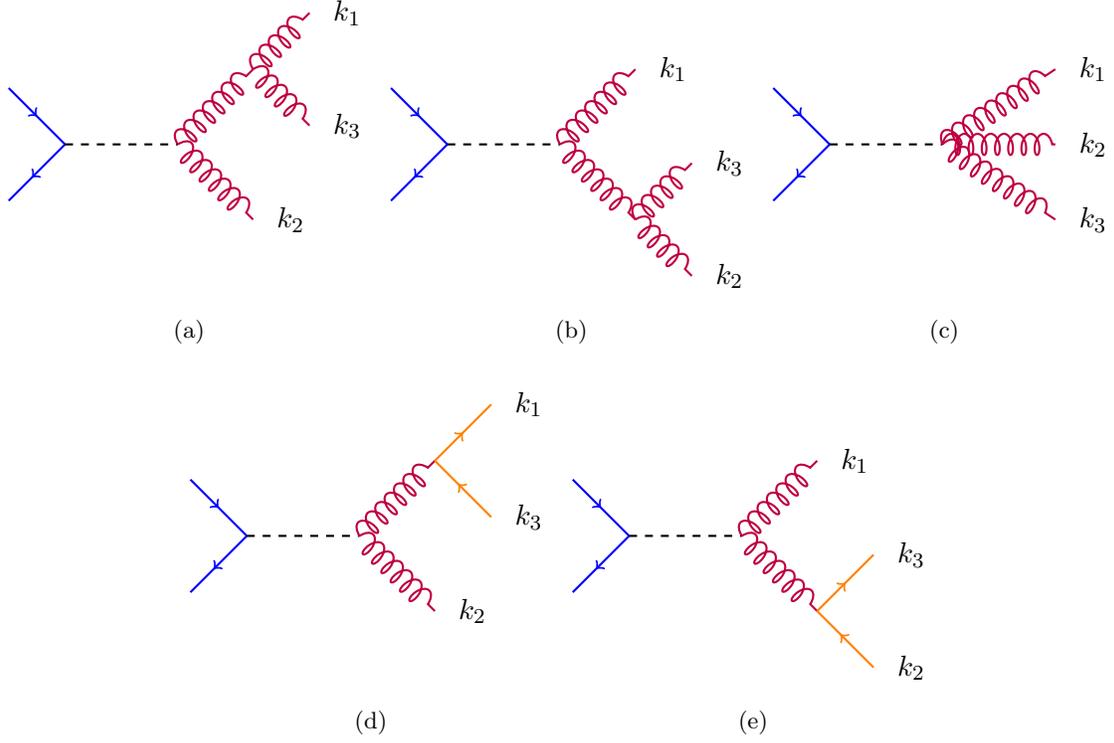

The diagrams contributing to the $e^+e^-\rightarrow H \rightarrow ggg(gq\bar{q})$ cross section are shown in \Fig{fig:gluon-me-diags}. 
Interference with the tree-level process $e^+e^- \rightarrow Z/\gamma^* \rightarrow gq\overline{q}$ is chirally suppressed at high energy and can be neglected.
We use the completeness relation for massless spin-one particles in the final state, with the fixed light-like vector $n^\mu = (1,0,0,0)$ and $k_i\cdot n = E_i$ to project onto only the two physical polarizations,
\begin{equation}
\sum_{\lambda} \epsilon^*_\mu (k,\lambda) \epsilon_{\nu}(k,\lambda) = -g_{\mu \nu} + \frac{k_\mu n_\nu + k_\nu n_\mu}{k\cdot n} - \frac{k_\mu k_\nu}{(k\cdot n)^2}\,.
\end{equation}
We rewrite the final-state momenta in terms of the energy fractions $y_i = 2E_i/\sqrt{s}$, 
\begin{align}
\begin{split}
k_i \cdot n &= \frac{1}{2}Qy_i\,, \\
k_i \cdot k_j &= \frac{Q^2}{2}(1-y_k)\,, \\
2 &= y_1 + y_2 + y_3\,. 
\end{split}
\end{align}
Using these variables, we write the squared, spin-summed matrix elements.
First define
\begin{align}
\nonumber
f(y_1,y_2) = 256\pi^2 \bigg(& y_1^5 + y_1^4(3y_2-5) + y_1^3(5y_2^2 - 12y_2 + 10) + y_1^2(18y_2 - 10)\\ \nonumber
&- 15 y_1^2y_2^2 -14y_1y_2+ 6y_1 - 2 +6y_2 + y_2^2(18y_1-10)\\
& + y_2^3(5y_1^2 - 12y_1 + 10) + y_2^4(3y_1-5) + y_2^5\bigg) \,.
\end{align}
Then the result for the $e^+e^-\rightarrow H \rightarrow ggg$ cross section is
\begin{align}
\begin{split}
\sum \vert \mathcal{M}_1\vert^2 &=  \sigma_0 \alpha_s C_A\left(\frac{1-y_1}{y_1^2(1-y_2)(2-y_1-y_2)^2}\right)f(y_1,y_2)\,,\\
\sum \vert \mathcal{M}_3\vert^2 &= \sigma_0\alpha_s  C_A\left( \frac{4(1-y_1)(1-y_2)}{y_1^2y_2^2(2-y_1-y_2)^2}\right)f(y_1,y_2)\,,\\
\mathcal{M}_1\mathcal{M}_2^* &= \sigma_0 \alpha_s C_A \left(\frac{1 }{y_1y_2(2-y_1-y_2)^2}\right)f(y_1,y_2)\,,\\
\mathcal{M}_1\mathcal{M}_3^* &= -\sigma_0 \alpha_s C_A\left( \frac{2(1-y_1)}{y_1^2y_2(2-y_1-y_2)^2}\right)f(y_1,y_2)\,,
\end{split}
\end{align}
and for the $e^+e^-\rightarrow H\rightarrow gq\bar{q}$ cross section,
\begin{equation}
\sum \vert \mathcal{M}_4\vert^2 =32\pi^2 \sigma_0  \left(\frac{(y_1+y_2-1)^2+(1-y_1)^2}{1-y_2}\right) \,.
\end{equation}
The remaining matrix elements can be obtained by exchanging $y_1 \leftrightarrow y_2$.


\subsection{Comparison of Matching Schemes}
\label{app:details-matching-comp}

There are multiple possible schemes that could be used to match resummed and fixed-order cross sections~\cite{Catani:1992ua,Catani:1998sf,Jones:2003yv,Gehrmann:2008kh}. 
In \Sec{sec:calculation-matching-fo}, we chose the log-$R$ matching scheme defined by \Eq{eq:matching-logR}. 

Another possible choice is a simple multiplicative matching scheme for the (normalized) differential distributions~\cite{Marzani:2017mva,Marzani:2017kqd},
\begin{equation}
\label{eq:matching-mult}
\frac{1}{\sigma}\frac{\text{d}\sigma^{\rm mult}_{\mbox{\tiny NLL+LO}}}{\text{d}\rho} = \left(\frac{1}{\sigma}\frac{\text{d}\sigma_{\mbox{\tiny NLL}}}{\text{d}\rho}\right)  \left(\frac{1}{\sigma_0}\frac{\text{d}\sigma_{\mbox{\tiny LO}}}{\text{d}\rho}\right) \left(\frac{1}{\sigma} \frac{\text{d}\sigma_{\mbox{\tiny NLL,}\alpha}}{\text{d}\rho}\right)^{-1}  \,.
\end{equation}
This scheme has the advantage that it automatically enforces that the $\rho$ distribution vanish as $\rho\rightarrow 0$ and above the LO kinematic limit \Eq{eq:rho-max}, $\rho_{\mbox{\scriptsize max,\tiny LO}} \approx 0.23$. 

Another common matching scheme is the additive $R$ matching scheme for the cumulative distributions~\cite{Catani:1992ua},
\begin{equation}
\label{eq:matching-add}
\Sigma^{\rm add}_{\mbox{\tiny NLL+LO}} = \Sigma_{\mbox{\tiny NLL}} + \Sigma_{\mbox{\tiny LO}} - \Sigma_{\mbox{\tiny NLL},\alpha}  \,.
\end{equation}
In order to get the kinematic endpoints correct, we must use the modified $R$ matching scheme. 
We first rewrite the resummed distribution in the form \cite{Catani:1992ua}
\begin{equation}
\label{eq:nll-cumulative-dist-alternate}
\Sigma(\rho) = C(\alpha_s) \exp \left( Lg_1\left(\alpha_s L\right) + g_2\left(\alpha_s L\right)+ \alpha_s g_3\left(\alpha_s L\right)+ \ldots\right) + D(\rho,\alpha_s)\,.
\end{equation}
Here $L = \log(1/\rho)$, $C(\alpha_s)$ is a perturbatively calculable $\rho$-independent coefficient, and the function $D(\rho,\alpha_s) \rightarrow 0$ as $\rho\rightarrow 0$. 
The function $g_1$ resums leading logarithms, $g_2$ resums next-to-leading logarithms, etc. 
These functions have the perturbative expansions
\begin{align}
\label{eq:factorization-expansion}
C(\alpha_s) &= 1+\sum_{n=1}^\infty C_n\left(\frac{\alpha_s}{2\pi}\right)^n\,,\\
D(\rho,\alpha_s) &= \sum_{n=1}^\infty \left(\frac{\alpha_s}{2\pi}\right)^n D_n(\rho)\,,\\
g_n(\alpha_s L) &= \sum_{k=1}^\infty G_{k,k+2-n}\left(\frac{\alpha_s}{2\pi}\right)^k L^{k+2-n}\,.
\end{align}
In order to enforce the constraint that the matched cross section vanish at the upper kinematic boundary, in the $R$ scheme we must again make the change of variables
\begin{equation}
\frac{1}{\rho} \rightarrow \frac{1}{\rho} - \frac{1}{\rho_{\mbox{\scriptsize max,\tiny LO}}} + e^{-B_i}\,,
\end{equation}
in addition to the replacements \cite{Catani:1992ua}
\begin{align}
G_{11} &\rightarrow G_{11}\left(1-\frac{\rho}{\rho_{\mbox{\scriptsize max,\tiny LO}}}\right)\,,\\
\exp(Lg_1 + g_2 + \ldots) &\rightarrow \exp(Lg_1 + g_2 + \ldots)  \times \exp\left(-\frac{\rho}{\rho_{\mbox{\scriptsize max,\tiny LO}}}G_{11}\alpha_s \log\left(\frac{1}{\rho}\right) \right)\,,
\\
D(\rho,\alpha_s) &\rightarrow  D(\rho,\alpha_s) + \left(1-\frac{\rho}{\rho_{\mbox{\scriptsize max,\tiny LO}}}G_{11}\alpha_s \log\left(\frac{1}{\rho}\right) \right)\,.
\end{align}
The term $G_{11}$ can be read off from \Eq{eq:approx-fixed-coupling-NLL-cumulative},
\begin{equation}
G_{11} = \frac{C_i}{\pi} \int_0^1 \text{d}x_j\, T_j(x_j) \,\left[ f^{g,0}(x_j\rho,1) \left(B_i - \log(x_j)\right) +f^{g,1}(x_j\rho,1)\right]\,.
\end{equation}
Due to the $\rho$ dependence in the lower endpoints of the track function logarithmic moments, this term technically includes power-suppressed terms beyond NLL order. 

\begin{figure}[t]
	\centering
	\subfloat[]{
		\includegraphics[width=0.45\textwidth]{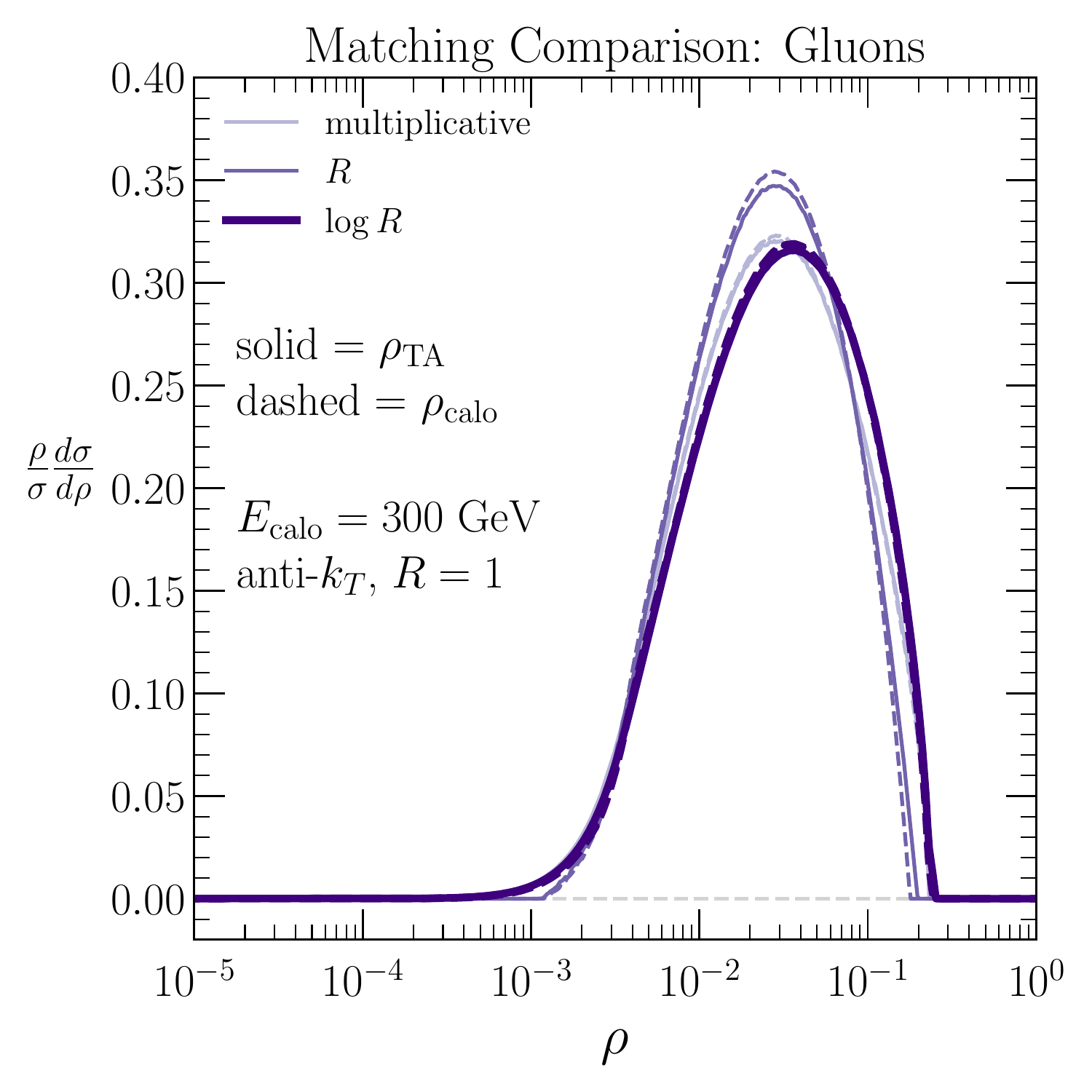}
		\label{fig:matching-gluon-compare}
	}
	\subfloat[]{
		\includegraphics[width=0.45\textwidth]{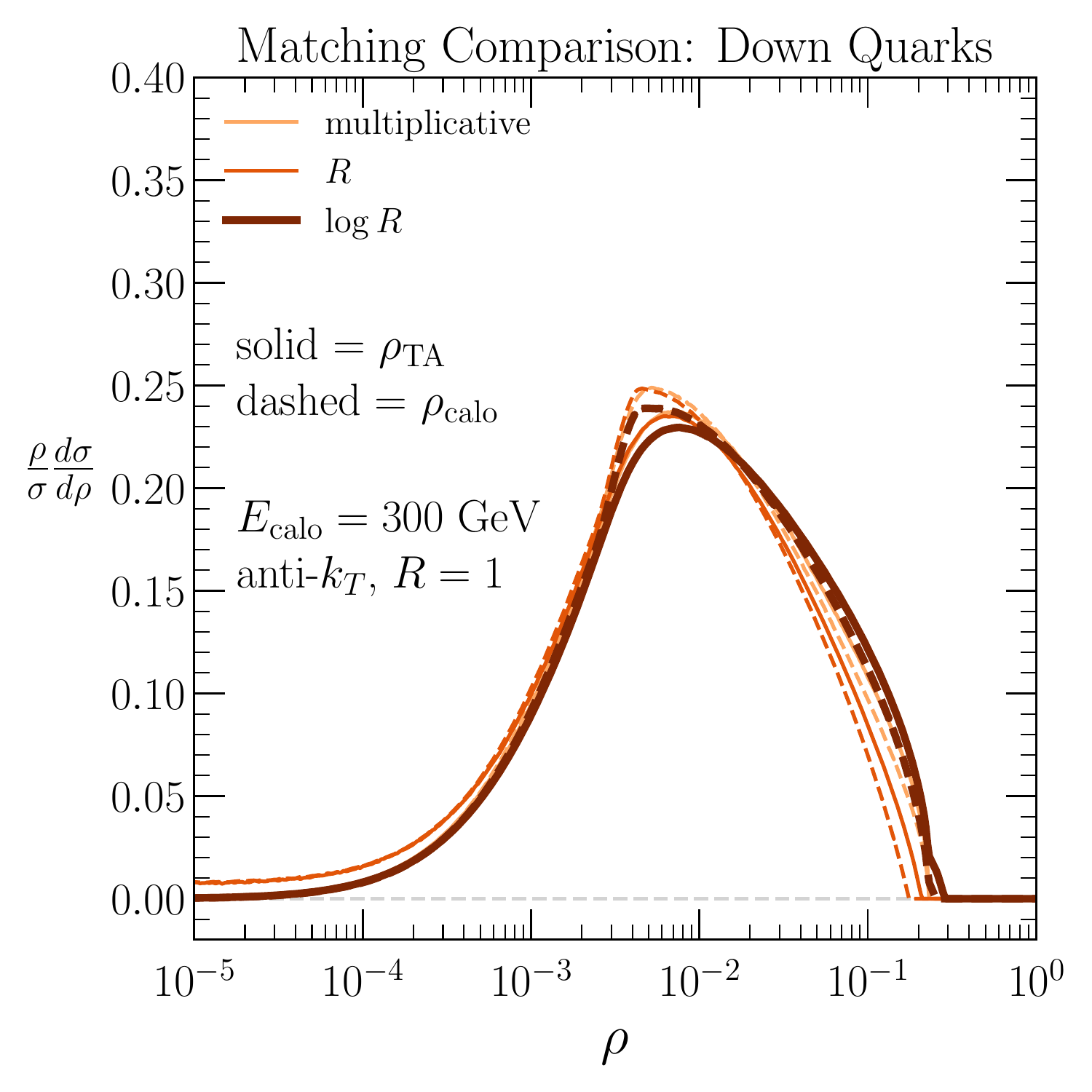}
		\label{fig:matching-quark-compare}
	}
	\caption{\label{fig:matching-compare} Comparison of fixed-order/resummed matching schemes for (a) gluon jets and (b) down-quark jets.}
\end{figure}

In \Fig{fig:matching-compare}, we show results for the calorimeter and track-assisted mass at NLL+LO order computed using three different matching schemes: the log-$R$ scheme, the multiplicative scheme, and the additive $R$ scheme. 
The log-$R$ scheme produces distribution with a pronounced bulge on the high-$\rho$ side as compared to the other two schemes. 
This feature is responsible for the corresponding high-$\rho$ bulge in the NLL+LO+NP distributions, which were compared to results from $\textsc{Vincia}$ in \Fig{fig:np-shape-fcn}. 

\section{Alternative Soft-Drop Implementation}
\label{app:alternate-sd-implementation}


Since soft drop changes the jet mass spectrum by design, we argued in \Sec{sec:soft-drop} that the appropriate distribution to compare with the soft-drop groomed GTAM distribution was the soft-drop groomed jet mass distribution.
To get the closest match in these distributions, the factor of $p_{T,\rm calo}$ in \Eq{eq:tam-generalized} should then be the total calorimeter $p_T$ of the soft-drop groomed jet.
This is the reason why we applied soft drop before computing the track mass in \Sec{sec:soft-drop}, such that the charged particles removed by grooming are the same in $p_{T,\rm calo}$ and $p_{T,\rm track}$.

On the other hand, applying soft-drop grooming to the calorimeter jet introduces angular resolution issues for neutral particles in the soft-drop procedure.
Thus, an alternate way to calculate GTAM with soft-drop grooming is to first recluster just the charged particles into a jet and then groom this charged-only jet.
The advantage of this approach is that $M_{\rm track}$ is calculated using only charged particles from beginning to end, including the soft-drop declustering step.
In this appendix, we describe analytic calculations for this alternative possibility.

At parton level, the soft-drop condition becomes
\begin{equation}
\label{eq:sd-groom-tracked}
\frac{\min(x_1 p_{T,1},x_2p_{T,2})}{x_1p_{T,1}+x_2p_{T,2}} > z_{\rm cut} \left(\frac{\Delta R_{12}}{R_0}\right)^\beta\,,
\end{equation}
\begin{equation}
\label{eq:sd-groom-tracked-sc}
\overset{s.c.}{\Longrightarrow} \frac{x_kz}{x_j(1-z)+x_kz} \approx \frac{x_kz}{x_j}> z_{\rm cut} \left(\frac{\theta}{R}\right)^\beta\,. 
\end{equation}
\Eq{eq:sd-groom-tracked-sc} follows from \Eq{eq:sd-groom-tracked} in the soft-collinear limit for a splitting $i\rightarrow jk$, where parton $k$ carries a fraction $z\ll \tfrac{1}{2}$ of parton $i$'s original momentum. 
For a NLL calculation, only the leading order in an expansion in $z$ is required. 
This changes the grooming $\Theta$-function in the radiator, \Eq{eq:sd-radiator}, to
\begin{equation}
\Theta\left(\frac{x_kz}{x_j} - z_{\rm cut} \left(\frac{\theta}{R}\right)^\beta \right) \,.
\end{equation}
Using this $\Theta$-function in the radiator, we can derive the explicit form of the radiators for $\beta > 0$ and $\beta = 0$ in the fixed-coupling approximation. 
If $\rho > z_{\rm cut}$, soft-drop grooming is not active, and the result is the same as the ungroomed case. For $\rho < z_{\rm cut}$:
\begin{align}
\label{eq:track-first-radiator-betagt0}
(\beta > 0) &:\\ \nonumber
R_{\mbox{\tiny TA}}(\rho,x_j) &= \frac{\alpha_sC_i}{\pi} \bigg\{ \tfrac{1}{2}f^{g,0}(x^*,1) \tfrac{\beta}{2+\beta} \ln^2\left(\tfrac{1}{\rho}\right) + \ln\left(\tfrac{1}{\rho}\right)f^{g,0}(x^*,1) \big[B_i-\ln(x_j)\\ \nonumber
&\hspace{1.5cm} - \tfrac{2}{2+\beta}\ln(z_{\rm cut})  \big] + \ln\left(\tfrac{1}{\rho}\right)f^{g,1}(x^*,1)\bigg\}\,, \\
\label{eq:track-first-radiator-betaeq0}
(\beta = 0) &:\\ \nonumber
R_{\mbox{\tiny TA}}(\rho,x_j) &= \frac{\alpha_sC_i}{\pi} \ln\left(\tfrac{1}{\rho}\right) \big\{f^{g,0}(x_jz_{\rm cut},1) \left[B_i - \ln(x_j) - \ln(z_{\rm cut})\right]\\
&\hspace{2.7cm} + f^{g,1}(x_jz_{\rm cut},1) + G_i(x_jz_{\rm cut}) \big\}
\end{align}
where we define,
\begin{align}
x^* &= x_j\rho^{\tfrac{\beta}{2+\beta}}z_{\rm cut}^{\tfrac{2}{2+\beta}} \\
G_q(a) &= -\tfrac{1}{4}\tilde{g}(a,2) + \tilde{g}(a,1)\,,\\
G_g(a) &=  - \tfrac{3}{4}\tilde{g}(a,3) + \tfrac{9}{8}\tilde{g}(a,2) - \tfrac{3}{8}\tilde{g}(a,1) - B_g \big[\tilde{g}(a,3) - \tfrac{3}{2}\tilde{g}(a,2) + \tfrac{3}{2}\tilde{g}(a,1)\big]\,,\\
\tilde{g}(a,n) &= a^n \int_{a}^1\text{d}x\, x^{-n}\,T_g(x), \ \ \ \lim_{a\rightarrow 0} \tilde{g}(a,n) = 0\,.
\end{align}
Note that the quantity $\tilde{g}(a,n)$ is well defined as $a\rightarrow 0$, so there is no divergence as $x_j\rho \rightarrow 0$. 
The radiator derivatives appearing in the multiple-emissions prefactor are
\begin{align}
\label{eq:track-first-rderiv-betagt0}
(\beta > 0):& \ \ R_{\mbox{\tiny TA}}'(\rho,x_j) = \frac{\alpha_sC_i}{\pi}f^{g,0}\left(x^*,1\right)\tfrac{\beta}{2+\beta} \big[\ln(\tfrac{1}{\rho})- \tfrac{2}{\beta}\ln(z_{\rm cut})\big]\,, \\ 
\label{eq:track-first-rderiv-betaeq0}
(\beta = 0):& \ \ R_{\mbox{\tiny TA}}'(\rho,x_j) = \frac{\alpha_sC_i}{\pi}\big[ - f^{g,0}\left(x_jz_{\rm cut},1\right)\ln(z_{\rm cut})\big] \,.
\end{align}
As in \Sec{sec:soft-drop}, the terms proportional to $\ln z_{\rm cut}$ are formally beyond NLL order, but they make an important numerical contribution in the region $\rho \approx z_{\rm cut}$. 

\begin{figure}
	\centering
	\subfloat[]{
		\includegraphics[width=0.45\textwidth]{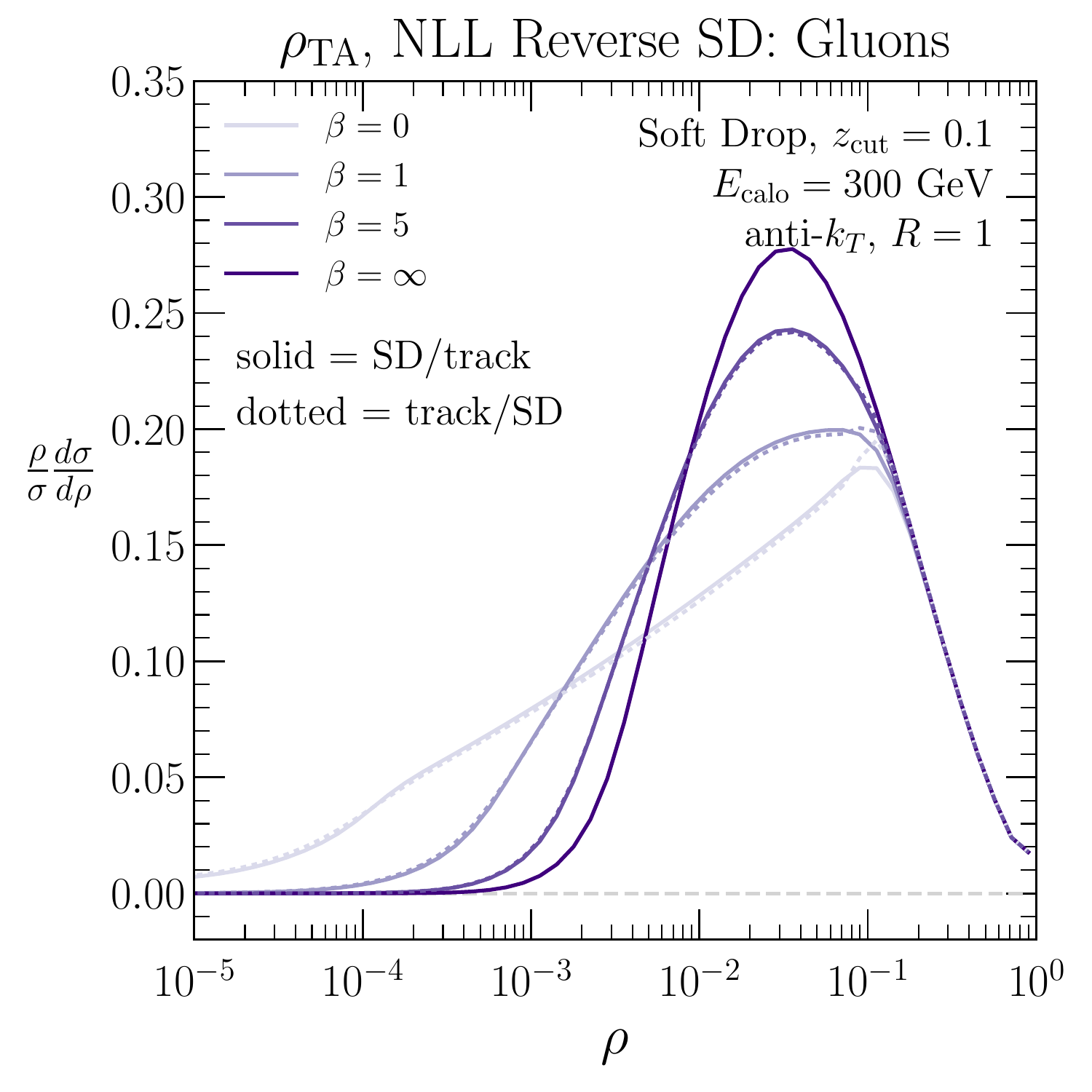}
		\label{fig:sdcomp-a}
	}
	\subfloat[]{
		\includegraphics[width=0.45\textwidth]{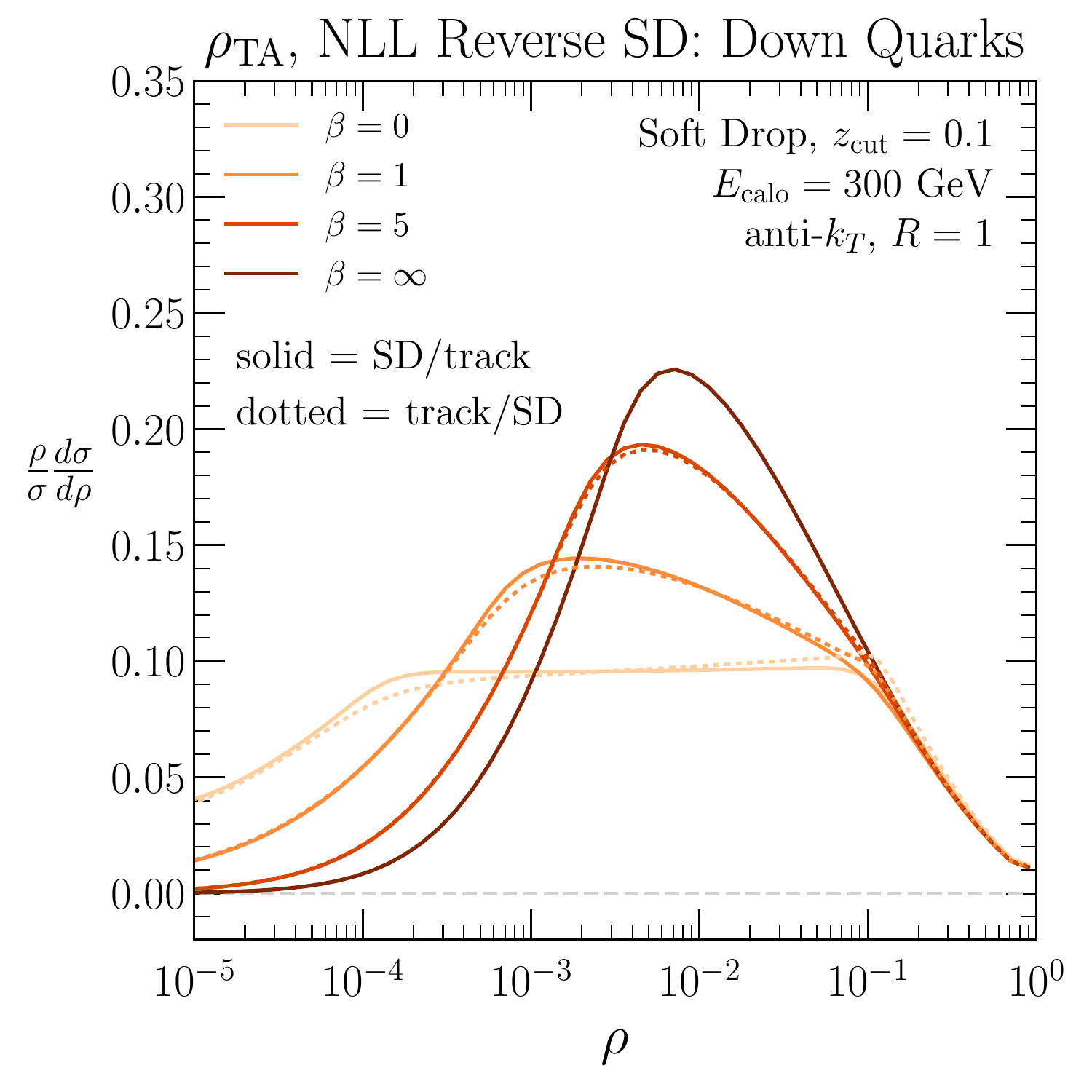}
		\label{fig:sdcomp-b}
	}
	\caption{\label{fig:sdcomp}Comparison between track-assisted mass with soft-drop grooming for (a) gluon jets and (b) down-quark jets. The solid curves were computed by first grooming the jet and then restricting to charged particles. The dotted lines were computed by reclustering only the charged particles and then grooming.}
\end{figure}

NLL resummed distributions for this alternate soft drop implementation are plotted in \Fig{fig:sdcomp} for $\rho_{\mbox{\tiny TA}}$ of groomed jets with $z_{\rm cut}= 0.1$ and $\beta = 0,1,5,$ and $\infty$. 
It is clear that changing the order of grooming and restricting to charged particles has very little quantitative impact. 

\bibliography{gtam}
\bibliographystyle{JHEP}


\end{document}